\documentclass[acmsmall]{acmart}
\acmJournal{PACMPL}
\acmVolume{1}
\acmNumber{Rutgers Computer Science Technical Report} % CONF = POPL or ICFP or OOPSLA
\acmArticle{753}
\acmYear{2020}
\acmMonth{7}
\acmDOI{} % \acmDOI{10.1145/nnnnnnn.nnnnnnn}
\startPage{1}
\setcopyright{rightsretained}

\usepackage{booktabs}
\usepackage{subcaption}
\usepackage{xspace}
\usepackage{amsthm}
\usepackage[linesnumbered]{algorithm2e}
\usepackage{empheq}
\usepackage{pifont}

\newcommand{\myblue}[1]{{#1}}
\newcommand{\tool}{\textsc{RLibm}\xspace}
\newcommand{\ourlibm}{\textsc{RLibm}\xspace}

\newcommand{\eg}{\emph{e.g.}\xspace}
\newcommand{\ie}{\emph{i.e.}\xspace}

\newcommand{\cmark}{\ding{51}}%
\newcommand{\xmark}{\ding{55}}%

\newtheorem{theorem}{Theorem}[section]

\theoremstyle{definition}
\newtheorem{definition}[theorem]{Definition}

\begin{document}

%% Title information
\title[A Novel Approach to Generate Correctly Rounded Math Libraries for
  New Floating Point Representations]{A Novel Approach to Generate Correctly
  Rounded Math Libraries for New Floating Point Representations}

\author{Jay P. Lim}
\affiliation{
  \institution{Rutgers University}
}
\email{jpl169@cs.rutgers.edu}

\author{Mridul Aanjaneya}
\affiliation{
  \institution{Rutgers University}
}
\email{mridul.aanjaneya@rutgers.edu}

\author{John Gustafson}
\affiliation{
  \institution{National University of Singapore}
}
\email{john.gustafson@nus.edu.sg}

\author{Santosh Nagarakatte}
\affiliation{
  \institution{Rutgers University}
}
\email{santosh.nagarakatte@cs.rutgers.edu}

\begin{abstract}
  Given the importance of floating point~(FP) performance in numerous
  domains, several new variants of FP and its alternatives have been
  proposed (\eg, \texttt{Bfloat16}, \texttt{TensorFloat32}, and
  \texttt{posits}). These representations do not have correctly
  rounded math libraries. Further, the use of existing FP libraries
  for these new representations can produce incorrect results.
  This paper proposes a novel approach for generating polynomial
  approximations that can be used to implement correctly rounded math
  libraries.
\myblue{Existing methods generate polynomials that approximate the real
  value of an elementary function $f(x)$ and produce wrong results due
  to approximation errors and rounding errors in the implementation.
  In contrast, our approach generates polynomials that approximate the
  correctly rounded value of $f(x)$ (\ie, the value of $f(x)$ rounded
  to the target representation). It provides more margin to identify
  efficient polynomials that produce correctly rounded results for all
  inputs. We frame the problem of generating efficient polynomials
  that produce correctly rounded results as a linear programming
  problem.
  Using our approach, we have developed correctly rounded, yet faster,
  implementations of elementary functions for multiple target
  representations.}
\end{abstract}

\maketitle
\section{Introduction}

\textbf{Approximating real numbers.} Every programming language has
primitive data types to represent numbers. The floating point~(FP)
representation, which was standardized with the IEEE-754
standard~\cite{ieee754}, is widely used in mainstream languages to
approximate real numbers. For example, every number in JavaScript is a
FP number! There is an ever-increasing need for improved FP
performance in domains such as machine learning and high performance
computing~(HPC). Hence, several new variants and alternatives to FP
have been proposed recently such as
\texttt{Bfloat16}~\cite{Tagliavini:bfloat:date:2018},
posits~\cite{Gustafson:online:2017:posit, Gustafson:sfi:2017:beating},
and \texttt{TensorFloat32}~\cite{nvidia:tensorfloat:online:2020}.

\texttt{Bfloat16}~\cite{Tagliavini:bfloat:date:2018} is a 16-bit FP
representation with 8-bits of exponent and 7-bits for the fraction. It
is already available in Intel FPGAs~\cite{intel:nervana:online:2019}
and Google TPUs~\cite{Wang:tpu:online:2019}.  \texttt{Bfloat16}'s
dynamic range is similar to a 32-bit \texttt{float} but has lower
memory traffic and footprint, which makes it appealing for neural
networks~\cite{Kalamkar:bfloatai:arxiv:2019}.  Nvidia's
\texttt{TensorFloat32}~\cite{nvidia:tensorfloat:online:2020} is a
19-bit FP representation with 8-bits of exponent and 10-bits for the
fraction, which is available with Nvidia's Ampere
architecture. \texttt{TensorFloat32} provides the dynamic range of a
32-bit \texttt{float} and the precision of \texttt{half} data type
(\ie, 16-bit float), which is intended for machine learning and HPC
applications.  In contrast to FP,
posit~\cite{Gustafson:online:2017:posit, Gustafson:sfi:2017:beating}
provides tapered precision with a fixed number of bits. Depending on
the value, the number of bits available for
representing the fraction can vary.  Inspired by posits, a
tapered precision log number system has been shown to be effective
with neural
networks~\cite{Johnson:online:2018:facebook,Bernstein:learning:arxiv:2020}.

\textbf{Correctly rounded math libraries.} Any number system that
approximates real numbers needs a math library that provides
implementations for elementary
functions~\cite{Muller:elemfunc:book:2005} (\ie, $log(x)$, $exp(x)$,
$sqrt(x)$, $sin(x)$).
The recent IEEE-754 standard recommends (although it does not require)
that the programming language standards define a list of math library
functions and implement them to produce the correctly rounded
result~\cite{ieee754}. Any application using an erroneous math library
will produce erroneous results.

A correctly rounded result of an elementary function $f$ for an input
$x$ is defined as the value produced by computing the value of $f(x)$
with real numbers and then rounding the result according to the
rounding rule of the target representation. Developing a correct math
library is a challenging task.
Hence, there is a large body of work on accurately approximating
elementary functions~\cite{Lefevre:toward:tc:1998,
  Chevillard:sollya:icms:2010, Brisebarre:maceffi:toms:2006,
  Chevillard:infnorm:qsic:2007, Chevillard:ub:tcs:2011,
  Olga:metalibm:icms:2014, Brunie:metalibm:ca:2015,
  Jeannerod:sqrt:tc:2011, Bui:exp:ccece:1999,
  Gustafson:unum:2020:online, Lim:cordic:cf:2020}, verifying the
correctness of math libraries~\cite{Dinechin:gappaverify:sac:2006,
  Dinechin:verify:tc:2011, Daumas:proofs:arith:2005,
  Lee:verify:popl:2018, Harrison:expproof:amst:1997,
  Harrison:verifywithHOL:tphol:1997, Boldo:reduction:toc:2009,
  Sawada:verify:acl:2002}, and repairing math libraries to increase
the accuracy ~\cite{Xin:repairmlib:popl:2019}.
\myblue{There are a few correctly rounded math libraries for
  \texttt{float} and \texttt{double} types in the IEEE-754
  standard~\cite{IBM:MathLib:online:2008,
    Abraham:fastcorrect:toms:1991,Sun:libmcr:online:2008,
    Daramy:crlibm:spie:2003,Fousse:toms:2007:mpfr}. Widely used math
  libraries (\eg, \texttt{libm} in \texttt{glibc} or Intel's math
  library) do not produce correctly rounded results for all inputs.}

\textbf{New representations lack math libraries.} The new FP
representations currently do not have math libraries specifically
designed for them. One stop-gap alternative is to promote values from
new representations to a \texttt{float}/\texttt{double} value and use
existing FP libraries for them. For example, we can convert a
\texttt{Bfloat16} value to a 32-bit \texttt{float} and use the FP math
library. However, this approach can produce wrong results for the
\texttt{Bfloat16} value even when we use the correctly rounded
\texttt{float} library~(see Section~\ref{sec:background:motivation}
for a detailed example). This approach also has suboptimal performance
as the math library for \texttt{float}/\texttt{double} types probably
uses a polynomial of a large degree with many more terms than
necessary to approximate these functions.

\textbf{Prior approaches for creating math libraries.} \myblue{Most prior
approaches use minimax approximation methods~(\ie, Remez
algorithm~\cite{Remes:algorithm:1934} or Chebyshev
approximations~\cite{Trefethen:chebyshev:book:2012}) to generate
polynomials that have the smallest error compared to the real value of
an elementary function.}
Typically, range reduction techniques
are used to reduce the input domain such that the polynomial only
needs to approximate the elementary function for a small input domain.
Subsequently, the result of the polynomial evaluation on the small
input domain is adjusted to produce the result for the entire input
domain, which is known as output compensation. Polynomial evaluation,
range reduction, and output compensation are implemented in some
finite representation that has higher precision than the target
representation. The approximated result is finally rounded to the
target representation.

When the result of an elementary function $f(x)$ with reals is
extremely close to the rounding-boundary (\ie, $f(x)$ rounds to a
value $v_1$ but $f(x) + \epsilon$ rounds to a different value $v_2$
for very small value $\epsilon$), then the error of the polynomial
must be smaller than $\epsilon$ to ensure that the result of the
polynomial produces the correctly rounded
value~\cite{Lefevre:worstcase:arith:2001}. This probably necessitates
a polynomial of a large degree with many terms. \myblue{Further, there can be
round-off errors in polynomial evaluation with a finite precision
representation. Hence, the result produced may not be the correctly
rounded result.}

\begin{figure}
  \centerline{\includegraphics[width=0.95\linewidth]{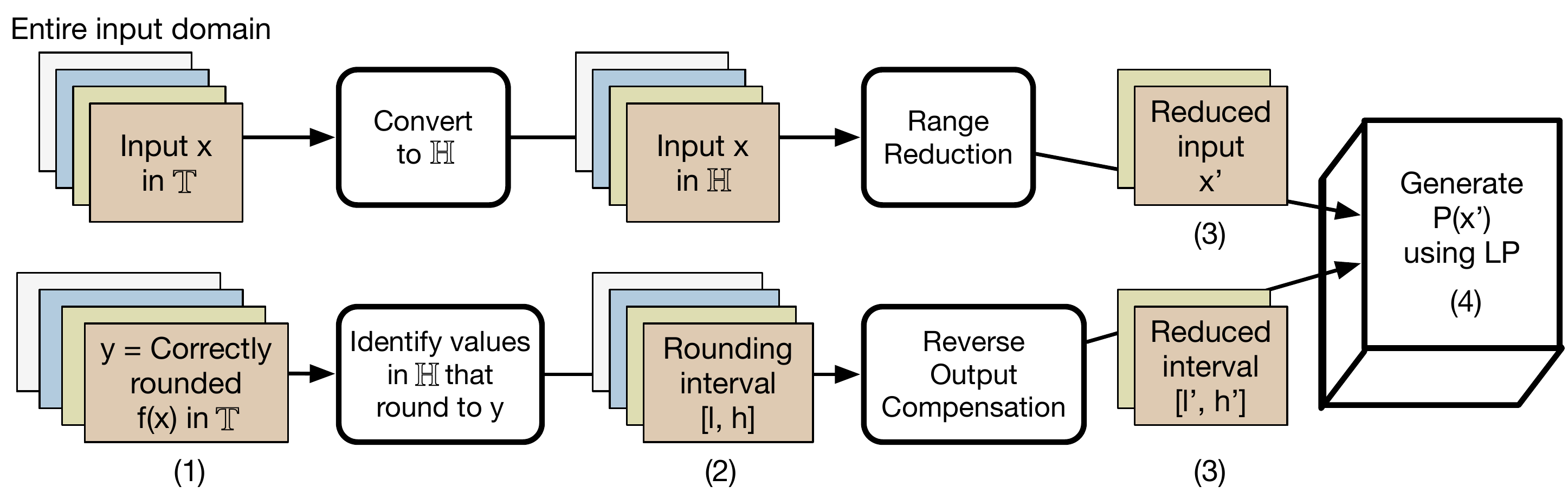}}
  \caption{Our approach to generate correctly rounded elementary
    functions for a target representation~($\mathbb{T}$). The math
    library is implemented in representation~$\mathbb{H}$. The goal is
    to synthesize a polynomial $P(x')$ using linear programming such
    that the final result after range reduction and output
    compensation is the correctly rounded result of $f(x)$ in
    $\mathbb{T}$. (1) For each input $x$ in $\mathbb{T}$, we compute
    the correctly rounded value of $f(x)$ (denoted as $y$) using an
    oracle. (2) Based on $y$, we identify an interval ($[l, h]$) where
    all values in the interval round to $y$. (3) Then, we compute the
    reduced input $x'$ using range reduction and the reduced interval
    ($[l', h']$) such that when the output of the polynomial on the
    reduced input $x'$ is adjusted (\ie, output compensation), it
    produces the result for the original input and it is in $[l,
      h]$. (4) Finally, we synthesize $P(x')$ that produces a value in
    the reduced interval $[l', h']$ for each reduced input $x'$.}
  \label{fig:workflow}
\end{figure}

\textbf{Our approach.} This paper proposes a novel approach to
generate correctly rounded implementations of elementary functions by
framing it as a linear programming problem. In contrast to prior
approaches that generate polynomials by minimizing the error compared
to the real value of an elementary function $f(x)$, we propose to
generate polynomials that directly approximate the correctly rounded
value of $f(x)$ inspired by the Minefield
approach~\cite{Gustafson:unum:2020:online}.  Specifically, we identify
an interval of values for each input that will result in a correctly
rounded output and use that interval to generate the polynomial
approximation.
For each input $x_i$, we use an oracle to generate an interval $[l_i,
  h_i]$ such that all real values in this interval round to the
correctly rounded value of $f(x_i)$. Using these intervals, we can
subsequently generate a set of constraints, which is given to a linear
programming solver, to generate a polynomial that computes the
correctly rounded result for all inputs.
The interval $[l_i, h_i]$ for correctly rounding the output of input
$x_i$ is larger than $[f(x_i) - \epsilon, f(x_i) + \epsilon]$ where
$\epsilon$ is the maximum error of the polynomial generated using
prior methods. Hence, our approach has larger freedom to generate
polynomials that produce correctly rounded results and also provide
better performance.

\textbf{Handling range reduction.} Typically, generating polynomials
for a small input domain is easier than a large input domain. Hence,
the input is reduced to a smaller domain with range
reduction. Subsequently, polynomial approximation is used for the
reduced input. The resulting value is adjusted with output
compensation to produce the final output. For example, the input
domain for $log_{2}(x)$ is $(0, \infty)$.
Approximating this function with a polynomial is much easier over the
domain $[1, 2)$ when compared to the entire input domain $(0,
  \infty)$.
Hence, we range reduce the input $x$ into $z$ using $x = z * 2^{e}$,
where $z\in[1, 2)$ and $e$ is an integer.
We compute $y' = log_{2}(z)$ using our polynomial for the domain $[1,
  2)$. We compute the final output $y$ using the range reduced output
  $y'$ and the output compensation function, which is $y$ = $y' + e$.
  Polynomial evaluation, range reduction, and output compensation are
  performed with a finite precision representation (\eg,
  \texttt{double}) and can experience numerical errors. Our approach
  for generating correctly rounded outputs has to consider the
  numerical error with output compensation.
To account for rounding errors with range reduction and output
compensation, we constrain the output intervals that we generated for
each input $x$ in the entire input domain~(see
Section~\ref{sec:approach}).
When our approach generates a polynomial, it is guaranteed that the
polynomial evaluation along with the range reduction and output
compensation can be implemented with finite precision to produce a
correctly rounded result for all inputs of an elementary function
$f(x)$. Figure~\ref{fig:workflow} pictorially provides an overview of
our methodology.
  
\myblue{\textbf{\tool.} We have developed a collection of correctly
  rounded math library functions, which we call \tool, for
  \texttt{Bfloat16}, posits, and floating point using our
  approach. \tool is open source~\cite{rlibm,rlibmgenerator}.
  Concretely, \tool contains twelve elementary functions for
  \texttt{Bfloat16}, eleven elementary functions for 16-bit posits,
  and $log_{2}(x)$ function for a 32-bit \texttt{float} type. We have
  validated that our implementation produces the correctly rounded
  result for all inputs. In contrast, \texttt{glibc}'s $log_{2}(x)$
  function for a 32-bit float produces wrong results for more than
  fourteen million inputs. Similarly, Intel's math library also
  produces wrong results for 276 inputs.  We also observed that
  re-purposing \texttt{glibc}'s and Intel's float library for
  \texttt{Bfloat16} produces a wrong result for $10^x$.}

\myblue{Our library functions for \texttt{Bfloat16} are on average
  $2.02\times$ faster than the \texttt{glibc}'s \texttt{double}
  library and $1.39\times$ faster than the \texttt{glibc}'s
  \texttt{float} library. Our library functions for \texttt{Bfloat16}
  are also $1.44\times$ and $1.30\times$ faster than the Intel's
  \texttt{double} and \texttt{float} math libraries, respectively.}

\textbf{Contributions.}  This paper makes the following contributions.
\begin{itemize}
\item Proposes a novel approach that generates polynomials based on
  the correctly rounded value of an elementary function rather than
  minimizing the error between the real value and the approximation.

\item Demonstrates that the task of generating polynomials with
  correctly rounded results can be framed as a linear programming
  problem while accounting for range reduction.

\item Demonstrates \tool, a library of elementary functions that
  produce correctly rounded results for all inputs for various new
  alternatives to floating point such as \texttt{Bfloat16} and
  posits. Our functions are faster than state-of-the-art libraries.
\end{itemize}

\section{Background and Motivation}
\label{sec:background}
We provide background on the FP representation and its variants (\ie,
\texttt{Bfloat16}), the posit representation, the state-of-the-art for
developing math libraries, and a motivating example illustrating how
the use of existing libraries for new representations can result in
wrong results.

\begin{figure}
  \begin{subfigure}[t]{0.45\linewidth}
    \centerline{\includegraphics[width=1.0\linewidth]{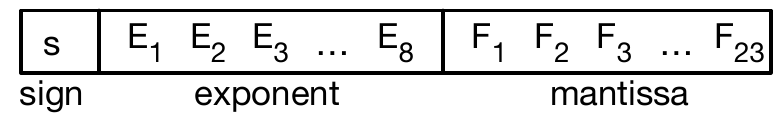}}
    \caption{Float}
  \end{subfigure}
  \hspace{3em}
  \begin{subfigure}[t]{0.45\linewidth}
    \centerline{\includegraphics[width=1.0\linewidth]{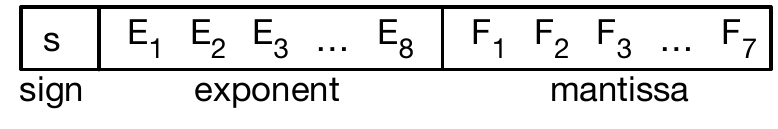}}
    \caption{Bfloat16}
  \end{subfigure}
  
  \begin{subfigure}[t]{0.26\linewidth}
    \centerline{\includegraphics[width=1.0\linewidth]{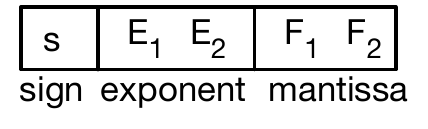}}
    \caption{5-bit floating point (FP5)}
  \end{subfigure}
  \hspace{3em}
  \begin{subfigure}[t]{0.63\linewidth}
    \centerline{\includegraphics[width=1.0\linewidth]{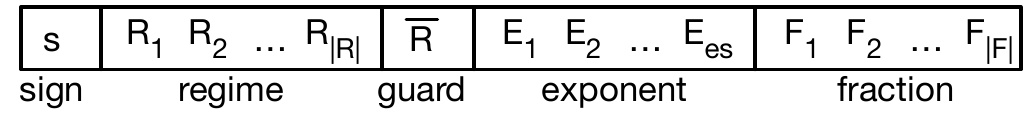}}
    \caption{Posit}
  \end{subfigure}
  \caption{(a) The bit-string for a 32-bit FP format (\texttt{float}). (b) The
    bit-string for the \texttt{Bfloat16} representation. (c) a 5-bit FP format
    used for illustration in the paper. It has 2 bits for the exponent
    and 2 bits for the fraction. (d) The bit pattern for a posit representation.}
  \label{fig:representations}
\end{figure}

\subsection{Floating Point and Its Variants}
\label{sec:finiteprecision}
The FP representation $\mathbb{F}_{n, |E|}$, which is specified in the
IEEE-754 standard~\cite{ieee754}, is parameterized by the total number
of bits $n$ and the number of bits for the exponent $|E|$. There are
three components in a FP bit-string: a sign bit $s$, $|E|$-bits to
represent the exponent, and $|F|$-bits to represent the mantissa $F$
where $|F| = n - 1 - |E|$. Figure~\ref{fig:representations}(a) shows
the FP format.
If $s = 0$, then the value is positive. If $s = 1$, then the value is
negative.
The value represented by the FP bit-string is a normal value if the
bit-string $E$, when interpreted as an unsigned integer, satisfies $0
< E < 2^{|E|} - 1$.  The normal value represented with this
bit-string is $(1 + \frac{F}{2^{|F|}})\times 2^{E-bias}$, where bias
is $2^{|E| - 1} - 1$.  If $E = 0$, then the FP value is a denormal
value. The value of the denormal value is 
$(\frac{F}{2^{|F|}})\times 2^{1-bias}$.
When $E=2^{|E|} - 1$, the FP bit-strings represent special values. If
$F = 0$, then the bit-string represents $\pm \infty$ depending on the
value of $s$ and in all other cases, it represents
\textit{not-a-number} (NaN).

IEEE-754 specifies a number of default FP types: 16-bit
($\mathbb{F}_{16, 5}$ or \texttt{half}), 32-bit ($\mathbb{F}_{32, 8}$
or \texttt{float}), and 64-bit ($\mathbb{F}_{64, 11}$ or
\texttt{double}).  Beyond the types specified in the IEEE-754
standard, recent extensions have increased the dynamic range and/or
precision.  \texttt{Bfloat16}~\cite{Tagliavini:bfloat:date:2018},
$\mathbb{F}_{16, 8}$, provides increased dynamic range compared to
FP's \texttt{half} type. Figure~\ref{fig:representations}(b)
illustrates the \texttt{Bfloat16} format. Recently proposed
\texttt{TensorFloat32}~\cite{nvidia:tensorfloat:online:2020},
$\mathbb{F}_{19, 8}$, increased both the dynamic range and precision
compared to the \texttt{half} type.

\subsection{The Posit Representation}
Posit~\cite{Gustafson:online:2017:posit, Gustafson:sfi:2017:beating}
is a new representation that provides tapered precision with a fixed
number of bits. A posit representation, $\mathbb{P}_{n, es}$, is
defined by the total number of bits $n$ and the maximum number of bits
for the exponents $es$. A posit bit-string consists of five components
(see Figure~\ref{fig:representations}(d)): a sign bit $s$, a number of
regime bits $R$, a regime guard bit $\overline{R}$, up to $es$-bits of
the exponent $E$, and fraction bits $F$. When the regime bits are not
used, they can be re-purposed to represent the fraction, which
provides tapered precision.

\textbf{Value of a posit bit-string.} The first bit is a sign bit. If
$s = 0$, then the value is positive.  If $s = 1$, then the value is
negative and the bit-string is decoded after taking the two's
complement of the remaining bit-string after the sign bit.  Three
components $R$, $\overline{R}$, and $E$ together are used to represent
the exponent of the final value. After the sign bit, the next $1 \leq
|R| \leq n - 1$ bits represent the regime $R$. Regime bits consist of
consecutive $1$'s (or $0$'s) and are only terminated if $|R| = n - 1$
or by an opposite bit $0$ (or $1$), which is known as the regime guard
bit ($\overline{R}$). The regime bits represent the super
exponent. Regime bits contribute $useed^{r}$ to the value of the
number where $useed = 2^{2^{es}}$ and $r = |R| - 1$ if $R$ consists of
1's and $r = -|R|$ if $R$ consists of 0's.

If $2 + |R| < n$, then the next $min\{es, n - 2 - |R|\}$ bits
represent the exponent bits. If $|E| < es$, then $E$ is padded with
$0$'s to the right until $|E| = es$. These $|es|$-bits contribute
$2^E$ to the value of the number. Together, the regime and the
exponent bits of the posit bit-string contribute $useed^{r}\times 2^{E}$
to the value of the number. If there are any remaining bits after the
$es$-exponent bits, they represent the fraction bits $F$. The fraction
bits are interpreted like a normal FP value, except the length of $F$
can vary depending on the number of regime bits. They contribute $1
+\frac{F}{2^{|F|}}$. Finally, the value $v$ represented by a posit
bit-string is,
\[
v = (-1)^s \times (1 + \frac{F}{2^{|F|}}) \times useed^{r} \times 2^{E} =
(-1)^s \times (1 + \frac{F}{2^{|F|}}) \times 2^{2^{es} \times r + E}
\]

There are two special cases. A bit-string of all $0$'s represents
$0$. A bit-string of $1$ followed by all $0s$'s represents
\textit{Not-a-Real} (NaR). 

\textbf{Example.}  Consider the bit-string \texttt{0000011011000000}
in the $\mathbb{P}_{16, 1}$ configuration. Here, $useed = 2^{2^{1}} =
2^2$. Also $s = $ \texttt{0}, $R =$ \texttt{0000}, $\overline{R} =$
\texttt{1}, $E =$ \texttt{1}, and $F =$ \texttt{011000000}. Hence, $r
= -|R| = -4$. The final exponent resulting from the regime and the
exponent bits is $(2^2)^{-4} \times 2^{1} = 2^{-7}$. The fraction
value is $1.375$. The value represented by this posit bit-string is
$1.375 \times 2^{-7}$.

\subsection{Rounding and Numerical Errors}
\label{sec:error_example}
When a real number $x$ cannot be represented in a target
representation $\mathbb{T}$, it has to be rounded to a value $v \in
\mathbb{T}$. The FP standard defines a number of rounding modes but
the default rounding mode is the
\textit{round-to-nearest-tie-goes-to-even} (RNE) mode. The posit
standard also specifies RNE rounding mode with a minor difference that
any non-zero value does not underflow to \texttt{0} or overflow to
\texttt{NaR}. We describe our approach with RNE mode but it is
applicable to other rounding modes.

\begin{figure}
  \centerline{\includegraphics[width=0.95\linewidth]{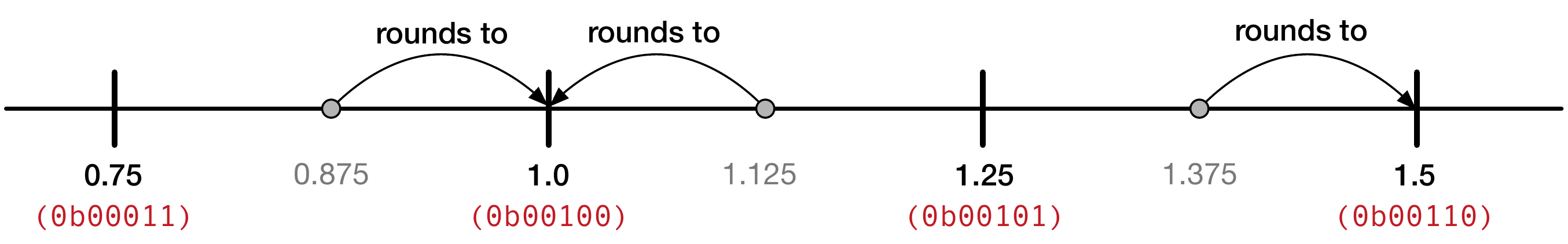}}
  \caption{Illustration of Round to Nearest with ties to Even (RNE)
    rounding mode with our 5-bit FP representation (\texttt{FP5}).
    There are two \texttt{FP5} values ($0.75$ and $1.0$) adjacent to
    the real number $0.875$, but both $0.75$ and $1.0$ are equidistant
    from $0.875$. In this case, RNE mode specifies that $0.875$ should
    round to $1.0$ because the bit representation of $1.0$
    (\texttt{0b00100}) is an even number when interpreted as an
    integer. Similarly, the real number $1.125$ rounds to $1.0$ and
    $1.375$ rounds to $1.5$.}
  \label{fig:FP5rne}
\end{figure}

In the RNE mode, the rounding function $v = RN_{\mathbb{T}}(x)$,
rounds $x \in \mathbb{R}$ (Reals) to $v \in \mathbb{T}$, such that $x$
is rounded to the nearest representable value in $\mathbb{T}$, \ie
$\forall_{v' \in \mathbb{T}} |x - v| \leq |x - v'|$. In the case of a
tie, where $\exists{v_1, v_2 \in \mathbb{T}, v_1 \neq v_2}$ such that
$|x - v_1| = |x - v_2|$ and $\forall_{v' \in \mathbb{T}} |x - v_1| \leq
|x - v'|$, then $x$ is rounded to $v_1$ if the bit-string encoding the
value $v_1$ is an even number when interpreted as an integer and to
$v_2$ otherwise. Figure~\ref{fig:FP5rne} illustrates the RNE mode with
a 5-bit FP representation from Figure~\ref{fig:representations}(c).

The result of primitive operations in FP or any other representation
experiences rounding error when it cannot be exactly represented.
Modern hardware and libraries produce correctly rounded results for
primitive operations.  However, this rounding error can get amplified
with a series of primitive operations because the intermediate result
of each primitive operation must be rounded. As math libraries are
also implemented with finite precision, numerical errors in the
implementation should also be carefully addressed.

\subsection{Background on Approximating Elementary Functions}
\label{sec:math_library}
The state-of-the-art methods to approximate an elementary function
$f(x)$ for a target representation ($\mathbb{T}$) involves two
steps. First, approximation theory (\eg, minimax methods) is used to
develop a function $A_{\mathbb{R}}(x)$ that closely approximates
$f(x)$ using real numbers. Second, $A_{\mathbb{R}}(x)$ is implemented
in a finite precision representation that has higher precision than
$\mathbb{T}$.

\textbf{Generating $A_{\mathbb{R}}(x)$.} Mathematically deriving
$A_{\mathbb{R}}(x)$ can be further split into three steps. First,
identify inputs that exhibit special behavior (\eg, $\pm
\infty$). Second, reduce the input domain to a smaller interval, $[a',
  b']$, with range reduction techniques and perform any other function
transformations. Third, generate a polynomial $P(x)$ that approximates
$f(x)$ in the domain $[a', b']$.

There are two types of special cases. The first type includes inputs
that produce undefined values or $\pm \infty$ when mathematically
evaluating $f(x)$. For example, in the case of $f(x) = 10^{x}$, $f(x)
= \infty$ if $x = \infty$. The second type consists of interesting
inputs for evaluating $RN_{\mathbb{T}}(f(x))$. These cases include a
range of inputs that produce interesting outputs such as
$RN_{\mathbb{T}}(f(x)) \in \{\pm \infty, 0\}$. For example, while
approximating $f(x) = 10^{x}$ for \texttt{Bfloat16}~($\mathbb{B}$), all values
$x \in (-\infty, -40.5]$ produce $RN_{\mathbb{B}}(10^{x}) = 0$, inputs
  $x \in [-8.46\dots \times 10^{-4}, 1.68\dots \times 10^{-3}]$
  produce $RN_{\mathbb{B}}(10^{x}) = 1$, and $x \in [38.75, \infty)$
    produces $RN_{\mathbb{B}}(10^{x}) = \infty$.
These properties are specific to each $f(x)$ and $\mathbb{T}$.

\textbf{Range reduction.}  It is mathematically simpler to approximate
$f(x)$ for a small domain of inputs. Hence, most math libraries use
range reduction to reduce the entire input domain into a smaller
domain before generating the polynomial.
Given an input $x \in [a, b]$ where $[a, b] \subseteq \mathbb{T}$, the
goal of range reduction is to reduce the input $x$ to $x' \in [a',
  b']$, where $[a', b'] \subset [a, b]$. We represent this process of
range reduction with $x' = RR(x)$. Then, the polynomial $P$
approximates the output $y'$ for the range reduced input (\ie, $y' =
P(x')$). The output ($y'$) of the range reduced input ($x'$) has to be
compensated to produce the output for the original input ($x$). The
output compensation function, $OC (y', x)$, produces the final result
by compensating the range reduced output $y'$ based on the range
reduction performed for input $x$.

For example, consider the function $f(x) = log_2(x)$ where the input
domain is defined over $(0, \infty)$. One way to range reduce the
original input is to use the mathematical property $log_2(a \times
2^b) = log_2(a) + b$. We decompose the input $x$ as $x = x’ \times
2^e$ where $x’ \in [1, 2)$ and $e$ is an integer. Approximating
  $log_2(x)$ is equivalent to approximating $log_2(x’ \times 2^e) =
  log_2(x’) + e$. Thus, we can range reduce the original input $x \in
  (0, \infty)$ into $x’ \in [1, 2)$. Then, we approximate $log_2(x’)$
    using $P(x’)$, which needs to only approximate $log_2(x)$ for the
    input domain $[1, 2)$. To produce the output of $log_2(x)$, we
      compensate the output of the reduced input by computing $P(x') +
      e$, where $e$ is dependent on the range reduction of $x$.

\textbf{Polynomial approximation $P(x)$.}  A common method to
approximate an elementary function $f(x)$ is with a polynomial
function, $P(x)$, which can be implemented with addition, subtraction,
and multiplication operations. Typically, $P(x)$ for math libraries is
generated using the minimax approximation technique, which aims to
minimize the maximum error, or $L_{\infty}$-norm,
\[
||P(x) - f(x)||_{\infty} = \sup_{x \in [a, b]} |P(x) - f(x)|
\]
where $\sup$ represents the supremum of a set.  The minimax approach
is attractive because the resulting $P(x)$ has a bound on the
error~(\ie, $|P(x) - f(x)|$). The most well-known minimax
approximation method is the Remez
algorithm~\cite{Remes:algorithm:1934}. Both
CR-LIBM~\cite{Daramy:crlibm:spie:2003} and
Metalibm~\cite{Olga:metalibm:icms:2014} use a modified Remez algorithm
to produce polynomial approximations~\cite{Brisebarre:epl:arith:2007}.

\textbf{Implementation of $A_{\mathbb{R}}(x)$ with finite precision.}
Finally, mathematical approximation $A_{\mathbb{R}}(x)$ is implemented
in finite precision to approximate $f(x)$. This implementation
typically uses a higher precision than the intended target
representation. We use $A_{\mathbb{H}}(x)$ to represent that
$A_{\mathbb{R}}(x)$ is implemented in a representation with higher
precision ($\mathbb{H}$) where $\mathbb{T} \subset \mathbb{H}$.
Finally, the result of the implementation $A_{\mathbb{H}}(x)$ is
rounded to the target representation $\mathbb{T}$.

\subsection{Challenges in Building Correctly Rounded Math Libraries}
An approximation of an elementary function $f(x)$ is defined to be a
correctly rounded approximation if for all inputs $x_i \in
\mathbb{T}$, it produces $RN_{\mathbb{T}}(f(x_i))$. There are two
major challenges in creating a correctly rounded approximation. First,
$A_{\mathbb{H}}(x)$ incurs error because $P(x)$ is an approximation of
$f(x)$. Second, the evaluation of $A_{\mathbb{H}}(x)$ has numerical
error because it is implemented in a representation with finite
precision (\ie, $\mathbb{H}$). Hence, the rounding of
$RN_{\mathbb{T}}(A_{\mathbb{H}}(x))$ can result in a value different
from $RN_{\mathbb{T}}(f(x))$, even if $A_{\mathbb{H}}(x)$ is
arbitrarily close to $f(x)$ for some $x \in \mathbb{T}$.

As $A_{\mathbb{R}}(x)$ uses a polynomial approximation of $f(x)$,
there is an inherent error of $|f(x) - A_{\mathbb{R}}(x)| >
0$. Further, the evaluation of $A_{\mathbb{H}}(x)$ experiences an
error of $|A_{\mathbb{H}}(x) - A_{\mathbb{R}}(x)| > 0$. It is not
possible to reduce both errors to 0. The error in approximating the
polynomial can be reduced by using a polynomial of a higher degree or
a piece-wise polynomial. The numerical error in the evaluation of
$A_{\mathbb{H}}(x)$ can be reduced by increasing the precision of
$\mathbb{H}$. Typically, library developers make trade-offs between
error and the performance of the implementation.

Unfortunately, there is no known general method to analyze and predict
the bound on the error for $A_{\mathbb{H}}(x)$ that guarantees
$RN_{\mathbb{T}}(A_{\mathbb{H}}(x)) = RN_{\mathbb{T}}(f(x))$ for all
$x$ because the error may need to be arbitrarily small. This problem
is widely known as \textit{table-maker's
  dilemma}~\cite{Kahan:tablemaker:online:2004}. It states that there
is no general method to predict the amount of precision in
$\mathbb{H}$ such that the result is correctly rounded for
$\mathbb{T}$.

\subsection{Why Not Use Existing Libraries for New Representations?}
\label{sec:background:motivation}
\begin{figure}
  \centerline{\includegraphics[width=0.75\linewidth]{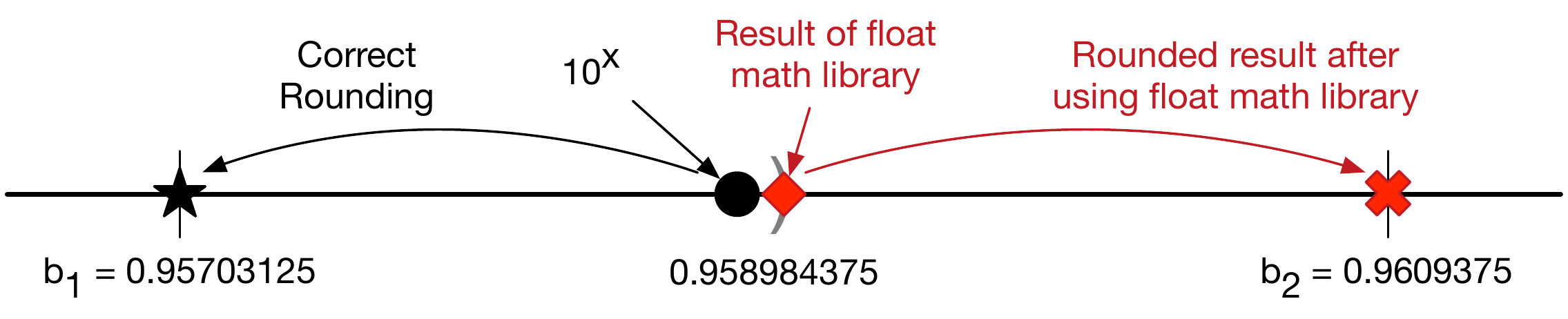}}
  \caption{Using a correctly rounded 32-bit FP math library to
    approximate $10^{x}$ for \texttt{Bfloat16} results in wrong
    results.  Horizontal axis represents a real number line. Given an
    input $x = -0.0181884765625$ that is exactly representable in
    \texttt{Bfloat16}, $b_1$ and $b_2$ represent the two closest
    \texttt{Bfloat16} values to the real value of $10^{x}$.  The
    correctly rounded \texttt{Bfloat16} value is $b_1$ (black
    star). When we use the 32-bit FP library to compute $10^{x}$, it
    produces the value shown with red diamond, which then rounds to
    $b_2$ producing an incorrect result.}
  \label{fig:exp10f_rounding}
\end{figure}

An alternative to developing math libraries for new representations is
to use existing libraries. We can convert the input $x \in
{\mathbb{T}}$ to $x' = RN_{\mathbb{T'}}(x)$, , where $\mathbb{T}$ is
the representation of interest and $\mathbb{T'}$ is the representation
that has a math library available (\eg,
\texttt{double}). Subsequently, we can use a math library for
$\mathbb{T'}$ and round the result back to $\mathbb{T}$.  This
strategy is appealing if a correctly rounded math library for
$\mathbb{T'}$ exists and $\mathbb{T'}$ has significantly more
precision bits than $\mathbb{T}$.

However, using a correctly rounded math library designed for
$\mathbb{T'}$ to approximate $f(x)$ for $\mathbb{T}$ can produce
incorrect results for values in $\mathbb{T}$.
%
\iffalse
More specifically, there exists an $x_i \in \mathbb{T}$ and a function
$C_{f, \mathbb{T'}}(x)$ that approximates an elementary function
$f(x)$ such that $RN_{\mathbb{T}}(C_{f,
  \mathbb{T'}}(RN_{\mathbb{T'}}(x_i))) \neq RN_{\mathbb{T}}(f(x))$
even if $\mathbb{T} \subset \mathbb{T'}$.
\fi
%
We illustrate this behavior by generating an approximation for the
function $f(x) = 10^{x}$ in the \texttt{Bfloat16}~($\mathbb{B}$)
representation (Figure~\ref{fig:exp10f_rounding}). Let's consider the
input $x = -0.0181884765625 \in \mathbb{B}$. The real value of $f(x)
\approx 0.95898435797\dots$ (black circle in
Figure~\ref{fig:exp10f_rounding}). This oracle result cannot be
exactly represented in \texttt{Bfloat16} and must be rounded.  There
are two \texttt{Bfloat16} values adjacent to $f(x)$, $b_1 =
0.95703125$ and $b_2 = 0.9609375$. Since $b_1$ is closer to $f(x)$,
the correctly rounded result is $RN_{\mathbb{B}}(10^{x}) = b_1$, which
is represented by a black star in Figure~\ref{fig:exp10f_rounding}.

If we use the correctly rounded \texttt{float} math library to
approximate $10^{x}$, we get the value, $y' = 0.958984375$,
represented by red diamond in Figure ~\ref{fig:exp10f_rounding}. From
the perspective of a 32-bit \texttt{float}, $y'$ is a correctly
rounded result, \ie $y' = RN_{\mathbb{F}_{32, 8}}(10^{x}) =
0.958984375$. Because $y' \notin \mathbb{B}$, we round $y'$ to
\texttt{Bfloat16} based on the rounding rule, $RN_{\mathbb{B}}(y') =
b_2$. Therefore, the \texttt{float} math library rounds the result to
$b_2$ but the correctly rounded result is $RN_{\mathbb{B}}(10^{x}) =
b_1$.

\textbf{Summary. }  Approximating an elementary function for
representation $\mathbb{T}$ using a math library designed for a higher
precision representation $\mathbb{T'}$ does not guarantee a correctly
rounded result. Further, the math library for $\mathbb{T'}$ probably
requires higher accuracy than the one for $\mathbb{T}$. Hence, it uses
a higher degree polynomial, which causes it to be slower than the math
library tailored for $\mathbb{T}$.

\section{High-Level Overview}
We provide a high-level overview of our methodology to generate
correctly rounded math libraries. We will illustrate this methodology
with an end-to-end example that creates correctly rounded results for
$ln(x)$ with \texttt{FP5} (\ie, a 5-bit FP type shown in
Figure~\ref{fig:representations}(c)).

\subsection{Our Methodology for Generating Correctly Rounded Elementary Functions}
Given an elementary function $f(x)$ and a target
representation~$\mathbb{T}$, our goal is to synthesize a polynomial
that when used with range reduction ($RR$) and output compensation
($OC$) function produces the correctly rounded result for all inputs
in $\mathbb{T}$. The evaluation of the polynomial, range reduction,
and output compensation are implemented in representation
$\mathbb{H}$, which has higher precision than $\mathbb{T}$.

Our methodology for generating correctly rounded elementary functions
is shown in Figure~\ref{fig:workflow}. Our methodology consists of
four steps. First, we use an oracle (\ie,
MPFR~\cite{Fousse:toms:2007:mpfr} with a large number of precision
bits) to compute the correctly rounded result of the function $f(x)$
for each input $x \in \mathbb{T}$. In this step, a small sample of the
entire input space can be used rather than using all inputs for a type
with a large input domain.

Second, we identify an interval $[l, h]$ around the correctly rounded
result such that any value in $[l, h]$ rounds to the correctly rounded
result in $\mathbb{T}$.  We call this interval the \textit{rounding
  interval}. Since the eventual polynomial evaluation happens in
$\mathbb{H}$, the rounding intervals are also in the $\mathbb{H}$
representation. The internal computations of the math library
evaluated in $\mathbb{H}$ should produce a value in the rounding
interval for each input $x$.

Third, we employ range reduction to transform input $x$ to $x'$. The
generated polynomial will approximate the result for
$x'$. Subsequently, we have to use an appropriate output compensation
code to produce the final correctly rounded output for $x$. Both range
reduction and output compensation happen in the $\mathbb{H}$
representation and can experience numerical errors. These numerical
errors should not affect the generation of correctly rounded results.
Hence, we infer intervals for the reduced domain so that the
polynomial evaluation over the reduced input domain produces the
correct results for the entire domain. Given $x$ and its rounding
interval $[l, h]$, we can compute the reduced input $x'$ with range
reduction. The next task before polynomial generation is identifying
the reduced rounding interval for $P(x')$ such that when used with
output compensation it produces the correctly rounded result. We use
the inverse of the output compensation function to identify the
reduced interval $[l', h']$. Any value in $[l', h']$ when used with
the implementation of output compensation in $\mathbb{H}$ produces the
correctly rounded results for the entire domain.

Fourth, we synthesize a polynomial of a degree $d$ using an arbitrary
precision linear programming (LP) solver that satisfies the
constraints (\ie, $l' \leq P(x') \leq h'$) when given a set of inputs
$x'$. Since the LP solver produces coefficients for the polynomial in
arbitrary precision, it is possible that some of the constraints will
not be satisfied when evaluated in $\mathbb{H}$.
In such cases, we refine the reduced intervals for those inputs whose
constraints are violated and repeat the above step. If the LP solver
is not able to produce a solution, then the developer of the library
has to either increase the degree of the polynomial or reduce the
input domain.

If the inputs were sampled in the first step, we check whether the
generated polynomial produces the correctly rounded result for all
inputs. If it does not, then the input is added to the sample and the
entire process is repeated. At the end of this process, the polynomial
along with range reduction and output compensation when evaluated in
$\mathbb{H}$ produces the correctly rounded outputs for all inputs in
$\mathbb{T}$.

\begin{figure}%
  \begin{subfigure}[b]{0.48\linewidth}
    \includegraphics[width=\textwidth]{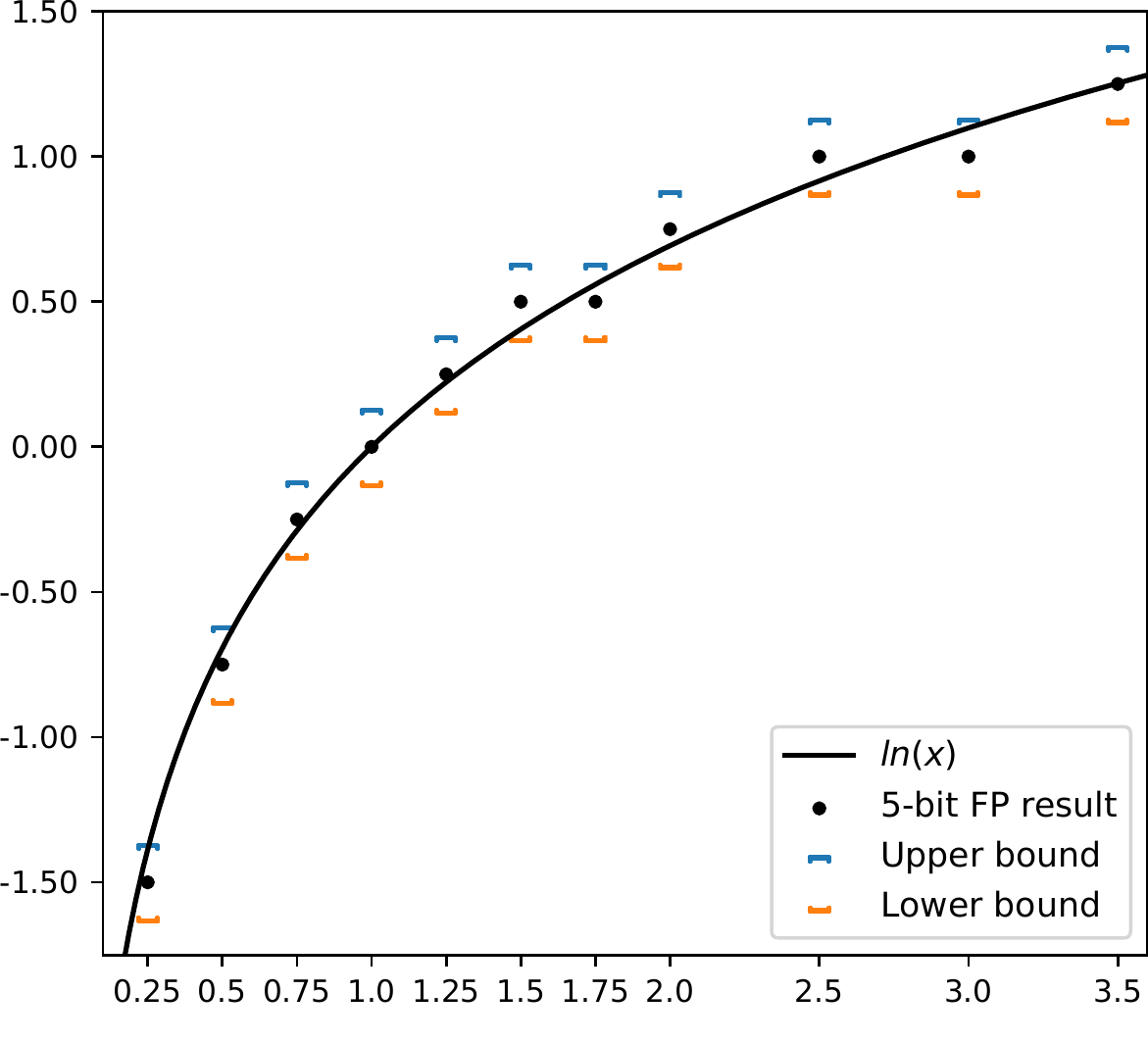}
    \caption{}
  \end{subfigure}
  \begin{subfigure}[b]{0.48\linewidth}
    \includegraphics[width=\textwidth]{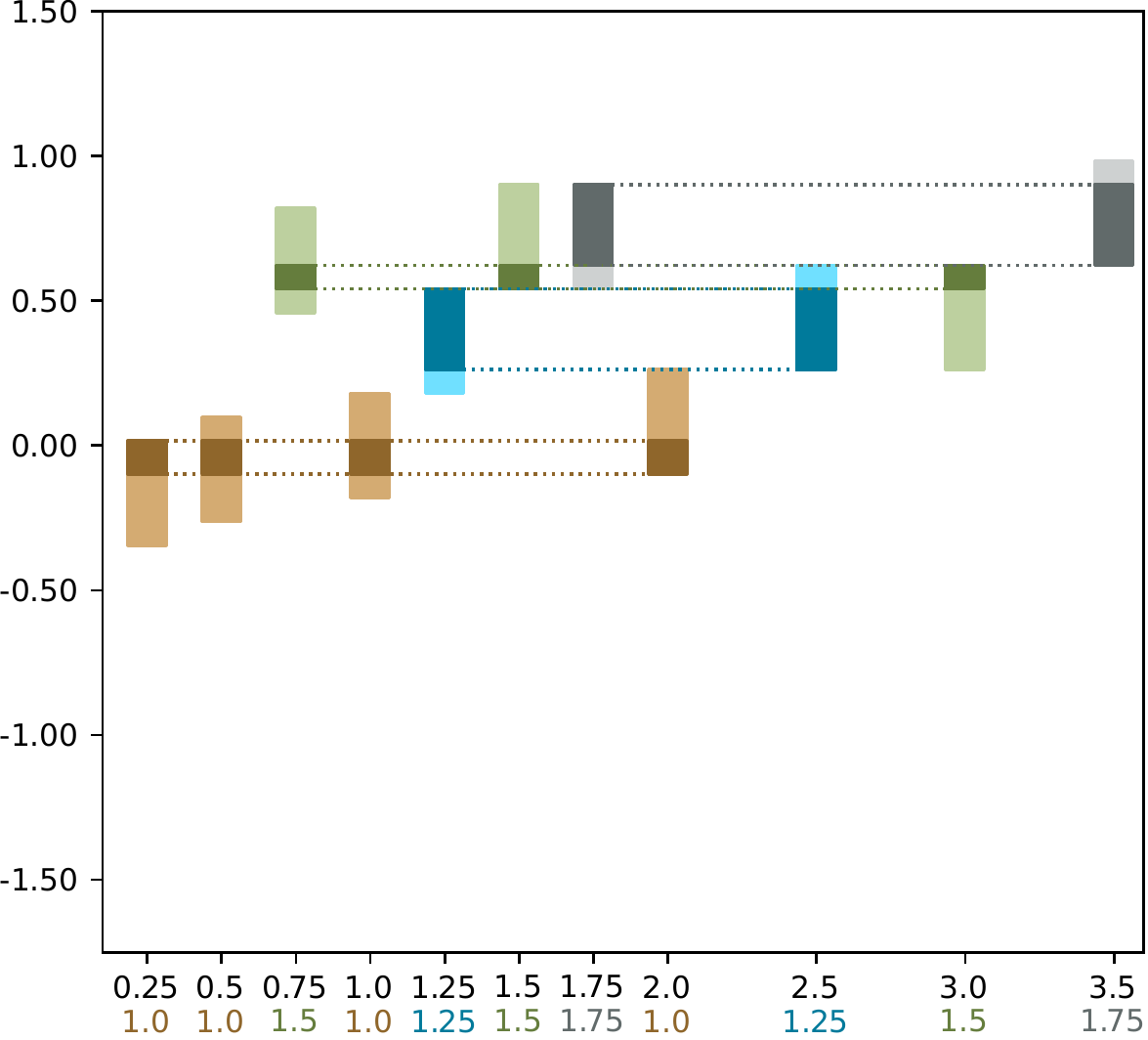}
    \caption{}
  \end{subfigure}
  
  \begin{subfigure}[b]{0.59\linewidth}
    \begin{subfigure}[b]{\linewidth} 
        \small
        \begin{empheq}[box=\fbox]{align}
            -0.098315\dots \leq P(1.00) \leq 0.016294\dots \nonumber \\
            0.262358\dots \leq P(1.25) \leq 0.541010\dots \nonumber \\
            0.541010\dots \leq P(1.50) \leq 0.623031\dots \nonumber \\
            0.623031\dots \leq P(1.75) \leq 0.901684\dots \nonumber
        \end{empheq}
        \caption{}
    \end{subfigure}
    
    \begin{subfigure}[b]{\linewidth} 
        \small
        \begin{empheq}[box=\fbox]{align}
        \begin{bmatrix}
        -0.09831\dots \\
        0.26235\dots \\
        0.54101\dots \\
        0.62303\dots
        \end{bmatrix}
        \leq
        \begin{bmatrix}
        1.0 & 1.0 \\
        1.0 & 1.25 \\
        1.0 & 1.5 \\
        1.0 & 1.75
        \end{bmatrix}
        \begin{bmatrix}
        c_0 \\
        c_1
        \end{bmatrix}
        \leq
        \begin{bmatrix}
        0.01629\dots \\
        0.54101\dots \\
        0.62303\dots \\
        0.90168\dots
        \end{bmatrix} \nonumber
        \end{empheq}
        \caption{}
    \end{subfigure}
    
    \begin{subfigure}[b]{\linewidth} 
      \small
      \begin{empheq}[box=\fbox]{align}
        P(x) & = c_0 + {c_1}{x} \nonumber \\
        c_0 & = -1.03313832433369645613652210158761590719\dots \nonumber \\
        c_1 & = 1.049432643111371854516278290248010307550\dots \nonumber 
      \end{empheq}
      \caption{}
    \end{subfigure}
  \end{subfigure}
  \begin{subfigure}[b]{0.36\linewidth}
    \includegraphics[width=\textwidth]{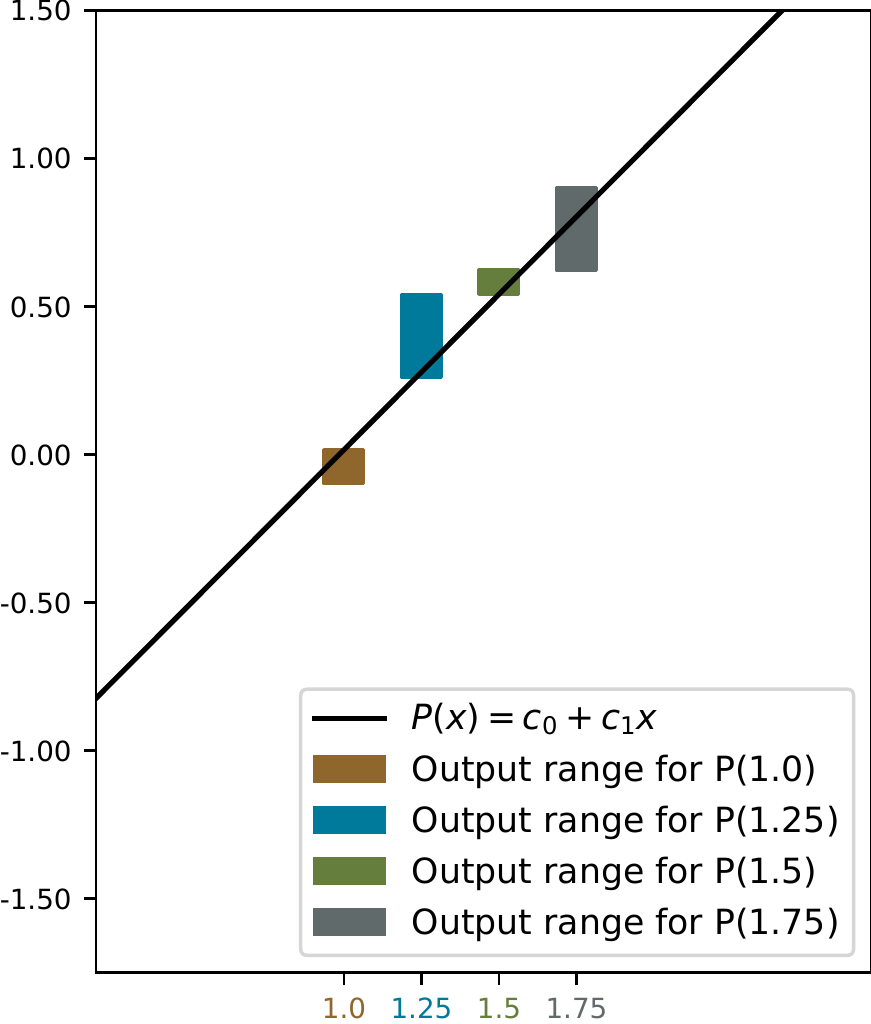}
    \caption{}
  \end{subfigure}
  \caption{Our approach for $ln(x)$ with \texttt{FP5}. (a) For each
    input $x$ in \texttt{FP5}, we accurately compute the correctly
    rounded result (black circle) and identify intervals around the
    result so that all values round to it. (b) For each input and
    corresponding interval computed in (a), we perform range reduction
    to obtain the reduced input. The number below a value on the
    x-axis represents the reduced input. The reduced interval to
    account for rounding errors in output compensation is also shown.
    Multiple distinct inputs can map to the same reduced input after
    range reduction (intervals with the same color). In such
    scenarios, we combine the reduce intervals by computing the common
    region in the intervals (highlighted in bold for each color with
    dotted lines).  (c) The set of constraints that must be satisfied
    by the polynomial for the reduced input. (d) LP formulation for
    the generation of a polynomial of degree one. (e) The
    coefficients generated by the LP solver for the polynomial. (f)
    Generated polynomial satisfies the combined intervals.}
  \label{fig:logxFloat5}
\end{figure}

\subsection{Illustration of Our Approach with $ln(x)$ for FP5}
We provide an end-to-end example of our approach by creating a
correctly rounded result of $ln(x)$ for the \texttt{FP5}
representation shown in Figure~\ref{fig:representations}(c) with the
RNE rounding mode. The $ln(x)$ function is defined over the input
domain $(0, \infty)$.  There are 11 values ranging from $0.25$ to
$3.5$ in \texttt{FP5} within $(0, \infty)$. We show the generation of
the polynomial with \texttt{FP5} for pedagogical reasons. With
\texttt{FP5}, it is beneficial to create a pre-computed table of
correctly rounded results for the 11 values.

Our strategy is to approximate $ln(x)$ by using $log_{2}(x)$.  Hence,
we perform range reduction and output compensation using the
properties of logarithm: $ln(x) = \frac{log_2(x)}{log_2(e)}$ and
$log_2(x \times y^z) = log_2(x) + zlog_2(y)$. We decompose the input
$x$ as $x = x' \times 2^n$ where $x'$ is the fractional value
represented by the mantissa, \ie $x' \in [1, 2)$, and $n$ is the
exponent of the value. We use $ln(x) = \frac{log_2(x') + m}{log_2(e)}$
for our range reduction. We construct the range reduction function
$RR(x)$ and the output compensation function $OC(y', x)$ as follows,
\[
RR(x) = fr(x), \quad \quad \quad \quad OC(y', x) = \frac{y' + exp(x)}{log_2(e)}
\]
where $fr(x)$ returns the fractional part of $x$ (\ie, $x' \in [1,
  2)$) and $exp(x)$ returns the exponent of $x$ (\ie, $n$).
Then, our polynomial approximation $P(x')$ should approximate the
function $log_2(x)$ for the reduced input domain $x' \in [1, 2)$.
The various steps of our approach are illustrated in
Figure~\ref{fig:logxFloat5}.

\textbf{Step 1: Identifying the correctly rounded result.}  There are
a total of 11 \texttt{FP5} values in the input domain of $ln(x)$,
$(0, \infty)$. These values are shown on the x-axis in
Figure~\ref{fig:logxFloat5}(a). Other values are special cases. They
are captured by the precondition for this function~(\ie, $x = 0$ or $x
=\infty$). Our goal is to generate the correctly rounded results for
these 11 \texttt{FP5} values. For each of these 11 inputs $x$, we use
an oracle (\ie, MPFR math library) to compute $y$, which is the
correctly rounded value of $ln(x)$. Figure~\ref{fig:logxFloat5}(a)
shows the correctly rounded result for each input as a black dot.

\iffalse Each input $x$ and its corresponding output $y$ specifies
what the input and the output of our math library function should be.
\fi

\begin{figure}
  \centerline{\includegraphics[width=0.95\linewidth]{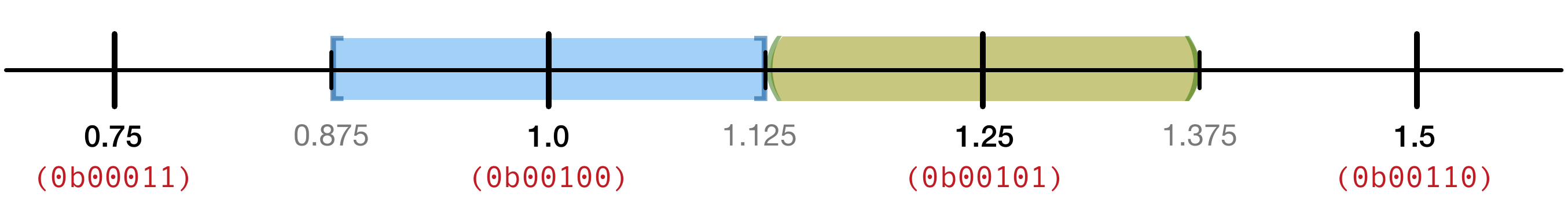}}
  \caption{This figure shows the real number line and a number of
    adjacent \texttt{FP5} values, 0.75, 1.0, 1.25, and 1.5.  Any real
    value in the blue interval $[0.875, 1.125]$, rounds to 1.0 in
    \texttt{FP5} with RNE rounding mode. Similarly, any value in the
    green interval $(1.125, 1.375)$ rounds to 1.25 in \texttt{FP5}. }
  \label{fig:FP5Interval}
\end{figure}
 \textbf{Step 2: Identifying the
  rounding interval $[l, h]$.}  The range reduction, output
compensation, and polynomial evaluation are performed with the
\texttt{double} type. The \texttt{double} result of the evaluation is
rounded to \texttt{FP5} to produce the final result. The next step is
to find a rounding interval $[l, h]$ in the \texttt{double} type for
each output. Figure~\ref{fig:logxFloat5}(a) shows the rounding
interval for each \texttt{FP5} output using the blue (upper bound) and
orange (lower bound) bracket.

Let us suppose that we want to compute the rounding interval for $y =
1.0$, which is the correctly rounded result of $ln(2.5)$.  To identify
the lower bound $l$ of the rounding interval for $y = 1.0$, we first
identify the preceding \texttt{FP5} value, which is $0.75$. Then we find a
value $v$ between $0.75$ and $1.0$ such that values greater than or
equal to $v$ rounds to $1.0$. In our case, $v = 0.875$, which is the
lower bound. Similarly, to identify the upper bound $h$, we identify
the \texttt{FP5} value succeeding $1.0$, which is $1.25$. We find a value $v$
such that any value less than or equal to $v$ rounds to $1.0$.  In our
case, the upper bound is $h =1.125$. Hence, the rounding interval for
$y=1.0$ is $[0.875, 1.125]$. Figure~\ref{fig:FP5Interval} shows the
intervals for a small subset of \texttt{FP5}.

\textbf{Step 3-a: Computing the reduced input $x'$ and the reduced
  interval $[l', h']$.} We perform range reduction and generate a
polynomial that computes $log_2(x)$ for all reduced inputs in $[1,
  2)$.
The next step is to identify the reduced input and the rounding
interval for the reduced input such that it accounts for any numerical
error in output compensation. Figure~\ref{fig:logxFloat5}(b) shows the
reduced input (number below the value on the x-axis) and the reduced
interval for each input.

To identify the reduced rounding interval, we use the inverse of the
output compensation function, which exists if $OC$ is continuous and
bijective over real numbers. For example, for the input $x = 3.5 =
1.75 \times 2^1$, the output compensation function is,
\[
OC(y', 3.5) = \frac{y' + 1}{log_2(e)}
\]
The inverse is
\[
OC^{-1}(y, 3.5) = {y}log_2(e) - 1
\]

Thus, we use the inverse output compensation function to compute the
candidate reduced interval $[l', h']$ by computing $l' = OC^{-1}(l,
x)$ and $h' = OC^{-1}(h, x)$. Then, we verify that the output
compensation result of $l'$ (\ie, $OC(l', x)$) and $h'$ (\ie, $OC(h',
x)$), when evaluated in \texttt{double} lies in $[l, h]$. If it does
not, then we iteratively refine the reduced interval by restricting
$[l', h']$ to a smaller interval until both $OC(l', x)$ and $OC(h',
x)$ evaluated in \texttt{double} results lie in $[l,h]$. The vertical
bars in Figure~\ref{fig:logxFloat5}(b) show the reduced input for each
$x$ and its corresponding reduced rounding interval.

\textbf{Step 3-b: Combining the reduced intervals.} Multiple inputs
from the original input domain can map to the same reduced input after
range reduction.  In our example, both $x_1 = 1.25$ and $x_2 = 2.5$
reduce to $x' = 1.25$. However, the reduced intervals that we compute
for $x_1$ and $x_2$ are $[l_1', h_1']$ and $[l_2', h_2']$,
respectively. They are not exactly the same. In
Figure~\ref{fig:logxFloat5}(b), the reduced intervals corresponding to
the original inputs that map to the same reduced input are colored
with the same color. The reduced intervals for $x_1=1.25$ and $x_2=2.5$
are colored in blue.

The reduced interval for $x_1$ indicates that $P(1.25)$ must produce a
value in $[l_1', h_1']$ such that the final result, after evaluating
the output compensation function in \texttt{double}, is the correctly
rounded value of $ln(1.25)$. The reduced interval for $x_2$ indicates
that $P(1.25)$ must produce a value in $[l_2', h_2']$ such that the
final result is the correct value of $ln(2.5)$. To produce the
correctly rounded result for both inputs $x_1$ and $x_2$, $P(1.25)$
must produce a value that is in both $[l_1', h_1']$ and $[l_2',
  h_2']$. Thus, we combine all reduced intervals that correspond to
the same reduced input by computing the common
interval. Figure~\ref{fig:logxFloat5}(b) shows the common interval for
a given reduced input using a darker shade.
At the end of this step, we are left with one combined interval for
each reduced input.

\textbf{Step 4: Generating the Polynomial for the reduced input.}  The
combined intervals specify the constraints on the output of the
polynomial for each reduced input, which when used with output
compensation in \texttt{double} results in a correctly rounded result
for the entire domain. Figure~\ref{fig:logxFloat5}(c) shows the
constraints for $P(x')$ for each reduced input.

To synthesize a polynomial $P(x')$ of a particular degree (the degree
is 1 in this example), we encode the problem as a linear programming
(LP) problem that solves for the coefficients of $P(x')$. We look for
a polynomial that satisfies constraints for each reduced
input~(Figure~\ref{fig:logxFloat5}(d)).
We use an LP solver to solve for the coefficients and find $P(x')$
with the coefficients in Figure~\ref{fig:logxFloat5}(e).  The
generated polynomial $P(x')$ satisfies all the linear constraints as
shown in Figure~\ref{fig:logxFloat5}(f).  Finally, we also verify that
the generated polynomial when used with range reduction and output
compensation produces the correctly rounded results for all inputs in
the original domain.

\section{Our Methodology for Generating Correctly Rounded Libraries}
\label{sec:approach}
Our goal is to create approximations for an elementary function $f(x)$
that produces correctly rounded results for all inputs in the target
representation~($\mathbb{T}$).
\begin{definition}{}
\label{def:cor_rounded}
A function that approximates an elementary function $f(x)$ is a
correctly rounded function for the target representation $\mathbb{T}$
if it produces $y = RN_{\mathbb{T}}(f(x))$ for all $x \in \mathbb{T}$.
\end{definition}

Intuitively, the result produced by the approximation should be same
as the result obtained when $f(x)$ is evaluated with infinite
precision and then rounded to the target representation.  It may be
beneficial to develop precomputed tables with correctly rounded
results of elementary functions for small data types~(\eg,
\texttt{FP5}). However, it is infeasible (due to memory overheads) to
store such tables for every elementary function even with modestly
sized data types.

We propose a methodology that produces polynomial approximation and
stores a few coefficients for evaluating the polynomial. There are
three main challenges in generating a correctly rounded result with
polynomial approximations. First, we have to generate polynomial
approximations that produce the correct result and are efficient to
evaluate. Second, the polynomial approximation should consider
rounding errors with range reduction and output compensation that are
implemented in some finite precision representation. Third, the
polynomial evaluation also is implemented with finite precision and
can experience numerical errors.

We will use $A_{\mathbb{H}}(x)$ to represent the approximation of the
elementary function $f(x)$ produced with our methodology while using a
representation $\mathbb{H}$ to perform polynomial evaluation, range
reduction, and output compensation. The result of $A_{\mathbb{H}}(x)$
is rounded to $\mathbb{T}$ to produce the final result.
Hence, $A_{\mathbb{H}}(x)$ is composed of three functions:
$A_{\mathbb{H}}(x) =
OC_{\mathbb{H}}(P_{\mathbb{H}}(RR_{\mathbb{H}}(x)), x)$ where $y' =
P_{\mathbb{H}}(x')$ is the polynomial approximation function, $x' =
RR_{\mathbb{H}}(x)$ is the range reduction function, and
$OC_{\mathbb{H}}(y', x)$ is the output compensation function. All
three functions, $RR_{\mathbb{H}}(x)$, $P_{\mathbb{H}}(x')$, and
$OC_{\mathbb{H}}(y', x)$ are evaluated in $\mathbb{H}$. Given
$RR_{\mathbb{H}}(x)$ and $OC_{\mathbb{H}}(y', x)$ for a particular
elementary function $f(x)$, the task of creating an approximation that
produces correctly rounded results involves synthesizing a polynomial
$P_{\mathbb{H}}(x)$ such that final result generated by
$A_{\mathbb{H}}(x)$ is a correctly rounded result for all inputs $x$.

Our methodology for identifying $A_{\mathbb{H}}(x)$ that produces
correctly rounded outputs is pictorially shown in
Figure~\ref{fig:workflow}. In our approach, we assume the existence of
an oracle, which generates the correct real result, to generate the
polynomial approximation for a target representation $\mathbb{T}$. We
can use existing MPFR libraries with large precision as an
oracle. Typically, the polynomial approximation is closely tied to
techniques used for range reduction and the resulting output
compensation.  We also require that the output compensation function
($OC$) is invertible (\ie, continuous and bijective). The degree of
the polynomial is an input provided by the developer of the math
library. The top-level algorithm shown in Figure~\ref{alg:main}
identifies a polynomial approximation of degree $d$. If it is unable
to find one, the developer of the math library should explore one with
a higher degree.

\begin{figure}
  \small
  \begin{minipage}{0.63\textwidth}
    \begin{algorithm}[H]
      \DontPrintSemicolon
      \SetKwFunction{FMain}{CorrectlyRoundedPoly}
      \SetKwFunction{FCalculateL}{CalcRndIntervals}
      \SetKwFunction{FCalculateLp}{CalcRedIntervals}
      \SetKwFunction{FCalculateLambda}{CombineRedIntervals}
      \SetKwFunction{FGeneratePoly}{GeneratePoly}
      \SetKwProg{Fn}{Function}{:}{}
      \Fn{\FMain{$f$, $\mathbb{T}$, $\mathbb{H}$, $X$, $RR_{\mathbb{H}}$, $OC_{\mathbb{H}}$, $d$}}{
        $L \leftarrow$ \FCalculateL{$f$, $\mathbb{T}$, $\mathbb{H}$, $X$}\;
        \lIf{$L = \emptyset$}{
          \Return{(false, DNE)}
        }
        $L' \leftarrow$ \FCalculateLp{$L$, $\mathbb{H}$, $RR_{\mathbb{H}}$, $OC_{\mathbb{H}}$}\;
        \lIf{$L' = \emptyset$}{
          \Return{(false, DNE)}
        }
        $\Lambda \leftarrow$ \FCalculateLambda{$L'$}\;
        \lIf{$\Lambda = \emptyset$}{
          \Return{(false, DNE)}
        }
        $S$, $P_{\mathbb{H}} \leftarrow$ \FGeneratePoly{$\Lambda$, $d$}\;
        \lIf{$S$ = true}{
          \Return{(true, $P_{\mathbb{H}}$)}\
        }
        \lElse{
          \Return{(false, DNE)}
        }
      }
    \end{algorithm}
  \end{minipage}
  \begin{minipage}{0.36\textwidth}
    \textbf{Input Description:}
    
    $f$: \textit{The oracle that computes the result of $f(x)$ in
      arbitrary precision.}

    $\mathbb{T}$: \textit{Target representation of math library.}

    $\mathbb{H}$: \textit{Higher precision representation.}

    $X$: \textit{Input domain of $A_{\mathbb{H}}(x)$.}

    $RR_{\mathbb{H}}$: \textit{The range reduction function.}

    $OC_{\mathbb{H}}$: \textit{The output compensation function.}

    $d$: The degree of polynomial to generate.
  \end{minipage}
  
  \caption{Our approach to generate a polynomial approximation
    $P_{\mathbb{H}}(x)$ that produces the correctly rounded result for
    all inputs. On successfully finding a polynomial, it returns
    (true, $P_{\mathbb{H}}$). Otherwise, it returns (false, DNE) where
    DNE means that the polynomial Does-Not-Exist. Functions,
    \texttt{CalcIntervals}, \texttt{CalcRedIntervals},
    \texttt{CombineRedIntervals}, and \texttt{GeneratePoly} are shown
    in Figure~\ref{alg:calculatel}, Figure~\ref{alg:calculatelp}, and
    Figure~\ref{alg:generatep}, respectively.}
\label{alg:main}
\end{figure}

Our approach has four main steps. First, we compute $y \in
\mathbb{T}$, the correctly rounded result of $f(x)$, \ie $y =
RN_{\mathbb{T}}(f(x))$ for each input $x$ (or a sample of the inputs
for a large data type) using our oracle. Then, we identify the
rounding interval $I = [l, h] \subseteq \mathbb{H}$ where all values
in the interval round to $y$. The pair $(x, I)$ specifies that
$A_{\mathbb{H}}(x)$ must produce a value in $I$ such that
$A_{\mathbb{H}}(x)$ rounds to $y$. The function
\texttt{CalcRndIntervals} in Figure~\ref{alg:main} returns a list $L$
that contains a pair $(x, I)$ for all inputs $x$.

Second, we compute the reduced input $x'$ using range reduction and a
reduced interval $I'=[l', h']$ for each pair $(x, I) \in L$.  The
reduced interval $I' = [l', h']$ ensures that any value in $I'$ when
used with output compensation code results in a value in $I$.  This
pair $(x', I')$ specifies the constraints for the output of the
polynomial approximation $P_{\mathbb{H}}(x')$ so $A_{\mathbb{H}}(x)$
rounds to the correctly rounded result. The function
\texttt{CalcRedIntervals} returns a list $L'$ with such reduced
constraints for all inputs $x$.

Third, multiple inputs from the original input domain will map to the
same input in the reduced domain after range reduction. Hence, there
will be multiple reduced constraints for each reduced input $x'$.  The
polynomial approximation, $P_{\mathbb{H}}(x')$, must produce a value
that satisfies all the reduced constraints to ensure that
$A_{\mathbb{H}}(x)$ produces the correct value for all inputs when
rounded. Thus, we combine all reduced intervals for each unique reduced
input $x'$ and produce the pair $(x', \Psi)$ where $\Psi$ represents
the combined interval. Function \texttt{CombineRedIntervals} in
Figure~\ref{alg:main} returns a list $\Lambda$ containing the
constraint pair $(x', \Psi)$ for each unique reduced input $x'$.
Finally, we generate a polynomial of degree $d$ using linear
programming so that all constraints $(x', \Psi) \in \Lambda$ are
satisfied. Next, we describe these steps in detail.

\subsection{Calculating the Rounding Interval}
The first step in our approach is to identify the values that
$A_{\mathbb{H}}(x)$ must produce so that the rounded value of
$A_{\mathbb{H}}(x)$ is equal to the correctly rounded result of $y =
f(x)$, \ie $RN_{\mathbb{T}}(A_{\mathbb{H}}(x)) = RN_{\mathbb{T}}(y)$,
for each input $x \in X$. \textit{Our key insight is that it is not
  necessary to produce the exact value of $y$ to produce a correctly
  rounded result}.  It is sufficient to produce any value in
$\mathbb{H}$ that round to the correct result.  For a given rounding
mode and an input, we are looking for an interval $I = [l, h]$ around
the oracle result that produces the correctly rounded result. We call
this the rounding interval.

\begin{figure}
\small
\begin{minipage}[t]{0.46\textwidth}
\begin{algorithm}[H]
\DontPrintSemicolon
\SetKwFunction{FMain}{Main}
\SetKwFunction{FCalculateL}{CalcRndIntervals}
\SetKwFunction{FRoundingInterval}{GetRndInterval}
\SetKwProg{Fn}{Function}{:}{}
\Fn{\FCalculateL{$f$, $\mathbb{T}$, $\mathbb{H}$, $X$}}{
    $L \leftarrow \emptyset$\;
    
    \ForEach{$x \in X$} {
        $y \leftarrow RN_{\mathbb{T}}(f(x))$\;
        $I \leftarrow $ \FRoundingInterval{$y$, $\mathbb{T}$, $\mathbb{H}$}\;
        \lIf{$I = \emptyset$} {
            \Return{$\emptyset$}
        }
        $L \leftarrow L \cup \{(x, I)\}$\;
    }
    \Return{$L$}\;
  }
\end{algorithm}
\end{minipage}
\begin{minipage}[t]{0.53\textwidth}
\begin{algorithm}[H]
\setcounter{AlgoLine}{9}
\DontPrintSemicolon
\SetKwFunction{FMain}{Main}
\SetKwFunction{FRoundingInterval}{GetRndInterval}
\SetKwFunction{FAdjLower}{GetPrecVal}
\SetKwFunction{FAdjHigher}{GetSuccVal}
\SetKwFunction{FHex}{Hex}
\SetKwProg{Fn}{Function}{:}{}
\Fn{\FRoundingInterval{$y$, $\mathbb{T}$, $\mathbb{H}$}}{
  $t_l \leftarrow$ \FAdjLower{$y$, $\mathbb{T}$}\;
  $l  \leftarrow min\{v \in \mathbb{H} | v \in [t_l, y] \text{ and } RN_{\mathbb{T}}(v) = y\}$\;
  $t_u \leftarrow$ \FAdjHigher{$y$, $\mathbb{T}$}\;
  $h  \leftarrow max\{v \in \mathbb{H} | v \in [y, t_u] \text{ and } RN_{\mathbb{T}}(v) = y\}$\;

    \Return $[l, h]$\;
  }
\end{algorithm}
\end{minipage}

\caption{For each input $x \in X$, \texttt{CalcRndIntervals}
  identifies the interval $I = [l, h]$ where all values in $I$ round
  to the correctly rounded result. The \texttt{GetRndInterval}
  function takes the correctly rounded result $y$ and returns the
  interval $I \subseteq \mathbb{H}$ where all values in $I$ round to
  $y$. \texttt{GetPrecValue}($y$, $\mathbb{T}$) returns the value
  preceeding $y$ in $\mathbb{T}$.  \texttt{GetSuccValue}($y$,
  $\mathbb{T}$) returns the value succeeding $y$ in $\mathbb{T}$.}
\label{alg:calculatel}
\end{figure}

Given an elementary function $f(x)$ and an input $x \in X$, define a
interval $I$ that is representable in $\mathbb{H}$ such that
$RN_{\mathbb{T}}(v) = RN_{\mathbb{T}}(f(x))$ for all $v \in I$. If
$A_{\mathbb{H}}(x) \in I$, then rounding the result of
$A_{\mathbb{H}}(x)$ to $\mathbb{T}$ produces the correctly rounded
result (\ie,
$RN_{\mathbb{T}}(A_{\mathbb{H}}(x)) = RN_{\mathbb{T}}(f(x))$).  For
each input $x$, if $A_{\mathbb{H}}(x)$ can produce a value that lies
within its corresponding rounding interval, then it will produce a
correctly rounded result. Thus, the pair $(x, I)$ for each input $x$
defines constraints on the output of $A_{\mathbb{H}}(x)$ such that
$RN_{\mathbb{T} }(A_{\mathbb{H}}(x))$ is a correctly rounded result.

Figure~\ref{alg:calculatel} presents our algorithm to compute
constraints $(x, I)$. For each input $x$ in our input domain $X$, we
compute the correctly rounded result of $f(x)$ using an oracle and
produce $y$. Next, we compute the rounding interval of $y$ where all
values in the interval round to $y$. The rounding interval can be
computed as follows. First, we identify $t_l$, the preceding value of
$y$ in $\mathbb{T}$ (line 11 in Figure ~\ref{alg:calculatel}). Then we
find the minimum value $l \in \mathbb{H}$ between $t_l$ and $y$ where
$l$ rounds to $y$ (line 12 in Figure ~\ref{alg:calculatel}). Similarly
for the upper bound, we identify $t_u$, the succeeding value of $y$ in
$\mathbb{T}$ (line 13 in Figure ~\ref{alg:calculatel}), and find the
maximum value $h \in \mathbb{H}$ between $y$ and $t_u$ where $h$
rounds to $y$ (line 14 in Figure ~\ref{alg:calculatel}). Then, $[l,
  h]$ is the rounding interval of $y$ and all values in $[l, h]$ round
to $y$.
Thus, the pair $(x, [l, h])$ specifies a constraint on the output of
$A_{\mathbb{H}}(x)$ to produce the correctly rounded result for input
$x$.  We generate such constraints for each input in the entire domain
(or for a sample of inputs) and produce a list of such constraints
(lines 7-9 in Figure ~\ref{alg:calculatel}).

\subsection{Calculating the Reduced Input and Reduced Interval}
After the previous step, we have a list of constraints, $(x, I)$, that
need to be satisfied by our approximation $A_{\mathbb{H}}(x)$ to
produce correctly rounded outputs. If we do not perform any range
reduction, then we can generate a polynomial that satisfies these
constraints. However, it is necessary to perform range
reduction~($RR$) in practice to reduce the complexity of the
polynomial and to improve performance. Range reduction is accompanied
by output compensation~($OC$) to produce the final output. Hence,
$A_{\mathbb{H}}(x) =
OC_{\mathbb{H}}(P_{\mathbb{H}}(RR_{\mathbb{H}}(x)), x)$.  Our goal is
to synthesize a polynomial $P_{\mathbb{H}}(x')$ that operates on the range
reduced input $x'$ and $A_{\mathbb{H}}(x) =
OC_{\mathbb{H}}(P_{\mathbb{H}}(RR_{\mathbb{H}}(x)), x)$ produces a
value in $I$ for each input $x$, which rounds to the correct output.

To synthesize this polynomial, we have to identify the reduced input
and the reduced interval for an input $x$ such that
$A_{\mathbb{H}}(x)$ produces a value in the rounding interval $I$
corresponding to $x$. The reduced input is available by applying range
reduction $x' = RR (x)$. Next, we need to compute the reduced interval
corresponding to $x'$.
The output of the polynomial on the reduced input will be fed to the
output compensation function to compute the output for the original
input.  For the reduced input $x'$ corresponding to the original input
$x$, $y' = P_{\mathbb{H}}(x')$, $A_{\mathbb{H}}(x) =
OC_{\mathbb{H}}(y', x)$, and $A_{\mathbb{H}}(x)$ must be within the
interval $I$ for input $x$ to produce a correct output. Hence, our
high-level strategy is to use the inverse of the output compensation
function to compute the reduced interval, which is feasible when the
output compensation function is continuous and bijective. In our
experience, all commonly used output compensation functions are
continuous and bijective.

\begin{figure}
\small
\begin{minipage}[t]{0.56\textwidth}
\begin{algorithm}[H]
\DontPrintSemicolon
\SetKwFunction{FMain}{Main}
\SetKwFunction{FCalculateLp}{CalcRedIntervals}
\SetKwFunction{FCalculateLambda}{CombineRedIntervals}
\SetKwFunction{FGeneratePoly}{GeneratePoly}
\SetKwFunction{FAdjLower}{GetPrecVal}
\SetKwFunction{FAdjHigher}{GetSuccVal}
\SetKwProg{Fn}{Function}{:}{}
\Fn{\FCalculateLp{$L$, $\mathbb{H}$, $RR_{\mathbb{H}}$, $OC_{\mathbb{H}}$}}{
    $L' \leftarrow \emptyset$\;
    \ForEach{$(x, [l, h]) \in L$} {
      $x' \leftarrow RR_{\mathbb{H}}(x)$\;
      \uIf{$OC_{\mathbb{H}}$ is an increasing function} {
        $[\alpha, \beta] \leftarrow [OC_{\mathbb{H}}^{-1}(l, x), OC_{\mathbb{H}}^{-1}(h, x)]$
      }
      \lElse{
        $[\alpha, \beta] \leftarrow [OC_{\mathbb{H}}^{-1}(h, x), OC_{\mathbb{H}}^{-1}(l, x)]$
      }
      \While{$OC_{\mathbb{H}}(\alpha, x) \notin [l, h]$}{
        $\alpha \leftarrow $ \FAdjHigher{$\alpha$, $\mathbb{H}$}\;
        \lIf{$\alpha > \beta$}{
          \Return{$\emptyset$}
        }
      }
      \While{$OC_{\mathbb{H}}(\beta, x) \notin [l, h]$}{
        $\beta \leftarrow $ \FAdjLower{$\beta$, $\mathbb{H}$}\;
        \lIf{$\alpha > \beta$}{
          \Return{$\emptyset$}
        }
      }
      $L' \leftarrow L' \cup \{(x', [\alpha, \beta])\}$\;
    }
    
    \Return{$L'$}\;
  }
\end{algorithm}
\end{minipage}
\begin{minipage}[t]{0.43\textwidth}
\begin{algorithm}[H]
  \DontPrintSemicolon
  \setcounter{AlgoLine}{18}
\SetKwFunction{FMain}{Main}
\SetKwFunction{FCalculateLp}{CalRedIntervals}
\SetKwFunction{FCalculateLambda}{CombineRedIntervals}
\SetKwFunction{FGeneratePoly}{GeneratePoly}
\SetKwFunction{FAdjLower}{GetPrecVal}
\SetKwFunction{FAdjHigher}{GetSuccVal}
\SetKwProg{Fn}{Function}{:}{}
\Fn{\FCalculateLambda{$L'$}}{
  $\hat{X} \leftarrow \{x' \mid (x',I') \in L'\}$\;
  $\Lambda \leftarrow \emptyset$\;
  
  \ForEach{$\hat{x} \in \hat{X}$} {
    $\Omega \leftarrow \{I' \mid (\hat{x}, I') \in L'$\}\;
    $\Psi \leftarrow \bigcap_{I' \in \Omega} I'$\;
    \lIf{$\Psi = \emptyset$} {
      \Return{$\emptyset$}
    }
    $\Lambda \leftarrow \Lambda \cup \{(\hat{x}, \Psi)\}$\;
    }
    \Return{$\Lambda$}\;
    
  }
\end{algorithm}
\end{minipage}

\caption{\texttt{CalcRedIntervals} computes the reduced input $x'$ and
  the reduced interval $I'$ for each constraint pair $(x, I)$ in
  $L$. The reduced constraint pair $(x', I')$ specifies the bound on
  the output of $P_{\mathbb{H}}(x')$ such that it produces the correct value for
  the input $x$. \texttt{CombineRedIntervals} combines any reduced
  constraints with the same reduced input, \ie $(x_1', I_1')$ and
  $(x_2', I_2')$ where $x_1' = x_2'$ into a single combined constraint
  $(x_1, \Psi)$ by computing the common interval range in $I_1'$ and
  $I_2'$.}
\label{alg:calculatelp}
\end{figure}

However, the output compensation function is evaluated in $\mathbb{H}$
, which necessitates us to take any numerical error in output
compensation with $\mathbb{H}$ into account.
Figure~\ref{alg:calculatelp} describes our algorithm to compute
reduced constraint $(x', I')$ for each $(x, I) \in L$ when the output
compensation is performed in $\mathbb{H}$.

To compute the reduced interval $I'$ for each constraint pair $(x, [l,
  h]) \in L$, we evaluate the values $v_1 = OC_{\mathbb{H}}^{-1}(l,
x)$ and $v_2 = OC_{\mathbb{H}}^{-1}(h, x)$ and create an interval
$[\alpha, \beta] = [v_1, v_2]$ if $OC_{\mathbb{R}}(y', x)$ is an
increasing function (lines 5-6 in Figure~\ref{alg:calculatelp}) or
$[v_2, v_1]$ if $OC_{\mathbb{R}}(y', x)$ is a decreasing function
(line 7 in Figure~\ref{alg:calculatelp}). The interval $[\alpha,
  \beta]$ is a candidate for $I'$. Then, we verify that the output
compensated value of $\alpha$ is in $[l, h]$ (\ie, $I$). If it is not,
we replace $\alpha$ with the succeeding value in $\mathbb{H}$ and
repeat the process until $OC_{\mathbb{H}}(\alpha, x)$ is in $I$ (lines
8-11 in Figure~\ref{alg:calculatelp}). Similarly, we verify that the
output compensated value of $\beta$ is in $[l, h]$ and repeatedly
replace $\beta$ with the preceding value in $\mathbb{H}$ if it is not
(lines 12-15 in Figure~\ref{alg:calculatelp}). If $\alpha > \beta$ at
any point during this process, then it indicates that there is no
polynomial $P(x')$ that can produce the correct result for all inputs.
As there are only finitely many values between $[\alpha, \beta]$ in
$\mathbb{H}$, this process terminates.  In the case when our algorithm
is not able to find a polynomial, the user can provide either a
different range reduction/output compensation function or increase the
precision to be higher than $\mathbb{H}$.

If the resulting interval $[\alpha, \beta] \neq \emptyset$, then $I' =
[\alpha, \beta]$ is our reduced interval. The reduced constraint pair,
$(x', [\alpha, \beta])$ created for each $(x, I) \in L$ specifies the
constraint on the output of $P_{\mathbb{H}}(x')$ such that
$A_{\mathbb{H}}(x) \in I$.  Finally, we create a list $L'$ containing
such reduced constraints.

\subsection{Combining the Reduced Constraints}
Each reduced constraint $(x_i', I_i') \in L'$ corresponds to a
constraint $(x_i, I_i) \in L$. It specifies the bound on the output of
$P_{\mathbb{H}}(x_i')$ (\ie, $P_{\mathbb{H}}(x_i') \in I_i'$ should be
satisfied), which ensures $A_{\mathbb{H}}(x_i)$ produces a value in
$I_i$.
Range reduction reduces the original input $x_i$ in the entire input
domain of $f(x)$ to a reduced input $x_i'$ in the reduced
domain. Hence, multiple inputs in the entire input domain can be range
reduced to the same reduced input. More specifically, there can exist
multiple constraints $(x_1, I_1), (x_2, I_2), \dots \in L$ such that
$RR_{\mathbb{H}}(x_1) = RR_{\mathbb{H}}(x_2) = \hat{x}$. Consequently,
$L'$ can contain reduced constraints ($\hat{x}, I_1'$), ($\hat{x},
I_2'$) $\dots \in L'$.
The polynomial $P_{\mathbb{H}}(\hat{x})$ must produce a value in
$I_1'$ to guarantee that $A_{\mathbb{H}}(x_1) \in I_1$. It must also
be within $I_2'$ to guarantee $A_{\mathbb{H}}(x_2) \in I_2$.  Hence,
for each unique reduced input $\hat{x}$, $P_{\mathbb{H}}(\hat{x})$
must satisfy all reduced constraints corresponding to $\hat{x}$, \ie
$P_{\mathbb{H}}(\hat{x}) \in I_1' \cap I2'$.

The function \texttt{CombineRedIntervals} in
Figure~\ref{alg:calculatelp} combines all reduced constraints with the
same reduced input by identifying the common interval ($\Psi$ in line
24 in Figure~\ref{alg:calculatelp}). If such a common interval
does not exist, then it is infeasible to find a single polynomial
$P_{\mathbb{H}}(x')$ that produces correct outputs for all inputs
before range reduction.  Otherwise, we create a pair $(\hat{x}, \Psi)$
for each unique reduced interval $\hat{x}$ and produce a list of
constraints $\Lambda$ (line 26 in Figure~\ref{alg:calculatelp}).

\subsection{Generating the Polynomial Using Linear Programming}

Each reduced constraint $(x', [l', h']) \in \Lambda$ requires that
$P_{\mathbb{H}}(x')$ satisfy the following condition: $l' \leq
P_{\mathbb{H}}(x') \leq h'$. This constraint ensures that when
$P_{\mathbb{H}}(x')$ is combined with range reduction and output
compensation, it produces the correctly rounded result for all inputs.
When we are trying to generate a polynomial of degree $d$, we can
express each of the above constraints in the form:
\[
l'\leq c_0 + c_1 x' + c_2 (x')^2 + ... + c_d (x')^d \leq h'
\]

The goal is to find coefficients for the polynomial evaluated in
$\mathbb{H}$. Here, $x'$, $l'$ and $h'$ are constants from perspective
of finding the coefficients. We can express all constraints $(x_i',
[l_i', h_i']) \in \Lambda$ in a single system of linear inequalities
as shown below, which can be solved using a linear programming (LP)
solver.
\[
\begin{bmatrix}
l_1' \\
l_2' \\
\vdots \\
l_{|\Lambda|}'
\end{bmatrix}
\leq
\begin{bmatrix}
1 && x_1' && \dots && (x_1')^d \\
1 && x_2' && \dots && (x_2')^d \\
\vdots && \vdots && \ddots && \vdots \\
1 && x_{|\Lambda|}' && \dots && (x_{|\Lambda|}')^d
\end{bmatrix}
\begin{bmatrix}
c_0 \\
c_1 \\
\vdots \\
c_d
\end{bmatrix}
\leq
\begin{bmatrix}
h_1' \\
h_2' \\
\vdots \\
h_{|\Lambda|}'
\end{bmatrix}
\]

Given a system of inequalities, the LP solver finds a solution for the
coefficients with real numbers. The polynomial when evaluated in real
(\ie $P_{\mathbb{R}}(x')$) satisfies all constraints in $\Lambda$.
However, numerical errors in polynomial evaluation in $\mathbb{H}$ can
cause the result to not satisfy $\Lambda$.  We propose a
\textit{search-and-refine} approach to address this problem. We use
the LP solver to solve for the coefficients of $P_{\mathbb{R}}(x')$
that satisfy $\Lambda$ and then check if $P_{\mathbb{H}}(x')$ that
evaluates $P_{\mathbb{R}}(x')$ in $\mathbb{H}$ satisfies the
constraints in $\Lambda$.  If $P_{\mathbb{H}}(x')$ does not satisfy a
constraint $(x', [l', h']) \in \Lambda$, then we refine the reduced
interval $[l', h']$ to a smaller interval. Subsequently, we use the LP
solver to generate the coefficients of $P_{\mathbb{R}}(x')$ for the
refined constraints. This process is repeated until either
$P_{\mathbb{H}}(x')$ satisfies all reduced constraints in $\Lambda$ or
the LP solver determines that there is no polynomial that satisfies
all the constraints.

\begin{figure}
\small
\begin{minipage}[t]{0.46\textwidth}
\begin{algorithm}[H]
\DontPrintSemicolon
\SetKwFunction{FGeneratePoly}{GeneratePoly}
\SetKwFunction{FAdjLower}{GetPrecVal}
\SetKwFunction{FAdjHigher}{GetSuccVal}
\SetKwFunction{FSolve}{LPSolve}
\SetKwFunction{FPoly}{CreateP}
\SetKwFunction{FVerify}{Verify}
\SetKw{Continue}{continue}
\SetKwProg{Fn}{Function}{:}{}
\Fn{\FGeneratePoly{$\Lambda$, $\mathbb{H}$ $d$}}{
    $\Upsilon \leftarrow \Lambda$\;
    \While{true} {
        $C \leftarrow$ \FSolve{$\Upsilon$, $d$}\;
        \lIf{$C = \emptyset$} {
            \Return{(false, DNE)}
        }
        $P_{\mathbb{H}} \leftarrow$ \FPoly{$C$, $d$, $\mathbb{H}$}\;
        $\Upsilon \leftarrow$ \FVerify{$P_{\mathbb{H}}$, $\Lambda$, $\Upsilon$,
          $\mathbb{H}$}\;
        \lIf{$\Upsilon =\emptyset$} {
          \Return{(true, $P_{\mathbb{H}}$)}
        }
    }
  }
\end{algorithm}
\end{minipage}
\begin{minipage}[t]{0.53\textwidth}
\begin{algorithm}[H]
\DontPrintSemicolon
\setcounter{AlgoLine}{9}
\SetKwFunction{FMain}{Main}
\SetKwFunction{FCalculateLp}{Calculate$L'$}
\SetKwFunction{FCalculateLambda}{Calculate$\Lambda$}
\SetKwFunction{FGeneratePoly}{GeneratePoly}
\SetKwFunction{FAdjLower}{GetPrecVal}
\SetKwFunction{FAdjHigher}{GetSuccVal}
\SetKwFunction{FHex}{Hex}
\SetKwFunction{FVerify}{Verify}
\SetKw{Continue}{continue}
\SetKwProg{Fn}{Function}{:}{}
\Fn{\FVerify{$P_{\mathbb{H}}$, $\Lambda$, $\Upsilon$, $\mathbb{H}$}}{
    
    $Z \leftarrow \{(x', \Psi, \psi) \mid (x', \Psi) \in \Lambda, (x', \psi) \in \Upsilon\}$\;
    \ForEach{$(x', [l', h'], [\sigma, \mu]) \in Z$} {
        \uIf{$P_{\mathbb{H}} (x') < l'$} {
            $\Upsilon \leftarrow \Upsilon - \{(x', [\sigma, \mu])\}$\;
            $\sigma' \leftarrow $ \FAdjHigher{$\sigma$, $\mathbb{H}$}\;
            \Return{$\Upsilon \cup \{(x', [\sigma', \mu])\}$}\;
        }
        \ElseIf{$P_{\mathbb{H}} (x') > h'$} {
            $\Upsilon \leftarrow \Upsilon - \{(x', [\sigma, \mu])\}$\;
            $\mu' \leftarrow $ \FAdjLower{$\mu$, $\mathbb{H}$}\;
            \Return{$\Upsilon \cup \{(x', [\sigma, \mu'])\}$}\;
        }
    }
    \Return{$\emptyset$}\;
  }
\end{algorithm}
\end{minipage}

\caption{The function \texttt{GeneratePoly} generates a polynomial
  $P_{\mathbb{H}}(x')$ of degree $d$ that satisfies all constraints in
  $\Lambda$ when evaluated in $\mathbb{H}$. If it cannot generate such
  a polynomial, then it returns \textit{false}. The function
  \texttt{LPSolve} solves for the real number coefficients of a
  polynomial $P_{\mathbb{R}}(x)$ using an LP solver where
  $P_{\mathbb{R}}(x)$ satisfies all constraints in $\Lambda$ when
  evaluated in real number. \texttt{CreateP} creates
  $P_{\mathbb{H}}(x)$ that evaluates the polynomial
  $P_{\mathbb{R}}(x)$ in $\mathbb{H}$. The \texttt{Verify} function
  checks whether the generated polynomial $P_{\mathbb{H}}(x)$
  satisfies all constraints in $\Lambda$ when evaluated in
  $\mathbb{H}$ and refines the constraints to a smaller interval for
  each constraint that $P_{\mathbb{H}}(x)$ does not satisfy.}
\label{alg:generatep}
\end{figure}

Figure~\ref{alg:generatep} provides the algorithm used for generating
the coefficients of the polynomial using the LP solver.
$\Upsilon$ tracks the refined constraints for $P_{\mathbb{H}}(x')$
during our search-and-refine process.  Initially, $\Upsilon$ is set to
$\Lambda$~(line 2 in Figure~\ref{alg:generatep}).
Here, $\Upsilon$ is used to generate the polynomial and $\Lambda$ is used to
  to verify that the generated polynomial satisfies all
  constraints. If the generated polynomial does not satisfy $\Lambda$,
  we restrict the intervals in $\Upsilon$.

We use an LP solver to solve for the coefficients of the
$P_{\mathbb{R}}(x')$ that satisfy all constraints in
$\Upsilon$~(line 4 in Figure~\ref{alg:generatep}). If the LP solver
cannot find the coefficients, our algorithm concludes that it is not
possible to generate a polynomial and terminates (line 5 in
Figure~\ref{alg:generatep}). Otherwise, we create $P_{\mathbb{H}}(x')$
that evaluates $P_{\mathbb{R}}(x')$ in $\mathbb{H}$ by rounding all
coefficients to $\mathbb{H}$ and perform all operations in
$\mathbb{H}$ (line 6 in Figure~\ref{alg:generatep}). The resulting
$P_{\mathbb{H}}(x')$ is a candidate for the correct polynomial for
$A_{\mathbb{H}}(x)$.

Next, we verify that $P_{\mathbb{H}}(x')$ satisfies all constraints in
$\Lambda$ (line 7 in Figure ~\ref{alg:generatep}). If
$P_{\mathbb{H}}(x')$ satisfies all constraints in $\Lambda$, then our
algorithm returns the polynomial. If there is a constraint $(x', [l',
  h']) \in \Lambda$ that is not satisfied by $P_{\mathbb{H}}(x')$,
then we further restrict the interval $(x', [\sigma, \mu])$ in
$\Upsilon$ corresponding to the reduced input $x'$. If
$P_{\mathbb{H}}(x')$ is smaller than the lower bound of the interval
constraint in $\Lambda$ (\ie $l'$), then we restrict the lower bound
of the interval constraint $\sigma$ in $\Upsilon$ to the value
succeeding $\sigma$ in $\mathbb{H}$ (lines 13-16 in
Figure~\ref{alg:generatep}). This forces the next coefficients for
$P_{\mathbb{R}}(x')$ that we generate using the LP solver to produce a
value larger than $l'$. Likewise, if $P_{\mathbb{H}}(x')$ produces a
value larger than the upper bound of the interval constraint in
$\Lambda$ (\ie $h'$), then we restrict the upper bound of the interval
constraint $\mu$ in $\Upsilon$ to the value preceding $\mu$ in
$\mathbb{H}$ (lines 17-20 in Figure~\ref{alg:generatep}).

We repeat this process of generating a new candidate polynomial with
the refined constraints $\Upsilon$ until it satisfies all constraints
in $\Lambda$ or the LP solver determines that it is infeasible.  If a
constraint $(x', [\sigma, \mu]) \in \Upsilon$ is restricted to the
point where $\sigma > \mu$ (or $[\sigma, \mu] = \emptyset$), then the
LP solver will determine that it is infeasible to generate the
polynomial. When we are successful in generating a polynomial, then
$P_{\mathbb{H}}(x)$ used in tandem with range reduction and the output
compensation in $\mathbb{H}$ is checked to ascertain that it produces
the correctly rounded results for all inputs.

\section{Experimental Evaluation}
\label{sec:evaluation}
This section describes our prototype for generating correctly rounded
elementary functions and the math library that we developed for
\texttt{Bfloat16}, posit, and \texttt{float} data types. We present
case studies for approximating elementary functions $10^{x}$, $ln(x)$,
$log_2(x)$, and $log_{10}(x)$ with our approach for various types. We
also evaluate the performance of our correctly rounded elementary
functions with state-of-the-art approximations.

\subsection{\tool Prototype and Experimental Setup}
\myblue{\textbf{Prototype.} We use \tool to refer to our prototype for
  generating correctly rounded elementary functions and the resulting
  math libraries generated from it. \tool supports \texttt{Bfloat16},
  \texttt{Posit16} (16-bit posit type in the Posit
  standard~\cite{Gustafson:online:2017:posit}), and the 32-bit
  \texttt{float} type in the FP representation.}
The user can provide custom range reduction and output compensation
functions. The prototype uses the MPFR
library~\cite{Fousse:toms:2007:mpfr} with $2,000$ precision bits as
the oracle to compute the real result of $f(x)$ and rounds it to the
target representation. Although there is no bound on the precision to
compute the oracle result~(\ie, Table-maker's dilemma), prior work has
shown around $160$ precision bits in the worst case is empirically
sufficient for the \texttt{double}
representation~\cite{Lefevre:worstcase:arith:2001}. Hence, we use
$2,000$ precision bits with the MPFR library to compute the oracle
result. The prototype uses SoPlex~\cite{Gleixner:soplex:issac:2012,
  Gleixner:soplex:tech:2015}, an exact rational LP solver as the
arbitrary precision LP solver for polynomial generation from
constraints.

\myblue{\tool's math library contains correctly rounded elementary
  functions for multiple data types. It contains twelve functions for
  \texttt{Bfloat16} and eleven functions for \texttt{Posit16}. The
  library produces the correctly rounded result for all inputs. To
  show that our approach can be used with large data types, \tool also
  includes a correctly rounded $log_2(x)$ for the 32-bit
  \texttt{float} type.}

\myblue{\tool performs range reduction and output compensation using
  the \texttt{double} type.  We use state-of-the-art range reduction
  techniques for various elementary functions. Additionally, we split
  the reduced domain into multiple disjoint smaller domains using the
  properties of specific elementary functions to generate efficient
  polynomials. We evaluate all polynomials using the Horner's method,
  \ie $P(x) = c_0 + x(c_1 + x (c_2 +
  \dots))$~\cite{borwein:polynomials:book:1995}, which reduces the
  number of operations in polynomial evaluation.
}

\myblue{The entire \tool prototype is written in C++. \tool is
  open-source~\cite{rlibm,rlibmgenerator}.  Although we have not
  optimized \tool for a specific target, it already has better
  performance than state-of-the-art approaches.}

\textbf{Experimental setup.} We describe our experimental setup to
check the correctness and performance of \tool. \myblue{There is no
  math library specifically designed for \texttt{Bfloat16}
  available. To compare the performance of our \texttt{Bfloat16}
  elementary functions, we convert the \texttt{Bfloat16} input to a
  \texttt{float} or a \texttt{double}, use \texttt{glibc}'s (and
  \texttt{Intel}'s) \texttt{float} or \texttt{double} math library
  function, and then convert the result back to \texttt{Bfloat16}.  We
  use SoftPosit-Math library~\cite{Leong:online:2019:positmath} to
  compare our \texttt{Posit16} functions.  We also compare our
  \texttt{float} $log_2(x)$ function to the one in
  \texttt{glibc}/\texttt{Intel}'s library.}

\myblue{For our performance experiments, we compiled the functions in
  \tool with g++ at the \texttt{O3} optimization level. All
  experiments were conducted on a machine with 4.20GHz Intel i7-7700K
  processor and 32GB of RAM, running the Ubuntu 16.04 LTS operating
  system.  We count the number of cycles taken to compute the
  correctly rounded result for each input using hardware performance
  counters. We use both the average number of cycles per input and
  total cycles for all inputs to compare performance.}

\begin{table}
  \small
  \caption{\myblue{(a) The list of \texttt{Bfloat16} functions used
      for our evaluation. The second column shows whether \tool
      produces the correct result for all inputs.  The third column
      and fourth column shows whether \texttt{glibc}'s float and
      \texttt{Intel}'s float library produces the correct result for
      all \texttt{Bfloat16} inputs. We use (\cmark) to indicate
      correctly rounded results and \xmark, otherwise. (b) The list of
      \texttt{Posit16} functions used. The second column shows whether
      \tool produces the correct results for all inputs. The third
      column shows whether the functions in SoftPosit-Math produces
      correctly rounded results for all inputs. \texttt{N/A} indicates
      that function is not available in SoftPosit-Math. (c) The
      \texttt{float} function used. First column indicates whether
      \tool produces the correctly rounded result for all inputs. In
      the second and third column, we show whether \texttt{glibc}'s
      float and \texttt{Intel}'s float math library produce the
      correct result for all inputs}.}
  \begin{minipage}{0.49\textwidth}
    \begin{tabular}{| c | c | c | c |} 
      \hline
      \begin{tabular}{@{}c@{}}\textbf{Bfloat16} \\ Functions\end{tabular}
      & \begin{tabular}{@{}c@{}}Using \\ \tool\end{tabular}
      & \begin{tabular}{@{}c@{}}Using \\ glibc float\end{tabular}
      & \begin{tabular}{@{}c@{}}Using \\ Intel float\end{tabular}\\
      \hline
      \hline
      $ln(x)$ & \cmark & \cmark & \cmark \\ \hline
      $log2(x)$ & \cmark & \cmark & \cmark \\ \hline
      $log10(x)$ & \cmark & \cmark & \cmark \\ \hline
      $exp(x)$ & \cmark & \cmark & \cmark \\ \hline
      $exp2(x)$ & \cmark & \cmark & \cmark \\ \hline
      $exp10(x)$ & \cmark & \xmark & \xmark \\ \hline
      $sinpi(x)$ & \cmark & N/A & \cmark \\ \hline
      $cospi(x)$ & \cmark & N/A & \cmark \\ \hline
      $sqrt(x)$ & \cmark & \cmark & \cmark \\  \hline 
      $cbrt(x)$ & \cmark & \cmark & \cmark \\  \hline
      $sinh(x)$ & \cmark & \cmark & \cmark \\  \hline
      $cosh(x)$ & \cmark & \cmark & \cmark \\  \hline
      \multicolumn{4}{c}{(a) Correctly rounded results with Bfloat16}
    \end{tabular}
  \end{minipage}
  \begin{minipage}{0.49\textwidth}
    \begin{minipage}{1.0\textwidth}
      \begin{tabular}{| c | c | c |} 
        \hline
        \begin{tabular}{@{}c@{}}\textbf{Posit16} \\ Functions\end{tabular}
        & \begin{tabular}{@{}c@{}}Using \\ \tool\end{tabular}
        & \begin{tabular}{@{}c@{}}Using \\ SoftPosit-Math\end{tabular}\\
        \hline
        \hline
        $ln(x)$ & \cmark & \cmark \\ \hline
        $log2(x)$ & \cmark & \cmark \\ \hline
        $log10(x)$ & \cmark & N/A \\ \hline
        $sinpi(x)$ & \cmark & \cmark \\ \hline
        $cospi(x)$ & \cmark & \cmark \\ \hline
        $sqrt(x)$ & \cmark & \cmark \\ \hline
        $exp(x)$ & \cmark & N/A \\ \hline
        $exp2(x)$ & \cmark & \cmark \\ \hline
        $exp10(x)$ & \cmark & \cmark \\ \hline
        $sinh(x)$ & \cmark & N/A \\ \hline
        $cosh(x)$ & \cmark & N/A \\ \hline
        \multicolumn{3}{c}{(b) Correctly rounded results with Posit16} 
      \end{tabular}
    \end{minipage}

    \begin{minipage}{1.0\textwidth}
      \begin{tabular}{| c | c | c | c |} 
        \hline
        \begin{tabular}{@{}c@{}}\textbf{float} \\ Functions\end{tabular}
        & \begin{tabular}{@{}c@{}}Using \\ \tool\end{tabular} 
        & \begin{tabular}{@{}c@{}}Using \\ glibc float\end{tabular}
        & \begin{tabular}{@{}c@{}}Using \\ Intel float\end{tabular}\\
        \hline
        \hline
        $log2(x)$ & \cmark & \xmark & \xmark \\ \hline
        \multicolumn{4}{c}{(c) Correctly rounded result with 32-bit float}
      \end{tabular}
    \end{minipage}
    
  \end{minipage}
\label{tbl:accuracy}
\end{table}

\subsection{Correctly Rounded Elementary Functions in \tool}
\myblue{Table~\ref{tbl:accuracy}(a) shows that \tool produces the
  correctly rounded result for all inputs with numerous elementary
  functions for the \texttt{Bfloat16} representation.}
\myblue{In contrast to \tool, we discovered that re-purposing existing
  \texttt{glibc}'s or \texttt{Intel}'s \texttt{float} library for
  \texttt{Bfloat16} did not produce the correctly rounded result for
  all inputs.  The case with input $x = -0.0181884765625$ for
  $exp10(x)$ was already discussed in
  Section~\ref{sec:background:motivation}. This case is interesting
  because both \texttt{glibc}'s and \texttt{Intel}'s \texttt{float}
  math library produces the correctly rounded result of $exp10(x)$
  with respect to the \texttt{float} type. However, the result for
  \texttt{Bfloat16} is wrong.
We found that both \texttt{glibc}'s and \texttt{Intel}'s
\texttt{double} library produce the correctly rounded result for all
inputs for \texttt{Bfloat16}.  Our experience during this evaluation
illustrates that a correctly rounded function for $\mathbb{T'}$ does
not necessarily produce a correctly rounded library for $\mathbb{T}$
even if $\mathbb{T'}$ has more precision that $\mathbb{T}$.}

\myblue{Table~\ref{tbl:accuracy}(b) reports that \tool produces
  correctly rounded results for all inputs with elementary functions
  for \texttt{Posit16}.  We found that SoftPosit-Math functions also
  produce the correctly rounded result for the available
  functions. However, functions $log10(x)$, $exp10(x)$, $sinh(x)$, and
  $cosh(x)$ are not available in the SoftPosit-Math library.}

\myblue{Table~\ref{tbl:accuracy}(c) reports that \tool produces the
  correctly rounded results for $log2(x)$ for all inputs with the
  32-bit float data type. The corresponding function
  in \texttt{glibc}'s and \texttt{Intel}'s \texttt{double} library
  produces the correct result for all inputs. However,
  \texttt{glibc}'s and \texttt{Intel}'s \texttt{float} math library
  does not produce the correctly rounded result for all inputs.
We found approximately fourteen million inputs where \texttt{glibc}'s
\texttt{float} library produces the wrong result and $276$ inputs
where \texttt{Intel}'s \texttt{float} library produces the wrong
result. In summary, we are able to generate correctly rounded results
for many elementary functions for various representations using our
proposed approach.}

\begin{table}
  \small
  \caption{\myblue{Details about the generated polynomials. For each
    elementary function, we report the total number of inputs in the
    target representation, number of special inputs, total number of
    reduced intervals, the number of intervals that we encoded in the
    LP query, the total time taken to generate the polynomials, the
    number of polynomials generated, the degree of the generated
    polynomial, and the number of terms in the polynomial}.}
    \begin{tabular}{| c | c | c | c | c | c | c | c | c |} 
      \hline
      \begin{tabular}{@{}c@{}}Elementary \\ Functions \end{tabular}
      & \begin{tabular}{@{}c@{}}Total \# \\ of Inputs \end{tabular}
      & \begin{tabular}{@{}c@{}}Special \\ Inputs\end{tabular}
      & \begin{tabular}{@{}c@{}}Reduced \\ Intervals\end{tabular}
      & \begin{tabular}{@{}c@{}}Intervals \\ Used in LP\end{tabular}
      & \begin{tabular}{@{}c@{}c@{}}Total \\ Time \\ (Seconds)\end{tabular}
      & \begin{tabular}{@{}c@{}c@{}}\# of \\ Poly- \\ nomials\end{tabular}
      & Degree
      & \begin{tabular}{@{}c@{}}\# of \\ Terms\end{tabular} \\
      \hline
      \hline
      \multicolumn{9}{| c |}{\textbf{Bfloat16 functions}} \\ \hline
      $ln(x)$  & $2^{16}$ & 32897 & 128 & 128 & 0.84 & 1 & 7 & 4  \\ \hline
      $log2(x)$  & $2^{16}$ & 32897 & 128 & 128 & 8.65 & 1 & 5 & 3  \\ \hline
      $log10(x)$  & $2^{16}$ & 32897 & 128 & 128 & 1.63 & 1 & 5 & 3  \\ \hline
      $exp(x)$  & $2^{16}$ & 61716 & 3820 & 3820 & 2.9 & 1 & 4 & 5 \\ \hline
      $exp2(x)$  & $2^{16}$ & 61548 & 1937 & 1937 & 0.89 & 1 & 4 & 5 \\ \hline
      $exp10(x)$  & $2^{16}$ & 61696 & 3840 & 3840 & 3 & 1 & 4 & 5 \\ \hline
      $sinpi(x)$  & $2^{16}$ & 30976 & 16129 & 16129 & 32 & 2 & \begin{tabular}{@{}c@{}} 1 \\ 7\end{tabular}  & \begin{tabular}{@{}c@{}} 1 \\ 4\end{tabular} \\ \hline
      $cospi(x)$  & $2^{16}$ & 30976 & 16129 & 16129 & 32.2 & 3 & \begin{tabular}{@{}c@{}c@{}} 0 \\ 6 \\ 0 \end{tabular} & \begin{tabular}{@{}c@{}c@{}} 1 \\ 4 \\ 1 \end{tabular}  \\ \hline
      $sqrt(x)$  & $2^{16}$ & 32897 & 256 & 256 & 0.07 & 1 & 4 & 5 \\  \hline 
      $cbrt(x)$  & $2^{16}$ & 257 & 384 & 384 & 0.16 & 1 & 6 & 7 \\  \hline
      $sinh(x)$  & $2^{16}$ & 63084 & 422 & 422 & 0.27 & 3 & \begin{tabular}{@{}c@{}c@{}} 5 \\ 0 \\ 6\end{tabular} & \begin{tabular}{@{}c@{}c@{}} 3 \\ 1 \\ 4\end{tabular}  \\  \hline
      $cosh(x)$  & $2^{16}$ & 62980 & 471 & 471 & 0.27 & 2 & \begin{tabular}{@{}c@{}} 5 \\ 6\end{tabular} & \begin{tabular}{@{}c@{}} 3 \\ 4\end{tabular} \\  \hline
      
      \hline
      \hline
      \multicolumn{9}{| c |}{\textbf{Posit16 functions}} \\ \hline
      $ln(x)$  & $2^{16}$ & 32769 & 4096 & 4096 & 3.32 & 1 & 9 & 5  \\ \hline
      $log2(x)$  & $2^{16}$ & 32769 & 4096 & 4096 & 5.69 & 1 & 9 & 5  \\ \hline
      $log10(x)$  & $2^{16}$ & 32769 & 4096 & 4096 & 6.51 & 1 & 9 & 5  \\ \hline
      $exp(x)$  & $2^{16}$ & 8165 & 57371 & 1740 & 6.17 & 1 & 6 & 7 \\ \hline
      $exp2(x)$  & $2^{16}$ & 7160 & 24201 & 805 & 5.15 & 1 & 6 & 7 \\ \hline
      $exp10(x)$  & $2^{16}$ &12430 & 53106 & 1879 & 11.97 & 1 & 6 & 7 \\ \hline
      $sinpi(x)$  & $2^{16}$ & 1 & 12289 & 12289 & 37.99 & 2 & \begin{tabular}{@{}c@{}} 1 \\ 9\end{tabular}  & \begin{tabular}{@{}c@{}} 1 \\ 5\end{tabular} \\ \hline
      $cospi(x)$  & $2^{16}$ & 1 & 12289 & 12289 & 85.74 & 3 & \begin{tabular}{@{}c@{}c@{}} 0 \\ 8 \\ 0\end{tabular} & \begin{tabular}{@{}c@{}c@{}} 1 \\ 5 \\ 1\end{tabular}  \\ \hline
      $sqrt(x)$  & $2^{16}$ & 32769 & 8192 & 8192 & 77.08 & 2 & \begin{tabular}{@{}c@{}} 6 \\ 6\end{tabular} & \begin{tabular}{@{}c@{}} 7 \\ 7\end{tabular} \\  \hline 
      $sinh(x)$  & $2^{16}$ & 14804 & 13044 & 13044 & 37.44 & 2 & \begin{tabular}{@{}c@{}} 7 \\ 6\end{tabular} & \begin{tabular}{@{}c@{}} 4 \\ 4\end{tabular}  \\  \hline
      $cosh(x)$  & $2^{16}$ & 11850 & 14400 & 14400 & 391.94 & 4 & \begin{tabular}{@{}c@{}c@{}c@{}} 1 \\ 7 \\ 6 \\ 6\end{tabular}  & \begin{tabular}{@{}c@{}c@{}c@{}} 1 \\ 4 \\ 4 \\ 4\end{tabular} \\  \hline
      
      \hline
      \hline
      \multicolumn{9}{| c |}{\textbf{32-bit float function}} \\ \hline
      $log2(x)$  & $2^{32}$ & 2155872257 & 7165657 & 7775 & 220.59 & 1 & 5 & 5 \\ \hline
      
    \end{tabular}
  
\label{tbl:statistics}
\end{table}

\myblue{Table~\ref{tbl:statistics} provides details on the polynomials
  for each elementary function and for each data type. For some
  elementary functions, we had to generate piece-wise polynomials
  using a trial-and-error approach. As the degree of the generated
  polynomials and the number of terms in the polynomial are small, the
  resulting libraries are faster than the state-of-the-art
  libraries. The time taken by our tool to generate the resulting
  polynomials depends on the bit-width and the degree of the
  polynomial. It ranges from a few seconds to a few minutes.}

\iffalse It took us about $3.6$ minutes to produce the correct
polynomial for $log2(x)$ for a 32-bit float.  \fi

\subsection{Performance Evaluation of Elementary Functions in \tool}
\myblue{We empirically compare the performance of the functions in
  \tool for \texttt{Bfloat16}, \texttt{Posit16}, and a 32-bit
  \texttt{float} type to the corresponding ones in \texttt{glibc},
  \texttt{Intel}, and SoftPosit-Math libraries.}

\subsubsection{Performance of \texttt{Bfloat16} Functions in \tool}
\begin{figure}
    \begin{subfigure}[b]{0.49\linewidth}
    \includegraphics[width=\textwidth]{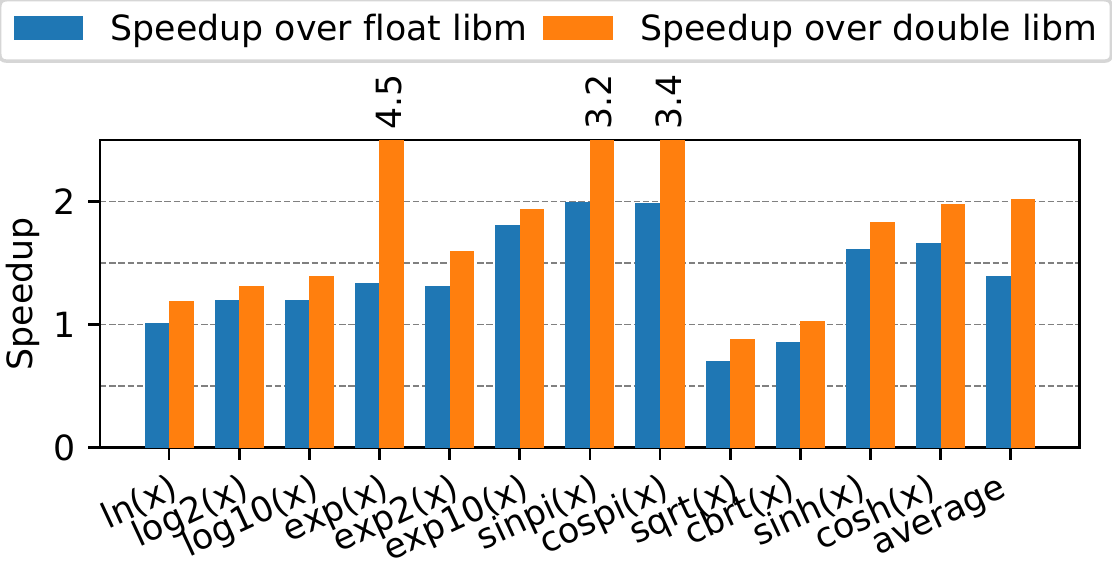}
    \caption{Speedup over \texttt{Glibc}'s math library}
    \end{subfigure}
    \begin{subfigure}[b]{0.49\linewidth}
    \includegraphics[width=\textwidth]{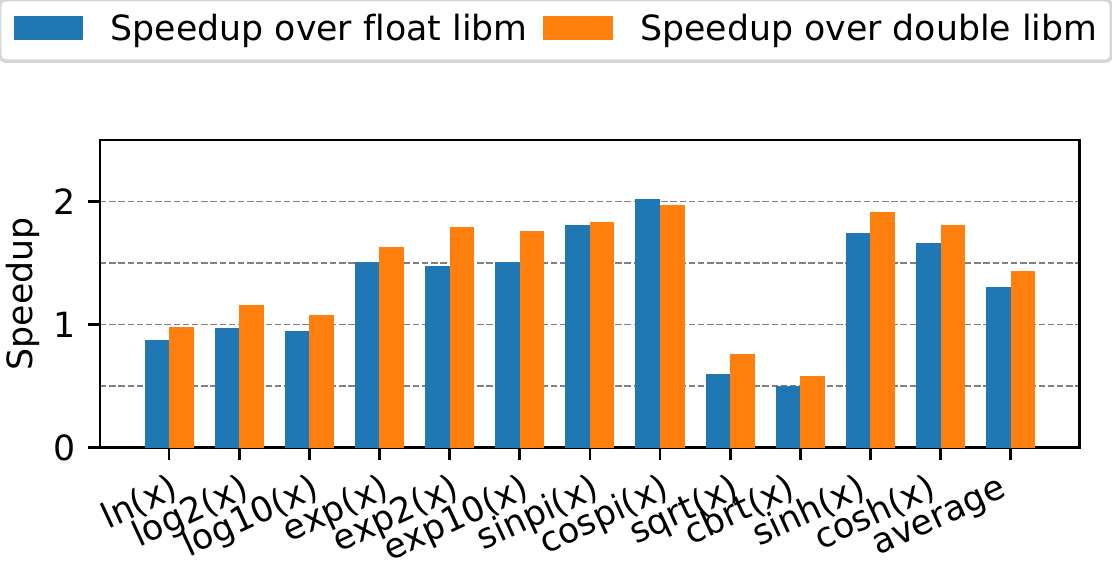}
    \caption{Speedup over \texttt{Intel}'s math library}
    \end{subfigure}
    \caption{(a) Speedup of \tool's elementary functions compared to a
      baseline using \texttt{Glibc}'s \texttt{float} math library
      (left bar) and \texttt{Glibc}'s \texttt{double} math library
      (right bar). (b) Speedup of \tool's elementary functions
      compared to a baseline using \texttt{Intel}'s \texttt{float}
      math library (left bar) and \texttt{Intel}'s \texttt{double}
      math library (right bar). These functions take a
      \texttt{Bfloat16} input and produce a \texttt{Bfloat16} output.}
    \label{fig:bfloatOverhead}
\end{figure}

To measure performance, we measure the amount of time it takes for
\tool to produce a \texttt{Bfloat16} result given a \texttt{Bfloat16}
input for all inputs.
\myblue{Similarly, we measure the time taken by \texttt{glibc} and
  \texttt{Intel} libraries to produce a \texttt{Bfloat16} output given
  a \texttt{Bfloat16} input. }
As $sinpi(x)$ and $cospi(x)$ are not available in \texttt{glibc}'s
libm, we transform $sinpi(x) = sin(\pi{x})$ and $cospi(x) =
cos(\pi{x})$ before using \texttt{glibc}'s $sin$ and $cos$ functions.
\myblue{\texttt{Intel}'s libm provides implementations of $sinpi(x)$
  and $cospi(x)$.}

Figure~\ref{fig:bfloatOverhead}(a) shows the speedup of \tool's
functions for \texttt{Bfloat16} compared to \texttt{glibc}'s
\texttt{float} math library (left bar in the cluster) and the
\texttt{double} library (right bar in the cluster).
\myblue{On average, \tool's functions are $1.39\times$ faster when
  compared to \texttt{glibc}'s \texttt{float} library and $2.02\times$
  faster over \texttt{glibc}'s \texttt{double} math
  library. Figure~\ref{fig:bfloatOverhead}(b) shows the speedup of
  \tool's functions for \texttt{Bfloat16} compared to \texttt{Intel}'s
  \texttt{float} math library (left boar in the cluster) and the
  \texttt{double} library (right bar in the cluster). On average,
  \tool's functions are $1.30\times$ faster when compared to
  \texttt{Intel}'s \texttt{float} library and $1.44\times$ faster
  compared to \texttt{Intel}'s \texttt{double} math library.}

\myblue{For $sqrt(x)$, \ourlibm's version has a slowdown because both
  \texttt{glibc} and \texttt{Intel} math library likely utilize the
  hardware instruction, \texttt{FSQRT}, to compute $sqrt(x)$ whereas
  \tool performs polynomial evaluation. Our $cbrt(x)$ function is
  slower than both the \texttt{glibc} and \texttt{Intel}'s math
  library and our logarithm functions are slower than \texttt{Intel}'s
  float math library. It is likely that they use sophisticated range
  reduction and has a lower degree polynomial. Overall, \tool's
  functions for \texttt{Bfloat16} not only produce correct results for
  all inputs but also are faster than the existing libraries
  re-purposed for \texttt{Bfloat16}.}

\subsubsection{Performance of \texttt{Posit16} Elementary Functions in \tool}
\begin{figure}
    \includegraphics[width=0.49\textwidth]{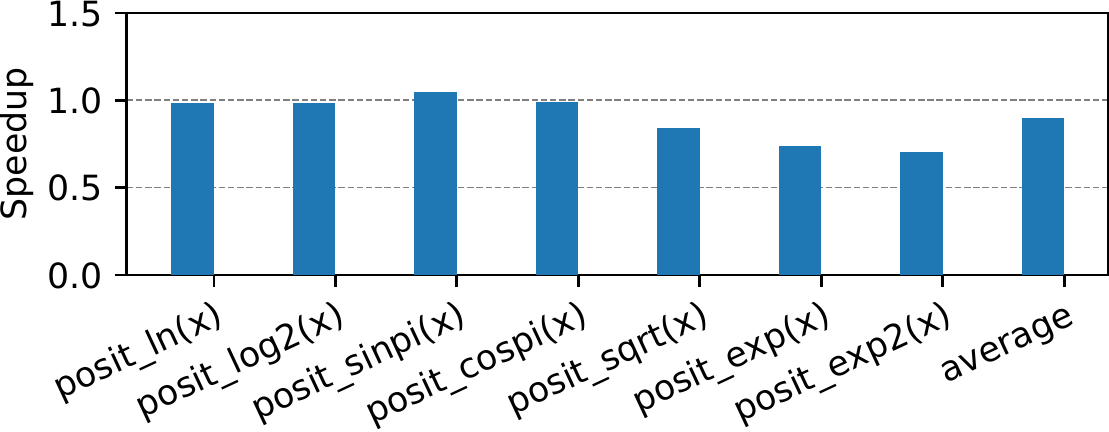}
    \caption{Performance speedup of \tool's functions compared to
      SoftPosit-Math library when the input is available as a
      \texttt{double}.  It avoids the cast from \texttt{Posit16} to
      \texttt{double} with \tool. SoftPosit-Math takes as input a
      \texttt{Posit16} value that is internally represented as an
      integer.}
    \label{fig:positOverhead}
\end{figure}
 Figure~\ref{fig:positOverhead} shows
the speedup of \tool's functions when compared to a baseline that uses
SoftPosit-Math functions. The \texttt{Posit16} input is cast to the
\texttt{double} type before using \tool. We did not measure the cost
of this cast, which can incur additional overhead.
\myblue{SoftPosit-Math library does not have an implementation for
  $log10(x)$, $exp10(x)$, $sinh(x)$, and $cosh(x)$ functions. Hence,
  we do not report them. On average, \ourlibm has $11\%$ slowdown
  compared to SoftPosit-Math.  \ourlibm's $log(x)$, $log2(x)$,
  $cospi(x)$, and $sinpi(x)$ have similar performance compared to
  SoftPosit-Math, while the super-optimized implementations of
  SoftPosit-Math show higher performance for $exp(x)$ and $exp2(x)$
  even though both libraries use polynomials of similar
  degree. Finally, SoftPosit-Math library computes $sqrt(x)$ using the
  Newton-Raphson refinement method and produces a more efficient
  function. 
We plan to explore integer operations for internal computation to
further improve \tool's performance.}

\subsubsection{Performance Evaluation of Elementary Functions for
  \texttt{Float}}
\myblue{\tool's $log2(x)$ function for the 32-bit \texttt{floating
    point} type has a $1.32\times$ speedup over \texttt{glibc}'s
  \texttt{float} math library, which produces wrong results for $14$
  million inputs. Compared to \texttt{glibc}'s \texttt{double} math
  library which produces the correctly rounded result for all
  \texttt{float} inputs, \tool has $1.36\times$ speedup. \tool's
  $log2(x)$ function for \texttt{float} has $1.1\times$ and
  $1.2\times$ speedup over \texttt{Intel}'s \texttt{float} and
  \texttt{double} math library, respectively. \texttt{Intel}'s
  \texttt{float} math library produces wrong results for $276$
  inputs.}

\subsection{Case Studies of Correctly Rounded Elementary Functions}
We provide case studies to show that our approach (1) has more
freedom in generating better polynomials, (2) generates different
polynomials for the same underlying elementary function to account for
numerical errors in range reduction and output compensation, and (3)
generates correctly rounded results even when the polynomial evaluation is
performed with the \texttt{double} type.

\subsubsection{Case Study with $10^{x}$ for \texttt{Bfloat16}}

The $10^{x}$ function is defined over the input domain $(-\infty,
\infty)$. There are four classes of special cases:
\[
  \text{Special cases of } 10^{x} =
  \begin{cases}
    0.0 & \text{if } x \leq -40.5 \\
    1.0 & \text{if } -8.4686279296875 \times 10^{-4} \leq x \leq
    1.68609619140625 \times 10^{-3} \\
    \infty & \text{if } x \geq 38.75 \\
    NaN & \text{if } x = NaN
  \end{cases}
\]
A quick initial check returns their result and reduces the overall
input that we need to approximate.

We approximate $10^x$ using $2^x$, which is easier to compute. We use
the property, $10^{x} = 2^{x log_2(10)}$ to approximate $10^{x}$ using
$2^{x}$. Subsequently, we perform range reduction by decomposing $x
log_2(10)$ as $x log_2(10) = i + x'$, where $i$ is an integer and $x'
\in [0, 1)$ is the fractional part.

Now, $10^{x}$ decomposes to
\[
  10^{x} = 2^{x log_2(10)} = 2^{i + x'} = 2^{i}2^{x'}
\]
The above decomposition requires us to approximate $2^{x'}$ where
$x'\in [0,1)$. Multiplication by $2^i$ can be performed using integer
  operations. The range reduction, output compensation, and the
  function we are approximating $g(x')$ is as follows:
\[
  RR(x) = x' = x log_2(10) - \lfloor x log_2(10) \rfloor \quad OC(y', x) =
  y' 2^i = y' 2^{\lfloor x log_2(10) \rfloor} \quad g(x') = 2^{x'}
\]

Our approach generated a $4^{th}$ degree polynomial that approximates
$2^{x'}$ in the input domain $[0, 1)$. Our polynomial produces the
  correctly rounded result for all inputs in the entire domain for
  $10^x$ when used with range reduction and output compensation.

\begin{figure}
  \centerline{\includegraphics[width=0.99\linewidth]{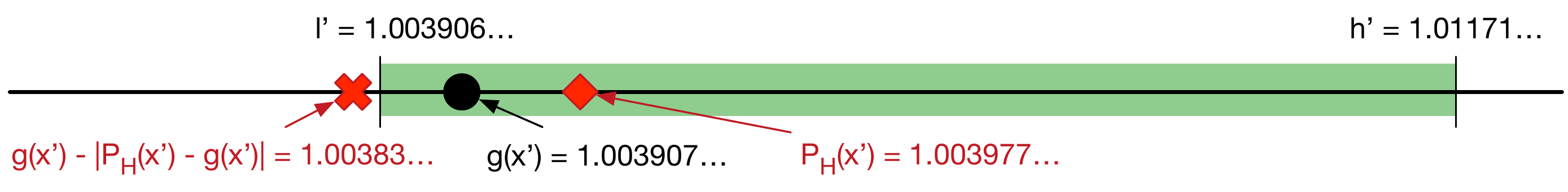}}
  \caption{More freedom in generating a polynomial for $10^x$ with our
    approach. The reduced interval $[l', h']$ (in green box)
    corresponds to the reduced input $x' = 0.0056264\dots$.  We show
    the real value of $g(x')$ (black circle) and the result produced
    by the polynomial generated with our approach (red diamond). If we
    approximated the real result $g(x')$ instead of the correctly
    rounded result, the margin of error for any such polynomial would
    be lower.}
  \label{fig:eval_error}
\end{figure}

We are able to generate a lower degree polynomial because our approach
provides more freedom to generate the correctly rounded results. We
illustrate this point with an example. Figure~\ref{fig:eval_error}
presents a reduced interval ($[l', h']$ in green region) for the
reduced input ($x' = 0.00562\dots$) in our approach. The real value of
$g(x')$ is shown in black circle.
In our approach, the polynomial that approximates $g(x')$ has to
produce a value in $[l', h']$ such that the output compensated value
produces the correctly rounded result of $10^{x}$ for all input $x$
that reduce to $x'$. The value of $g(x')$ is extremely close to $l'$
with a margin of error $\epsilon = |g(x') - l'| \approx
1.31\times10^{-6}$. In contrast to our approach, if we approximated
the real value of $g(x')$, then we must generate a polynomial with an
error of at most $\epsilon$, \ie the polynomial has to produce a value
in $[g(x') - \epsilon, g(x') + \epsilon]$, which potentially
necessitates a higher degree polynomial. The polynomial that we
generate produces a value shown in Figure~\ref{fig:eval_error} with
red diamond. This value has an error of $|P_{\mathbb{H}}(x') - g(x')|
\approx 7.05\times 10^{-5}$, which is much larger than
$\epsilon$. Still, the $4^{th}$ degree polynomial generated by our
approach produces the correctly rounded value when used with the
output compensation function for all inputs.

\subsubsection{Case Study with $ln(x)$, $log_2(x)$, and $log_{10}(x)$
  for \texttt{Bfloat16}}

While creating the \texttt{Bfloat16} approximations for functions $ln(x)$,
$log_{2}(x)$, and $log_{10}(x)$, we observed that our approach
generates different polynomials for the same underlying elementary
function to account for numerical errors in range reduction and output
compensation. We highlight this observation in this case study.

To approximate these functions, we use a slightly modified version of
the Cody and Waite range reduction technique~\cite{Cody:book:1980}.
As a first step, we use mathematical properties of
logarithms, $log_{b}(x) = \frac{log_{2}(x)}{log_{2}(b)}$ to
approximate all three functions $ln(x)$, $log_2(x)$, and $log_{10}(x)$
using the approximation for $log_{2}(x)$. As a second step, we perform
range reduction by decomposing the input $x$ as $x = t \times 2^e$
where $t \in [1,2)$ is the fractional value represented by the
  mantissa and $e$ is an integer representing the exponent of the
  value.
Then, we use the mathematical property of logarithms, $log_{b}(x
\times y^z) = log_{b}(x) + z log_{b}(y)$, to perform range reduction
and output compensation. Now, any logarithm function $log_b(x)$ can be
decomposed to $log_b(x) = \frac{log_2(t) + e}{log_2(b)}$.

As a third step, to ease the job of generating a polynomial for
$log_2(t)$, we introduce a new variable $x' = \frac{t - 1}{t + 1}$ and
transform the function $log_2(t)$ to a function with rapidly
converging polynomial expansion:
\[
  g(x') = log_2 \left( \frac{1 + x'}{1 - x'} \right)
\]
where the function $g(x')$ evaluates to $log_2(t)$.

The above input transformation, attributed to Cody and
Waite~\cite{Cody:book:1980}, enables the creation of a rapidly
convergent odd polynomial, $P(x) = c_1x + c_3x^3 ...$, which reduces
the number of operations. In contrast, the polynomial would be of the
form $P(x) = c_0 + c_1x + c_2x^2 ...$ in the absence of above input
transformation, which has terms with both even and odd degrees.

When the input $x$ is decomposed into $x = t * e$ where $t\in [1, 2)$
  and $e$ is an integer, the range reduction function $x' = RR(x)$ , the
  output compensation function $y = OC(y', x)$, and the function that
  we need to approximate, $y' = g(x')$ are as follows,
\[
  RR(x) = x' = \frac{t - 1}{t + 1},
  \quad\quad
  OC(y', x) = \frac{y' + e}{log_{2}(b)}
  \quad\quad
  g(x') = log_2 \left( \frac{1 + x'}{1 - x'} \right)
\]
Hence, we approximate the same elementary function for $ln(x)$,
$log_2(x)$ and $log_10(x)$ (\ie, $g(x')$).  However, the output
compensation functions are different for each of them.

We observed that our approach produced different polynomials that
produced correct output for $ln(x)$, $log_2(x)$, and $log_{10}(x)$
functions for \texttt{Bfloat16}, which is primarily to account for numerical
errors in each output compensation function.  We produced a $5^{th}$
degree odd polynomial for $log2(x)$, a $5^{th}$ degree odd polynomial
with different coefficients for $log_{10}(x)$, and a $7^{th}$ degree
odd polynomial for $ln(x)$. Our technique also determined that there
was no correct $5^{th}$ degree odd polynomial for $ln(x)$.  Although
these polynomials approximate the same function $g(x')$, they cannot
be used interchangeably.  For example, our experiment show that the
$5^{th}$ degree polynomial produced for $log_2(x)$ cannot be used to
produce the correctly rounded result of $ln(x)$ for all inputs.

\subsubsection{Case Study with $log_{2}(x)$ for a 32-bit
  \texttt{Float}}
\myblue{To show that our approach is scalable to data types with
  numerous inputs, we illustrate a correctly rounded $log_2(x)$
  function for a 32-bit \texttt{float} type. Even with
  state-of-the-art range reduction for
  $log_2(x)$~\cite{Tang:log:toms:1990}, there are roughly seven
  million reduced inputs and its corresponding intervals. Solving an
  LP problem with seven million constraints is infeasible with our LP
  solver. Hence, we sampled five thousand reduced inputs and generated
  a polynomial that produces correct result for the sampled inputs.
  Next, we validated whether the generated polynomial produces the
  correctly rounded result for all inputs. We added any input where
  the polynomial did not produce the correctly rounded result to the
  sample and re-generated the polynomial. We repeated the process
  until the generated polynomial produced the correctly rounded result
  for all inputs.}

\myblue{ We were able to generate a $5^{th}$ degree polynomial that
  produces the correct result for all inputs by using $7,775$ reduced
  inputs. This case study shows that our approach can be adapted for
  generating correctly rounded functions for data types with numerous
  inputs.}

\section{Discussion}
\myblue{We discuss alternatives to polynomial approximation for
  computing correctly rounded results for small data types, design
  considerations with our approach, and opportunities for future
  work.}

\myblue{\textbf{Look-up tables.}  A lookup table is an attractive
  alternative to polynomial approximation for data types with small
  bit-widths. However, it requires additional space to store these
  tables for each function (\ie, space versus latency tradeoff). In
  the case of embedded controllers, computing the function in a few
  cycles with polynomial approximation can be appealing because lookup
  tables can have non-deterministic latencies due to memory footprint
  issues. Further, lookup tables are likely infeasible for 32-bit
  float or posit values.}

\myblue{\textbf{Scalability with large data types.} Our goal is to
  eventually generate the correctly rounded math library for FP types
  with larger bit-widths. The LP solver can become a bottleneck when
  the domain is large. In the case of \texttt{Bfloat16} and posit16,
  we can use all inputs to generate intervals. We observed that it is
  not necessary to add every interval to the LP formulation. Only
  highly constrained intervals need to be added.  We plan to explore
  systematic sampling of intervals to generate polynomials for data
  types with larger bit-widths.  }

\myblue{When our approach cannot generate a single polynomial that
  produces correctly rounded results for all inputs, we currently use
  a trial-and-error approach to generate piece-wise polynomials (\eg,
  $sinpi$, $cospi$, $sinh$, and $cosh(x)$ in
  Section~\ref{sec:evaluation}). We plan to explore a systematic
  approach to generate piecewise polynomials as future work.}

\myblue{\textbf{Validation of correctness for all inputs.} In our
  approach, we enumerate each possible input and obtain the oracle
  result for each input using the same elementary function in the MPFR
  library that is computed with 2000 bits of precision. This MPFR
  result is rounded to the target representation.  We validate that
  the polynomial generated by our approach produces exactly the same
  oracle result by evaluating it with each input. Although it is
  possible to validate whether a particular polynomial produces the
  correctly rounded output for the \texttt{float} data type
  by enumeration, it is not possible for the \texttt{double}
  type. Validating the correctness of the result produced by a
  polynomial for the \texttt{double} type is an open research
  question.}

\myblue{\textbf{Importance of Range reduction.} Efficient range
  reduction is important when the goal is to produce correctly rounded
  results for all inputs with the best possible performance.
  The math library designer has to choose an appropriate range
  reduction technique for various elementary functions with our
  approach. Fortunately, there is a rich body of prior work on range
  reduction for many elementary functions, which we use. In the
  absence of such customized range reduction techniques, it is
  possible to generate polynomials that produce correctly rounded
  results with our approach. However, it will likely not be efficient.
  Further, effective range reduction techniques are important to
  decrease the condition number of the LP problem and to avoid
  overflows in polynomial evaluation. We plan to explore if we can
  automatically generate customized range reduction techniques as
  future work.}

\myblue{\textbf{Handling multivariate functions.}  Currently, our
  approach does not handle multivariate functions such as $pow(x,
  y)$. The key challenge lies in encoding the constraints of
  multivariate functions as linear constraints, which we are exploring
  as part of future work.}

\section{Related Work}

\textbf{Correctly rounded math libraries for FP.}  Since the
introduction of the floating point standard~\cite{ieee754}, a number
of correctly rounded math libraries have been proposed. For example,
the IBM LibUltim (or also known as
MathLib)~\cite{IBM:MathLib:online:2008,
  Abraham:fastcorrect:toms:1991}, Sun Microsystem's
LibMCR~\cite{Sun:libmcr:online:2008},
CR-LIBM~\cite{Daramy:crlibm:spie:2003}, and the MPFR math
library~\cite{Fousse:toms:2007:mpfr}. MPFR produces the correctly
rounded result for any arbitrary precision.

CR-LIBM ~\cite{Daramy:crlibm:spie:2003, Lefevre:toward:tc:1998} is a
correctly rounded double math library developed using
Sollya~\cite{Chevillard:sollya:icms:2010}. Given a degree $d$, a
representation $\mathbb{H}$, and the elementary function $f(x)$,
Sollya generates polynomials of degree $d$ with coefficients in
$\mathbb{H}$ that has the minimum infinity
norm~\cite{Brisebarre:epl:arith:2007}. Sollya uses a modified Remez
algorithm with lattice basis reduction to produce polynomials. It also
computes the error bound on the polynomial evaluation using interval
arithmetic~\cite{Chevillard:infnorm:qsic:2007, Chevillard:ub:tcs:2011}
and produces Gappa~\cite{Guillaume:Gappa:online:2019} proofs for the
error bound. Metalibm~\cite{Olga:metalibm:icms:2014,
  Brunie:metalibm:ca:2015} is a math library function generator built
using Sollya. MetaLibm is able to automatically identify range
reduction and domain splitting techniques for some transcendental
functions. It has been used to create correctly rounded elementary
functions for the \texttt{float} and \texttt{double} types.

A number of other approaches have been proposed to generate correctly
rounded results for different transcendental functions including
square root~\cite{Jeannerod:sqrt:tc:2011} and
exponentiation~\cite{Bui:exp:ccece:1999}. A modified Remez algorithm
has also been used to generate polynomials for approximating some
elementary functions~\cite{Arzelier:poly:arith:2019}. It generates a
polynomial that minimizes the infinity norm compared to an ideal
elementary function and the numerical error in the polynomial
evaluation. It can be used to produce correctly rounded results when
range reduction is not necessary.
Compared to prior techniques, our approach approximates the correctly
rounded value $RN_{\mathbb{T}}(f(x))$ and the margin of error is much
higher, which generates efficient polynomials.  Additionally, our
approach also takes into account numerical errors in range reduction,
output compensation, and polynomial evaluation.

\textbf{Posit math libraries.}
SoftPosit-Math~\cite{Leong:online:2019:positmath} has a number of
correctly rounded \texttt{Posit16} elementary functions, which are created
using the Minefield method~\cite{Gustafson:unum:2020:online}. The
Minefield method identifies the interval of values that the internal
computation should produce and declares all other regions as a
minefield. Then the goal is to generate a polynomial that avoids the
mines. The polynomials in the minefield method were generated by trial
and error. Our approach is inspired by the Minefield method. It
generalizes it to numerous representations, range reduction, and
output compensation. Our approach also automates the process of
generating polynomials by encoding the mines as linear constraints and
uses an LP solver.
In our prior work~\cite{Lim:cordic:cf:2020}, we have used the CORDIC
method to generate approximations to trigonometric functions for the
\texttt{Posit32} type. However, they do not produce the correctly
rounded result for all inputs.

\textbf{Verification of math libraries.}  As performance and
correctness are both important with math libraries, there is extensive
research to prove the correctness of math libraries.  Sollya verifies
that the generated implementations of elementary functions produce
correctly rounded results with the aid of
Gappa~\cite{Dinechin:gappaverify:sac:2006, Dinechin:verify:tc:2011,
  Daumas:proofs:arith:2005}.  It has been used to prove the
correctness of CR-LIBM. Recently, researchers have also verified that
many functions in Intel's \texttt{math.h} implementations have at most
1 ulp error~\cite{Lee:verify:popl:2018}. Various elementary function
implementations have also been proven correct using HOL
Light~\cite{Harrison:expproof:amst:1997,
  Harrison:verifywithHOL:tphol:1997,harrison:hollight:tphols:2009}.
Similarly, CoQ proof assistant has been used to prove the correctness
of argument reduction~\cite{Boldo:reduction:toc:2009}. Instruction
sets of mainstream processors have also been proven correct using
proof assistants (\eg, division and $sqrt(x)$ instruction in IBM
Power4 processor~\cite{Sawada:verify:acl:2002}).
\tool validates that the reported polynomial produces the correctly
rounded result for all inputs. We likely have to rely on prior
verification efforts to check the correctness of \tool's polynomials
for the \texttt{double} type.

\textbf{Rewriting tools.}  Mathematical rewriting tools are other
alternatives to create correctly rounded functions. If the rounding
error in the implementation is the root cause of an incorrect result,
we can use tools that detect numerical errors to diagnose
them~\cite{Daming:fpe:popl:2020, Xin:repairmlib:popl:2019,
  Chowdhary:positdebug:2020:pldi, Benz:pldi:2012:dynamic,
  Fu:weakdistance:pldi:2019, Goubalt:2001:sas,
  Sanchez:pldi:2018:herbgrind}. Subsequently, we can rewrite them
using tools such as Herbie~\cite{Panchekha:herbie:pldi:2015} or
Salsa~\cite{Damouche:salsa:afm:2018}. Recently, a repair tool was
proposed specifically for reducing the error of math
libraries~\cite{Xin:repairmlib:popl:2019}. It identifies the domain of
inputs that result in high error. Then, it uses piecewise linear or
quadratic equations to repair them for the specific domain. However,
currently, these rewriting tools do not guarantee correctly rounded
results for all inputs.

\section{Conclusion}
\myblue{A library to approximate elementary functions is a key
  component of any FP representation.  We propose a novel approach to
  generate correctly rounded results for all inputs of an elementary
  function. The key insight is to identify the amount of freedom
  available to generate the correctly rounded result. Subsequently, we
  use this freedom to generate a polynomial using linear programming
  that produces the correct result for all inputs. The resulting
  polynomial approximations are faster than existing libraries while
  producing correct results for all inputs.  Our approach can also
  allow designers of elementary functions to make pragmatic trade-offs
  with respect to performance and correctness. More importantly, it
  can enable standards to mandate correctly rounded results for
  elementary functions with new representations.}

%% Acknowledgments
\begin{acks}                            %% acks environment is optional
 This material is based upon work supported by the
 \grantsponsor{GS100000001}{National Science
   Foundation}{http://dx.doi.org/10.13039/100000001} under Grant
 No.~\grantnum{GS100000001}{1908798}, Grant
 No.~\grantnum{GS100000001}{1917897}, and Grant
 No.~\grantnum{GS100000001}{1453086}.  Any opinions, findings, and
 conclusions or recommendations expressed in this material are those
 of the authors and do not necessarily reflect the views of the
 National Science Foundation.
\end{acks}

%% Bibliography
\bibliography{reference}


\begin{thebibliography}{60}

%%% the Web address.

\ifx \showCODEN    \undefined \def \showCODEN     #1{\unskip}     \fi
\ifx \showDOI      \undefined \def \showDOI       #1{#1}\fi
\ifx \showISBNx    \undefined \def \showISBNx     #1{\unskip}     \fi
\ifx \showISBNxiii \undefined \def \showISBNxiii  #1{\unskip}     \fi
\ifx \showISSN     \undefined \def \showISSN      #1{\unskip}     \fi
\ifx \showLCCN     \undefined \def \showLCCN      #1{\unskip}     \fi
\ifx \shownote     \undefined \def \shownote      #1{#1}          \fi
\ifx \showarticletitle \undefined \def \showarticletitle #1{#1}   \fi
\ifx \showURL      \undefined \def \showURL       {\relax}        \fi
% invisible to TeX
\providecommand\bibfield[2]{#2}
\providecommand\bibinfo[2]{#2}
\providecommand\natexlab[1]{#1}
\providecommand\showeprint[2][]{arXiv:#2}

\bibitem[\protect\citeauthoryear{{Arzelier}, {Bréhard}, and
  {Joldes}}{{Arzelier} et~al\mbox{.}}{2019}]%
        {Arzelier:poly:arith:2019}
\bibfield{author}{\bibinfo{person}{Denis {Arzelier}}, \bibinfo{person}{Florent
  {Bréhard}}, {and} \bibinfo{person}{Mioara {Joldes}}.}
  \bibinfo{year}{2019}\natexlab{}.
\newblock \showarticletitle{Exchange Algorithm for Evaluation and Approximation
  Error-Optimized Polynomials}. In \bibinfo{booktitle}{\emph{2019 IEEE 26th
  Symposium on Computer Arithmetic (ARITH)}}. \bibinfo{pages}{30--37}.
\newblock
\urldef\tempurl%
\url{https://doi.org/10.1109/ARITH.2019.00014}
\showDOI{\tempurl}


\bibitem[\protect\citeauthoryear{Benz, Hildebrandt, and Hack}{Benz
  et~al\mbox{.}}{2012}]%
        {Benz:pldi:2012:dynamic}
\bibfield{author}{\bibinfo{person}{Florian Benz}, \bibinfo{person}{Andreas
  Hildebrandt}, {and} \bibinfo{person}{Sebastian Hack}.}
  \bibinfo{year}{2012}\natexlab{}.
\newblock \showarticletitle{A Dynamic Program Analysis to Find Floating-point
  Accuracy Problems}. In \bibinfo{booktitle}{\emph{Proceedings of the 33rd ACM
  SIGPLAN Conference on Programming Language Design and Implementation}}
  (Beijing, China) \emph{(\bibinfo{series}{PLDI '12})}.
  \bibinfo{publisher}{ACM}, \bibinfo{address}{New York, NY, USA},
  \bibinfo{pages}{453--462}.
\newblock
\showISBNx{978-1-4503-1205-9}
\urldef\tempurl%
\url{https://doi.org/10.1145/2345156.2254118}
\showDOI{\tempurl}


\bibitem[\protect\citeauthoryear{Bernstein, Zhao, Meister, Liu, Anandkumar, and
  Yue}{Bernstein et~al\mbox{.}}{2020}]%
        {Bernstein:learning:arxiv:2020}
\bibfield{author}{\bibinfo{person}{Jeremy Bernstein}, \bibinfo{person}{Jiawei
  Zhao}, \bibinfo{person}{Markus Meister}, \bibinfo{person}{Ming-Yu Liu},
  \bibinfo{person}{Anima Anandkumar}, {and} \bibinfo{person}{Yisong Yue}.}
  \bibinfo{year}{2020}\natexlab{}.
\newblock \bibinfo{title}{Learning compositional functions via multiplicative
  weight updates}.
\newblock
\newblock
\showeprint[arxiv]{2006.14560}~[cs.NE]


\bibitem[\protect\citeauthoryear{{Boldo}, {Daumas}, and {Li}}{{Boldo}
  et~al\mbox{.}}{2009}]%
        {Boldo:reduction:toc:2009}
\bibfield{author}{\bibinfo{person}{Sylvie {Boldo}}, \bibinfo{person}{Marc
  {Daumas}}, {and} \bibinfo{person}{Ren-Cang {Li}}.}
  \bibinfo{year}{2009}\natexlab{}.
\newblock \showarticletitle{Formally Verified Argument Reduction with a Fused
  Multiply-Add}. In \bibinfo{booktitle}{\emph{IEEE Transactions on Computers}},
  Vol.~\bibinfo{volume}{58}. \bibinfo{pages}{1139--1145}.
\newblock
\urldef\tempurl%
\url{https://doi.org/10.1109/TC.2008.216}
\showDOI{\tempurl}


\bibitem[\protect\citeauthoryear{Borwein and Erdelyi}{Borwein and
  Erdelyi}{1995}]%
        {borwein:polynomials:book:1995}
\bibfield{author}{\bibinfo{person}{Peter Borwein} {and} \bibinfo{person}{Tamas
  Erdelyi}.} \bibinfo{year}{1995}\natexlab{}.
\newblock \bibinfo{booktitle}{\emph{Polynomials and Polynomial Inequalities}}.
\newblock \bibinfo{publisher}{Springer New York}.
\newblock
\showISBNx{9780387945095}
\showLCCN{95008374}
\urldef\tempurl%
\url{https://doi.org/10.1007/978-1-4612-0793-1}
\showDOI{\tempurl}


\bibitem[\protect\citeauthoryear{{Brisebarre} and {Chevillard}}{{Brisebarre}
  and {Chevillard}}{2007}]%
        {Brisebarre:epl:arith:2007}
\bibfield{author}{\bibinfo{person}{Nicolas {Brisebarre}} {and}
  \bibinfo{person}{Sylvvain {Chevillard}}.} \bibinfo{year}{2007}\natexlab{}.
\newblock \showarticletitle{Efficient polynomial L-approximations}. In
  \bibinfo{booktitle}{\emph{18th IEEE Symposium on Computer Arithmetic (ARITH
  '07)}}.
\newblock
\urldef\tempurl%
\url{https://doi.org/10.1109/ARITH.2007.17}
\showDOI{\tempurl}


\bibitem[\protect\citeauthoryear{Brisebarre, Muller, and Tisserand}{Brisebarre
  et~al\mbox{.}}{2006}]%
        {Brisebarre:maceffi:toms:2006}
\bibfield{author}{\bibinfo{person}{Nicolas Brisebarre},
  \bibinfo{person}{Jean-Michel Muller}, {and} \bibinfo{person}{Arnaud
  Tisserand}.} \bibinfo{year}{2006}\natexlab{}.
\newblock \showarticletitle{Computing Machine-Efficient Polynomial
  Approximations}. In \bibinfo{booktitle}{\emph{ACM ACM Transactions on
  Mathematical Software}}, Vol.~\bibinfo{volume}{32}.
  \bibinfo{publisher}{Association for Computing Machinery},
  \bibinfo{address}{New York, NY, USA}, \bibinfo{pages}{236–256}.
\newblock
\urldef\tempurl%
\url{https://doi.org/10.1145/1141885.1141890}
\showDOI{\tempurl}


\bibitem[\protect\citeauthoryear{{Brunie}, de~{Dinechin}, {Kupriianova}, and
  {Lauter}}{{Brunie} et~al\mbox{.}}{2015}]%
        {Brunie:metalibm:ca:2015}
\bibfield{author}{\bibinfo{person}{Nicolas {Brunie}}, \bibinfo{person}{Florent
  de {Dinechin}}, \bibinfo{person}{Olga {Kupriianova}}, {and}
  \bibinfo{person}{Christoph {Lauter}}.} \bibinfo{year}{2015}\natexlab{}.
\newblock \showarticletitle{Code Generators for Mathematical Functions}. In
  \bibinfo{booktitle}{\emph{2015 IEEE 22nd Symposium on Computer Arithmetic}}.
  \bibinfo{pages}{66--73}.
\newblock
\urldef\tempurl%
\url{https://doi.org/10.1109/ARITH.2015.22}
\showDOI{\tempurl}


\bibitem[\protect\citeauthoryear{{Bui} and {Tahar}}{{Bui} and {Tahar}}{1999}]%
        {Bui:exp:ccece:1999}
\bibfield{author}{\bibinfo{person}{Hung~Tien {Bui}} {and}
  \bibinfo{person}{Sofiene {Tahar}}.} \bibinfo{year}{1999}\natexlab{}.
\newblock \showarticletitle{Design and synthesis of an IEEE-754 exponential
  function}. In \bibinfo{booktitle}{\emph{Engineering Solutions for the Next
  Millennium. 1999 IEEE Canadian Conference on Electrical and Computer
  Engineering}}, Vol.~\bibinfo{volume}{1}. \bibinfo{pages}{450--455 vol.1}.
\newblock
\urldef\tempurl%
\url{https://doi.org/10.1109/CCECE.1999.807240}
\showDOI{\tempurl}


\bibitem[\protect\citeauthoryear{Chevillard, Harrison, Joldes, and
  Lauter}{Chevillard et~al\mbox{.}}{2011}]%
        {Chevillard:ub:tcs:2011}
\bibfield{author}{\bibinfo{person}{Sylvain Chevillard}, \bibinfo{person}{John
  Harrison}, \bibinfo{person}{Mioara Joldes}, {and} \bibinfo{person}{Christoph
  Lauter}.} \bibinfo{year}{2011}\natexlab{}.
\newblock \showarticletitle{Efficient and accurate computation of upper bounds
  of approximation errors}.
\newblock \bibinfo{journal}{\emph{Theoretical Computer Science}}
  \bibinfo{volume}{412}.
\newblock
\urldef\tempurl%
\url{https://doi.org/10.1016/j.tcs.2010.11.052}
\showDOI{\tempurl}


\bibitem[\protect\citeauthoryear{Chevillard, Joldes, and Lauter}{Chevillard
  et~al\mbox{.}}{2010}]%
        {Chevillard:sollya:icms:2010}
\bibfield{author}{\bibinfo{person}{Sylvain Chevillard}, \bibinfo{person}{Mioara
  Joldes}, {and} \bibinfo{person}{Christoph Lauter}.}
  \bibinfo{year}{2010}\natexlab{}.
\newblock \showarticletitle{Sollya: An Environment for the Development of
  Numerical Codes}. In \bibinfo{booktitle}{\emph{Mathematical Software - ICMS
  2010}} \emph{(\bibinfo{series}{Lecture Notes in Computer Science},
  Vol.~\bibinfo{volume}{6327})}. \bibinfo{publisher}{Springer},
  \bibinfo{address}{Heidelberg, Germany}, \bibinfo{pages}{28--31}.
\newblock
\urldef\tempurl%
\url{https://doi.org/10.1007/978-3-642-15582-6_5}
\showDOI{\tempurl}


\bibitem[\protect\citeauthoryear{{Chevillard} and {Lauter}}{{Chevillard} and
  {Lauter}}{2007}]%
        {Chevillard:infnorm:qsic:2007}
\bibfield{author}{\bibinfo{person}{Sylvain {Chevillard}} {and}
  \bibinfo{person}{Christopher {Lauter}}.} \bibinfo{year}{2007}\natexlab{}.
\newblock \showarticletitle{A Certified Infinite Norm for the Implementation of
  Elementary Functions}. In \bibinfo{booktitle}{\emph{Seventh International
  Conference on Quality Software (QSIC 2007)}}. \bibinfo{pages}{153--160}.
\newblock
\urldef\tempurl%
\url{https://doi.org/10.1109/QSIC.2007.4385491}
\showDOI{\tempurl}


\bibitem[\protect\citeauthoryear{Chowdhary, Lim, and Nagarakatte}{Chowdhary
  et~al\mbox{.}}{2020}]%
        {Chowdhary:positdebug:2020:pldi}
\bibfield{author}{\bibinfo{person}{Sangeeta Chowdhary}, \bibinfo{person}{Jay~P.
  Lim}, {and} \bibinfo{person}{Santosh Nagarakatte}.}
  \bibinfo{year}{2020}\natexlab{}.
\newblock \showarticletitle{Debugging and Detecting Numerical Errors in
  Computation with Posits}. In \bibinfo{booktitle}{\emph{41st ACM SIGPLAN
  Conference on Programming Language Design and Implementation}}
  \emph{(\bibinfo{series}{PLDI'20})}.
\newblock
\urldef\tempurl%
\url{https://doi.org/10.1145/3385412.3386004}
\showDOI{\tempurl}


\bibitem[\protect\citeauthoryear{Cody and Waite}{Cody and Waite}{1980}]%
        {Cody:book:1980}
\bibfield{author}{\bibinfo{person}{William~J Cody} {and}
  \bibinfo{person}{William~M Waite}.} \bibinfo{year}{1980}\natexlab{}.
\newblock \bibinfo{booktitle}{\emph{{Software manual for the elementary
  functions}}}.
\newblock \bibinfo{publisher}{Prentice-Hall}, \bibinfo{address}{Englewood
  Cliffs, NJ}.
\newblock


\bibitem[\protect\citeauthoryear{Cowlishaw}{Cowlishaw}{2008}]%
        {ieee754}
\bibfield{author}{\bibinfo{person}{Mike Cowlishaw}.}
  \bibinfo{year}{2008}\natexlab{}.
\newblock \bibinfo{booktitle}{\emph{{IEEE} Standard for Floating-Point
  Arithmetic}}.
\newblock \bibinfo{type}{IEEE} 754-2008. \bibinfo{institution}{IEEE Computer
  Society}. \bibinfo{pages}{1--70} pages.
\newblock
\urldef\tempurl%
\url{https://doi.org/10.1109/IEEESTD.2008.4610935}
\showDOI{\tempurl}


\bibitem[\protect\citeauthoryear{Damouche and Martel}{Damouche and
  Martel}{2018}]%
        {Damouche:salsa:afm:2018}
\bibfield{author}{\bibinfo{person}{Nasrine Damouche} {and}
  \bibinfo{person}{Matthieu Martel}.} \bibinfo{year}{2018}\natexlab{}.
\newblock \showarticletitle{Salsa: An Automatic Tool to Improve the Numerical
  Accuracy of Programs}. In \bibinfo{booktitle}{\emph{Automated Formal
  Methods}} \emph{(\bibinfo{series}{Kalpa Publications in Computing},
  Vol.~\bibinfo{volume}{5})}, \bibfield{editor}{\bibinfo{person}{Natarajan
  Shankar} {and} \bibinfo{person}{Bruno Dutertre}} (Eds.).
  \bibinfo{pages}{63--76}.
\newblock
\urldef\tempurl%
\url{https://doi.org/10.29007/j2fd}
\showDOI{\tempurl}


\bibitem[\protect\citeauthoryear{Daramy, Defour, Dinechin, and Muller}{Daramy
  et~al\mbox{.}}{2003}]%
        {Daramy:crlibm:spie:2003}
\bibfield{author}{\bibinfo{person}{Catherine Daramy}, \bibinfo{person}{David
  Defour}, \bibinfo{person}{Florent Dinechin}, {and}
  \bibinfo{person}{Jean-Michel Muller}.} \bibinfo{year}{2003}\natexlab{}.
\newblock \showarticletitle{CR-LIBM: A correctly rounded elementary function
  library}. In \bibinfo{booktitle}{\emph{Proceedings of SPIE Vol. 5205:
  Advanced Signal Processing Algorithms, Architectures, and Implementations
  XIII}}, Vol.~\bibinfo{volume}{5205}.
\newblock
\urldef\tempurl%
\url{https://doi.org/10.1117/12.505591}
\showDOI{\tempurl}


\bibitem[\protect\citeauthoryear{{Daumas}, {Melquiond}, and {Munoz}}{{Daumas}
  et~al\mbox{.}}{2005}]%
        {Daumas:proofs:arith:2005}
\bibfield{author}{\bibinfo{person}{Marc {Daumas}}, \bibinfo{person}{Guillaume
  {Melquiond}}, {and} \bibinfo{person}{Cesar {Munoz}}.}
  \bibinfo{year}{2005}\natexlab{}.
\newblock \showarticletitle{Guaranteed proofs using interval arithmetic}. In
  \bibinfo{booktitle}{\emph{17th IEEE Symposium on Computer Arithmetic
  (ARITH'05)}}. \bibinfo{pages}{188--195}.
\newblock
\urldef\tempurl%
\url{https://doi.org/10.1109/ARITH.2005.25}
\showDOI{\tempurl}


\bibitem[\protect\citeauthoryear{{de Dinechin}, {Lauter}, and {Melquiond}}{{de
  Dinechin} et~al\mbox{.}}{2011}]%
        {Dinechin:verify:tc:2011}
\bibfield{author}{\bibinfo{person}{Florent {de Dinechin}},
  \bibinfo{person}{Christopher {Lauter}}, {and} \bibinfo{person}{Guillaume
  {Melquiond}}.} \bibinfo{year}{2011}\natexlab{}.
\newblock \showarticletitle{Certifying the Floating-Point Implementation of an
  Elementary Function Using Gappa}. In \bibinfo{booktitle}{\emph{IEEE
  Transactions on Computers}}, Vol.~\bibinfo{volume}{60}.
  \bibinfo{pages}{242--253}.
\newblock
\urldef\tempurl%
\url{https://doi.org/10.1109/TC.2010.128}
\showDOI{\tempurl}


\bibitem[\protect\citeauthoryear{de~Dinechin, Lauter, and
  Melquiond}{de~Dinechin et~al\mbox{.}}{2006}]%
        {Dinechin:gappaverify:sac:2006}
\bibfield{author}{\bibinfo{person}{Florent de Dinechin},
  \bibinfo{person}{Christoph~Quirin Lauter}, {and} \bibinfo{person}{Guillaume
  Melquiond}.} \bibinfo{year}{2006}\natexlab{}.
\newblock \showarticletitle{Assisted Verification of Elementary Functions Using
  Gappa}. In \bibinfo{booktitle}{\emph{Proceedings of the 2006 ACM Symposium on
  Applied Computing}} (Dijon, France) \emph{(\bibinfo{series}{SAC ’06})}.
  \bibinfo{publisher}{Association for Computing Machinery},
  \bibinfo{address}{New York, NY, USA}, \bibinfo{pages}{1318–1322}.
\newblock
\showISBNx{1595931082}
\urldef\tempurl%
\url{https://doi.org/10.1145/1141277.1141584}
\showDOI{\tempurl}


\bibitem[\protect\citeauthoryear{Fousse, Hanrot, Lef\`{e}vre, P{\'e}lissier,
  and Zimmermann}{Fousse et~al\mbox{.}}{2007}]%
        {Fousse:toms:2007:mpfr}
\bibfield{author}{\bibinfo{person}{Laurent Fousse}, \bibinfo{person}{Guillaume
  Hanrot}, \bibinfo{person}{Vincent Lef\`{e}vre}, \bibinfo{person}{Patrick
  P{\'e}lissier}, {and} \bibinfo{person}{Paul Zimmermann}.}
  \bibinfo{year}{2007}\natexlab{}.
\newblock \showarticletitle{MPFR: A Multiple-precision Binary Floating-point
  Library with Correct Rounding}.
\newblock \bibinfo{journal}{\emph{ACM Trans. Math. Software}}
  \bibinfo{volume}{33}, \bibinfo{number}{2}, Article \bibinfo{articleno}{13}
  (\bibinfo{date}{June} \bibinfo{year}{2007}).
\newblock
\showISSN{0098-3500}
\urldef\tempurl%
\url{https://doi.org/10.1145/1236463.1236468}
\showDOI{\tempurl}


\bibitem[\protect\citeauthoryear{Fu and Su}{Fu and Su}{2019}]%
        {Fu:weakdistance:pldi:2019}
\bibfield{author}{\bibinfo{person}{Zhoulai Fu} {and} \bibinfo{person}{Zhendong
  Su}.} \bibinfo{year}{2019}\natexlab{}.
\newblock \showarticletitle{Effective Floating-point Analysis via Weak-distance
  Minimization}. In \bibinfo{booktitle}{\emph{Proceedings of the 40th ACM
  SIGPLAN Conference on Programming Language Design and Implementation}}
  (Phoenix, AZ, USA) \emph{(\bibinfo{series}{PLDI 2019})}.
  \bibinfo{publisher}{ACM}, \bibinfo{address}{New York, NY, USA},
  \bibinfo{pages}{439--452}.
\newblock
\showISBNx{978-1-4503-6712-7}
\urldef\tempurl%
\url{https://doi.org/10.1145/3314221.3314632}
\showDOI{\tempurl}


\bibitem[\protect\citeauthoryear{Gleixner, Steffy, and Wolter}{Gleixner
  et~al\mbox{.}}{2015}]%
        {Gleixner:soplex:tech:2015}
\bibfield{author}{\bibinfo{person}{Ambros Gleixner}, \bibinfo{person}{Daniel~E.
  Steffy}, {and} \bibinfo{person}{Kati Wolter}.}
  \bibinfo{year}{2015}\natexlab{}.
\newblock \bibinfo{booktitle}{\emph{Iterative Refinement for Linear
  Programming}}.
\newblock \bibinfo{type}{{T}echnical {R}eport} 15-15.
  \bibinfo{institution}{ZIB}, \bibinfo{address}{Takustr. 7, 14195 Berlin}.
\newblock
\urldef\tempurl%
\url{https://doi.org/10.1287/ijoc.2016.0692}
\showDOI{\tempurl}


\bibitem[\protect\citeauthoryear{Gleixner, Steffy, and Wolter}{Gleixner
  et~al\mbox{.}}{2012}]%
        {Gleixner:soplex:issac:2012}
\bibfield{author}{\bibinfo{person}{Ambros~M. Gleixner},
  \bibinfo{person}{Daniel~E. Steffy}, {and} \bibinfo{person}{Kati Wolter}.}
  \bibinfo{year}{2012}\natexlab{}.
\newblock \showarticletitle{Improving the Accuracy of Linear Programming
  Solvers with Iterative Refinement}. In \bibinfo{booktitle}{\emph{Proceedings
  of the 37th International Symposium on Symbolic and Algebraic Computation}}
  (Grenoble, France) \emph{(\bibinfo{series}{ISSAC ’12})}.
  \bibinfo{publisher}{Association for Computing Machinery},
  \bibinfo{address}{New York, NY, USA}, \bibinfo{pages}{187–194}.
\newblock
\showISBNx{9781450312691}
\urldef\tempurl%
\url{https://doi.org/10.1145/2442829.2442858}
\showDOI{\tempurl}


\bibitem[\protect\citeauthoryear{Goubault}{Goubault}{2001}]%
        {Goubalt:2001:sas}
\bibfield{author}{\bibinfo{person}{Eric Goubault}.}
  \bibinfo{year}{2001}\natexlab{}.
\newblock \showarticletitle{Static Analyses of the Precision of Floating-Point
  Operations}. In \bibinfo{booktitle}{\emph{Proceedings of the 8th
  International Symposium on Static Analysis}} \emph{(\bibinfo{series}{SAS})}.
  \bibinfo{publisher}{Springer}, \bibinfo{pages}{234--259}.
\newblock
\showISBNx{978-3-540-47764-8}
\urldef\tempurl%
\url{https://doi.org/10.1007/3-540-47764-0_14}
\showDOI{\tempurl}


\bibitem[\protect\citeauthoryear{Gustafson}{Gustafson}{2017}]%
        {Gustafson:online:2017:posit}
\bibfield{author}{\bibinfo{person}{John Gustafson}.}
  \bibinfo{year}{2017}\natexlab{}.
\newblock \bibinfo{booktitle}{\emph{Posit Arithmetic}}.
\newblock
\urldef\tempurl%
\url{https://posithub.org/docs/Posits4.pdf}
\showURL{%
\tempurl}


\bibitem[\protect\citeauthoryear{Gustafson}{Gustafson}{2020}]%
        {Gustafson:unum:2020:online}
\bibfield{author}{\bibinfo{person}{John Gustafson}.}
  \bibinfo{year}{2020}\natexlab{}.
\newblock \bibinfo{booktitle}{\emph{The Minefield Method: A Uniformly Fast
  Solution to the Table-Maker’s Dilemma}}.
\newblock
\urldef\tempurl%
\url{https://bit.ly/2ZP4kHj}
\showURL{%
\tempurl}


\bibitem[\protect\citeauthoryear{Gustafson and Yonemoto}{Gustafson and
  Yonemoto}{2017}]%
        {Gustafson:sfi:2017:beating}
\bibfield{author}{\bibinfo{person}{John Gustafson} {and} \bibinfo{person}{Isaac
  Yonemoto}.} \bibinfo{year}{2017}\natexlab{}.
\newblock \showarticletitle{Beating Floating Point at Its Own Game: Posit
  Arithmetic}.
\newblock \bibinfo{journal}{\emph{Supercomputing Frontiers and Innovations: an
  International Journal}} \bibinfo{volume}{4}, \bibinfo{number}{2}
  (\bibinfo{date}{June} \bibinfo{year}{2017}), \bibinfo{pages}{71--86}.
\newblock
\showISSN{2409-6008}
\urldef\tempurl%
\url{https://doi.org/10.14529/jsfi170206}
\showDOI{\tempurl}


\bibitem[\protect\citeauthoryear{Harrison}{Harrison}{1997a}]%
        {Harrison:expproof:amst:1997}
\bibfield{author}{\bibinfo{person}{John Harrison}.}
  \bibinfo{year}{1997}\natexlab{a}.
\newblock \showarticletitle{Floating point verification in HOL light: The
  exponential function}. In \bibinfo{booktitle}{\emph{Algebraic Methodology and
  Software Technology}}, \bibfield{editor}{\bibinfo{person}{Michael Johnson}}
  (Ed.). \bibinfo{publisher}{Springer Berlin Heidelberg},
  \bibinfo{address}{Berlin, Heidelberg}, \bibinfo{pages}{246--260}.
\newblock
\urldef\tempurl%
\url{https://doi.org/10.1007/BFb0000475}
\showDOI{\tempurl}


\bibitem[\protect\citeauthoryear{Harrison}{Harrison}{1997b}]%
        {Harrison:verifywithHOL:tphol:1997}
\bibfield{author}{\bibinfo{person}{John Harrison}.}
  \bibinfo{year}{1997}\natexlab{b}.
\newblock \showarticletitle{Verifying the Accuracy of Polynomial Approximations
  in HOL}. In \bibinfo{booktitle}{\emph{International Conference on Theorem
  Proving in Higher Order Logics}}.
\newblock
\urldef\tempurl%
\url{https://doi.org/10.1007/BFb0028391}
\showDOI{\tempurl}


\bibitem[\protect\citeauthoryear{Harrison}{Harrison}{2009}]%
        {harrison:hollight:tphols:2009}
\bibfield{author}{\bibinfo{person}{John Harrison}.}
  \bibinfo{year}{2009}\natexlab{}.
\newblock \showarticletitle{{HOL} {L}ight: An Overview}. In
  \bibinfo{booktitle}{\emph{Proceedings of the 22nd International Conference on
  Theorem Proving in Higher Order Logics, TPHOLs 2009}}
  \emph{(\bibinfo{series}{Lecture Notes in Computer Science},
  Vol.~\bibinfo{volume}{5674})}, \bibfield{editor}{\bibinfo{person}{Stefan
  Berghofer}, \bibinfo{person}{Tobias Nipkow}, \bibinfo{person}{Christian
  Urban}, {and} \bibinfo{person}{Makarius Wenzel}} (Eds.).
  \bibinfo{publisher}{Springer-Verlag}, \bibinfo{address}{Munich, Germany},
  \bibinfo{pages}{60--66}.
\newblock
\urldef\tempurl%
\url{https://doi.org/10.1007/978-3-642-03359-9_4}
\showDOI{\tempurl}


\bibitem[\protect\citeauthoryear{IBM}{IBM}{2008}]%
        {IBM:MathLib:online:2008}
\bibfield{author}{\bibinfo{person}{IBM}.} \bibinfo{year}{2008}\natexlab{}.
\newblock \bibinfo{booktitle}{\emph{Accurate Portable MathLib}}.
\newblock
\urldef\tempurl%
\url{http://oss.software.ibm.com/mathlib/}
\showURL{%
\tempurl}


\bibitem[\protect\citeauthoryear{Intel}{Intel}{2019}]%
        {intel:nervana:online:2019}
\bibfield{author}{\bibinfo{person}{Intel}.} \bibinfo{year}{2019}\natexlab{}.
\newblock \bibinfo{booktitle}{\emph{Delivering a New Intelligence with AI at
  Scale}}.
\newblock
\urldef\tempurl%
\url{https://www.intel.com/content/www/us/en/artificial-intelligence/posts/nnp-aisummit.html}
\showURL{%
\tempurl}


\bibitem[\protect\citeauthoryear{Jeannerod, Knochel, Monat, and Revy}{Jeannerod
  et~al\mbox{.}}{2011}]%
        {Jeannerod:sqrt:tc:2011}
\bibfield{author}{\bibinfo{person}{Claude-Pierre Jeannerod},
  \bibinfo{person}{Hervé Knochel}, \bibinfo{person}{Christophe Monat}, {and}
  \bibinfo{person}{Guillaume Revy}.} \bibinfo{year}{2011}\natexlab{}.
\newblock \showarticletitle{Computing Floating-Point Square Roots via Bivariate
  Polynomial Evaluation}.
\newblock \bibinfo{journal}{\emph{IEEE Trans. Comput.}}  \bibinfo{volume}{60}.
\newblock
\urldef\tempurl%
\url{https://doi.org/10.1109/TC.2010.152}
\showDOI{\tempurl}


\bibitem[\protect\citeauthoryear{Johnson}{Johnson}{2018}]%
        {Johnson:online:2018:facebook}
\bibfield{author}{\bibinfo{person}{Jeff Johnson}.}
  \bibinfo{year}{2018}\natexlab{}.
\newblock \bibinfo{booktitle}{\emph{Rethinking floating point for deep
  learning}}.
\newblock
\urldef\tempurl%
\url{http://export.arxiv.org/abs/1811.01721}
\showURL{%
\tempurl}


\bibitem[\protect\citeauthoryear{Kahan}{Kahan}{2004}]%
        {Kahan:tablemaker:online:2004}
\bibfield{author}{\bibinfo{person}{William Kahan}.}
  \bibinfo{year}{2004}\natexlab{}.
\newblock \bibinfo{booktitle}{\emph{A Logarithm Too Clever by Half}}.
\newblock
\urldef\tempurl%
\url{https://people.eecs.berkeley.edu/~wkahan/LOG10HAF.TXT}
\showURL{%
\tempurl}


\bibitem[\protect\citeauthoryear{Kalamkar, Mudigere, Mellempudi, Das, Banerjee,
  Avancha, Vooturi, Jammalamadaka, Huang, Yuen, Yang, Park, Heinecke,
  Georganas, Srinivasan, Kundu, Smelyanskiy, Kaul, and Dubey}{Kalamkar
  et~al\mbox{.}}{2019}]%
        {Kalamkar:bfloatai:arxiv:2019}
\bibfield{author}{\bibinfo{person}{Dhiraj~D. Kalamkar},
  \bibinfo{person}{Dheevatsa Mudigere}, \bibinfo{person}{Naveen Mellempudi},
  \bibinfo{person}{Dipankar Das}, \bibinfo{person}{Kunal Banerjee},
  \bibinfo{person}{Sasikanth Avancha}, \bibinfo{person}{Dharma~Teja Vooturi},
  \bibinfo{person}{Nataraj Jammalamadaka}, \bibinfo{person}{Jianyu Huang},
  \bibinfo{person}{Hector Yuen}, \bibinfo{person}{Jiyan Yang},
  \bibinfo{person}{Jongsoo Park}, \bibinfo{person}{Alexander Heinecke},
  \bibinfo{person}{Evangelos Georganas}, \bibinfo{person}{Sudarshan
  Srinivasan}, \bibinfo{person}{Abhisek Kundu}, \bibinfo{person}{Misha
  Smelyanskiy}, \bibinfo{person}{Bharat Kaul}, {and} \bibinfo{person}{Pradeep
  Dubey}.} \bibinfo{year}{2019}\natexlab{}.
\newblock \bibinfo{title}{A Study of BFLOAT16 for Deep Learning Training}.
\newblock
\newblock
\showeprint[arxiv]{1905.12322}


\bibitem[\protect\citeauthoryear{Kupriianova and Lauter}{Kupriianova and
  Lauter}{2014}]%
        {Olga:metalibm:icms:2014}
\bibfield{author}{\bibinfo{person}{Olga Kupriianova} {and}
  \bibinfo{person}{Christoph Lauter}.} \bibinfo{year}{2014}\natexlab{}.
\newblock \showarticletitle{Metalibm: A Mathematical Functions Code Generator}.
  In \bibinfo{booktitle}{\emph{4th International Congress on Mathematical
  Software}}.
\newblock
\urldef\tempurl%
\url{https://doi.org/10.1007/978-3-662-44199-2_106}
\showDOI{\tempurl}


\bibitem[\protect\citeauthoryear{Lee, Sharma, and Aiken}{Lee
  et~al\mbox{.}}{2017}]%
        {Lee:verify:popl:2018}
\bibfield{author}{\bibinfo{person}{Wonyeol Lee}, \bibinfo{person}{Rahul
  Sharma}, {and} \bibinfo{person}{Alex Aiken}.}
  \bibinfo{year}{2017}\natexlab{}.
\newblock \showarticletitle{On Automatically Proving the Correctness of Math.h
  Implementations}.
\newblock \bibinfo{journal}{\emph{Proceedings of the ACM on Programming
  Languages}} \bibinfo{volume}{2}, \bibinfo{number}{POPL}, Article
  \bibinfo{articleno}{47} (\bibinfo{date}{Dec.} \bibinfo{year}{2017}),
  \bibinfo{numpages}{32}~pages.
\newblock
\urldef\tempurl%
\url{https://doi.org/10.1145/3158135}
\showDOI{\tempurl}


\bibitem[\protect\citeauthoryear{Lef\`{e}vre and Muller}{Lef\`{e}vre and
  Muller}{2001}]%
        {Lefevre:worstcase:arith:2001}
\bibfield{author}{\bibinfo{person}{Vincent Lef\`{e}vre} {and}
  \bibinfo{person}{Jean-Michel Muller}.} \bibinfo{year}{2001}\natexlab{}.
\newblock \showarticletitle{Worst Cases for Correct Rounding of the Elementary
  Functions in Double Precision}. In \bibinfo{booktitle}{\emph{15th IEEE
  Symposium on Computer Arithmetic}} \emph{(\bibinfo{series}{Arith '01})}.
  \bibinfo{pages}{111--118}.
\newblock
\urldef\tempurl%
\url{https://doi.org/10.1109/ARITH.2001.930110}
\showDOI{\tempurl}


\bibitem[\protect\citeauthoryear{Lef\`{e}vre, {Muller}, and
  {Tisserand}}{Lef\`{e}vre et~al\mbox{.}}{1998}]%
        {Lefevre:toward:tc:1998}
\bibfield{author}{\bibinfo{person}{Vincent Lef\`{e}vre},
  \bibinfo{person}{Jean-Michel {Muller}}, {and} \bibinfo{person}{Arnaud
  {Tisserand}}.} \bibinfo{year}{1998}\natexlab{}.
\newblock \showarticletitle{Toward correctly rounded transcendentals}.
\newblock \bibinfo{journal}{\emph{IEEE Trans. Comput.}} \bibinfo{volume}{47},
  \bibinfo{number}{11} (\bibinfo{year}{1998}), \bibinfo{pages}{1235--1243}.
\newblock
\urldef\tempurl%
\url{https://doi.org/10.1109/12.736435}
\showDOI{\tempurl}


\bibitem[\protect\citeauthoryear{Leong}{Leong}{2019}]%
        {Leong:online:2019:positmath}
\bibfield{author}{\bibinfo{person}{Cerlane Leong}.}
  \bibinfo{year}{2019}\natexlab{}.
\newblock \bibinfo{booktitle}{\emph{SoftPosit-Math}}.
\newblock
\urldef\tempurl%
\url{https://gitlab.com/cerlane/softposit-math}
\showURL{%
\tempurl}


\bibitem[\protect\citeauthoryear{Lim and Nagarakatte}{Lim and
  Nagarakatte}{2020a}]%
        {rlibm}
\bibfield{author}{\bibinfo{person}{Jay~P. Lim} {and} \bibinfo{person}{Santosh
  Nagarakatte}.} \bibinfo{year}{2020}\natexlab{a}.
\newblock \bibinfo{booktitle}{\emph{RLibm}}.
\newblock
\urldef\tempurl%
\url{https://github.com/rutgers-apl/rlibm}
\showURL{%
\tempurl}


\bibitem[\protect\citeauthoryear{Lim and Nagarakatte}{Lim and
  Nagarakatte}{2020b}]%
        {rlibmgenerator}
\bibfield{author}{\bibinfo{person}{Jay~P. Lim} {and} \bibinfo{person}{Santosh
  Nagarakatte}.} \bibinfo{year}{2020}\natexlab{b}.
\newblock \bibinfo{booktitle}{\emph{RLibm-generator}}.
\newblock
\urldef\tempurl%
\url{https://github.com/rutgers-apl/rlibm-generator}
\showURL{%
\tempurl}


\bibitem[\protect\citeauthoryear{Lim, Shachnai, and Nagarakatte}{Lim
  et~al\mbox{.}}{2020}]%
        {Lim:cordic:cf:2020}
\bibfield{author}{\bibinfo{person}{Jay~P. Lim}, \bibinfo{person}{Matan
  Shachnai}, {and} \bibinfo{person}{Santosh Nagarakatte}.}
  \bibinfo{year}{2020}\natexlab{}.
\newblock \showarticletitle{Approximating Trigonometric Functions for Posits
  Using the CORDIC Method}. In \bibinfo{booktitle}{\emph{Proceedings of the
  17th ACM International Conference on Computing Frontiers}} (Catania, Sicily,
  Italy) \emph{(\bibinfo{series}{CF ’20})}. \bibinfo{publisher}{Association
  for Computing Machinery}, \bibinfo{address}{New York, NY, USA},
  \bibinfo{pages}{19–28}.
\newblock
\urldef\tempurl%
\url{https://doi.org/10.1145/3387902.3392632}
\showDOI{\tempurl}


\bibitem[\protect\citeauthoryear{Melquiond}{Melquiond}{2019}]%
        {Guillaume:Gappa:online:2019}
\bibfield{author}{\bibinfo{person}{Guillaume Melquiond}.}
  \bibinfo{year}{2019}\natexlab{}.
\newblock \bibinfo{booktitle}{\emph{Gappa}}.
\newblock
\urldef\tempurl%
\url{http://gappa.gforge.inria.fr}
\showURL{%
\tempurl}


\bibitem[\protect\citeauthoryear{Microsystems}{Microsystems}{2008}]%
        {Sun:libmcr:online:2008}
\bibfield{author}{\bibinfo{person}{Sun Microsystems}.}
  \bibinfo{year}{2008}\natexlab{}.
\newblock \bibinfo{booktitle}{\emph{LIBMCR 3 "16 February 2008" "libmcr-0.9"}}.
\newblock
\urldef\tempurl%
\url{http://www.math.utah.edu/cgi-bin/man2html.cgi?/usr/local/man/man3/libmcr.3}
\showURL{%
\tempurl}


\bibitem[\protect\citeauthoryear{Muller}{Muller}{2005}]%
        {Muller:elemfunc:book:2005}
\bibfield{author}{\bibinfo{person}{Jean-Michel Muller}.}
  \bibinfo{year}{2005}\natexlab{}.
\newblock \bibinfo{booktitle}{\emph{Elementary Functions: Algorithms and
  Implementation}}.
\newblock \bibinfo{publisher}{Birkhauser}.
\newblock
\showISBNx{0817643729}
\urldef\tempurl%
\url{https://doi.org/10.1007/978-1-4899-7983-4}
\showDOI{\tempurl}


\bibitem[\protect\citeauthoryear{NVIDIA}{NVIDIA}{2020}]%
        {nvidia:tensorfloat:online:2020}
\bibfield{author}{\bibinfo{person}{NVIDIA}.} \bibinfo{year}{2020}\natexlab{}.
\newblock \bibinfo{booktitle}{\emph{TensorFloat-32 in the A100 GPU Accelerates
  AI Training, HPC up to 20x}}.
\newblock
\urldef\tempurl%
\url{https://blogs.nvidia.com/blog/2020/05/14/tensorfloat-32-precision-format/}
\showURL{%
\tempurl}


\bibitem[\protect\citeauthoryear{Panchekha, Sanchez-Stern, Wilcox, and
  Tatlock}{Panchekha et~al\mbox{.}}{2015}]%
        {Panchekha:herbie:pldi:2015}
\bibfield{author}{\bibinfo{person}{Pavel Panchekha}, \bibinfo{person}{Alex
  Sanchez-Stern}, \bibinfo{person}{James~R. Wilcox}, {and}
  \bibinfo{person}{Zachary Tatlock}.} \bibinfo{year}{2015}\natexlab{}.
\newblock \showarticletitle{Automatically Improving Accuracy for Floating Point
  Expressions}. In \bibinfo{booktitle}{\emph{Proceedings of the 36th ACM
  SIGPLAN Conference on Programming Language Design and Implementation}},
  Vol.~\bibinfo{volume}{50}. \bibinfo{publisher}{Association for Computing
  Machinery}, \bibinfo{address}{New York, NY, USA}, \bibinfo{pages}{1–11}.
\newblock
\showISSN{0362-1340}
\urldef\tempurl%
\url{https://doi.org/10.1145/2813885.2737959}
\showDOI{\tempurl}


\bibitem[\protect\citeauthoryear{Remes}{Remes}{1934}]%
        {Remes:algorithm:1934}
\bibfield{author}{\bibinfo{person}{Eugene Remes}.}
  \bibinfo{year}{1934}\natexlab{}.
\newblock \showarticletitle{Sur un proc{\'e}d{\'e} convergent
  d’approximations successives pour d{\'e}terminer les polyn{\^o}mes
  d’approximation}.
\newblock \bibinfo{journal}{\emph{Comptes rendus de l'Académie des Sciences}}
  \bibinfo{volume}{198} (\bibinfo{year}{1934}), \bibinfo{pages}{2063--2065}.
\newblock


\bibitem[\protect\citeauthoryear{Sanchez-Stern, Panchekha, Lerner, and
  Tatlock}{Sanchez-Stern et~al\mbox{.}}{2018}]%
        {Sanchez:pldi:2018:herbgrind}
\bibfield{author}{\bibinfo{person}{Alex Sanchez-Stern}, \bibinfo{person}{Pavel
  Panchekha}, \bibinfo{person}{Sorin Lerner}, {and} \bibinfo{person}{Zachary
  Tatlock}.} \bibinfo{year}{2018}\natexlab{}.
\newblock \showarticletitle{Finding Root Causes of Floating Point Error}. In
  \bibinfo{booktitle}{\emph{Proceedings of the 39th ACM SIGPLAN Conference on
  Programming Language Design and Implementation}} (Philadelphia, PA, USA)
  \emph{(\bibinfo{series}{PLDI 2018})}. \bibinfo{publisher}{ACM},
  \bibinfo{address}{New York, NY, USA}, \bibinfo{pages}{256--269}.
\newblock
\showISBNx{978-1-4503-5698-5}
\urldef\tempurl%
\url{https://doi.org/10.1145/3296979.3192411}
\showDOI{\tempurl}


\bibitem[\protect\citeauthoryear{Sawada}{Sawada}{2002}]%
        {Sawada:verify:acl:2002}
\bibfield{author}{\bibinfo{person}{Joe Sawada}.}
  \bibinfo{year}{2002}\natexlab{}.
\newblock \showarticletitle{Formal verification of divide and square root
  algorithms using series calculation}. In \bibinfo{booktitle}{\emph{3rd
  International Workshop on the ACL2 Theorem Prover and its Applications}}.
\newblock


\bibitem[\protect\citeauthoryear{{Tagliavini}, {Mach}, {Rossi}, {Marongiu}, and
  {Benin}}{{Tagliavini} et~al\mbox{.}}{2018}]%
        {Tagliavini:bfloat:date:2018}
\bibfield{author}{\bibinfo{person}{Giuseppe {Tagliavini}},
  \bibinfo{person}{Stefan {Mach}}, \bibinfo{person}{Davide {Rossi}},
  \bibinfo{person}{Andrea {Marongiu}}, {and} \bibinfo{person}{Luca {Benin}}.}
  \bibinfo{year}{2018}\natexlab{}.
\newblock \showarticletitle{A transprecision floating-point platform for
  ultra-low power computing}. In \bibinfo{booktitle}{\emph{2018 Design,
  Automation Test in Europe Conference Exhibition (DATE)}}.
  \bibinfo{pages}{1051--1056}.
\newblock
\urldef\tempurl%
\url{https://doi.org/10.23919/DATE.2018.8342167}
\showDOI{\tempurl}


\bibitem[\protect\citeauthoryear{Tang}{Tang}{1990}]%
        {Tang:log:toms:1990}
\bibfield{author}{\bibinfo{person}{Ping-Tak~Peter Tang}.}
  \bibinfo{year}{1990}\natexlab{}.
\newblock \showarticletitle{Table-Driven Implementation of the Logarithm
  Function in IEEE Floating-Point Arithmetic}.
\newblock \bibinfo{journal}{\emph{ACM Trans. Math. Software}}
  \bibinfo{volume}{16}, \bibinfo{number}{4} (\bibinfo{date}{Dec.}
  \bibinfo{year}{1990}), \bibinfo{pages}{378–400}.
\newblock
\urldef\tempurl%
\url{https://doi.org/10.1145/98267.98294}
\showDOI{\tempurl}


\bibitem[\protect\citeauthoryear{Trefethen}{Trefethen}{2012}]%
        {Trefethen:chebyshev:book:2012}
\bibfield{author}{\bibinfo{person}{Lloyd~N. Trefethen}.}
  \bibinfo{year}{2012}\natexlab{}.
\newblock \bibinfo{booktitle}{\emph{Approximation Theory and Approximation
  Practice (Other Titles in Applied Mathematics)}}.
\newblock \bibinfo{publisher}{Society for Industrial and Applied Mathematics},
  \bibinfo{address}{USA}.
\newblock
\showISBNx{1611972396}


\bibitem[\protect\citeauthoryear{Wang and Kanwar}{Wang and Kanwar}{2019}]%
        {Wang:tpu:online:2019}
\bibfield{author}{\bibinfo{person}{Shibo Wang} {and} \bibinfo{person}{Pankaj
  Kanwar}.} \bibinfo{year}{2019}\natexlab{}.
\newblock \bibinfo{booktitle}{\emph{BFloat16: The secret to high performance on
  Cloud TPUs}}.
\newblock
\urldef\tempurl%
\url{https://cloud.google.com/blog/products/ai-machine-learning/bfloat16-the-secret-to-high-performance-on-cloud-tpus}
\showURL{%
\tempurl}


\bibitem[\protect\citeauthoryear{Yi, Chen, Mao, and Ji}{Yi
  et~al\mbox{.}}{2019}]%
        {Xin:repairmlib:popl:2019}
\bibfield{author}{\bibinfo{person}{Xin Yi}, \bibinfo{person}{Liqian Chen},
  \bibinfo{person}{Xiaoguang Mao}, {and} \bibinfo{person}{Tao Ji}.}
  \bibinfo{year}{2019}\natexlab{}.
\newblock \showarticletitle{Efficient Automated Repair of High Floating-Point
  Errors in Numerical Libraries}.
\newblock \bibinfo{journal}{\emph{Proceedings of the ACM on Programming
  Languages}} \bibinfo{volume}{3}, \bibinfo{number}{POPL}, Article
  \bibinfo{articleno}{56} (\bibinfo{date}{Jan.} \bibinfo{year}{2019}),
  \bibinfo{numpages}{29}~pages.
\newblock
\urldef\tempurl%
\url{https://doi.org/10.1145/3290369}
\showDOI{\tempurl}


\bibitem[\protect\citeauthoryear{Ziv}{Ziv}{1991}]%
        {Abraham:fastcorrect:toms:1991}
\bibfield{author}{\bibinfo{person}{Abraham Ziv}.}
  \bibinfo{year}{1991}\natexlab{}.
\newblock \showarticletitle{Fast Evaluation of Elementary Mathematical
  Functions with Correctly Rounded Last Bit}.
\newblock \bibinfo{journal}{\emph{ACM Trans. Math. Software}}
  \bibinfo{volume}{17}, \bibinfo{number}{3} (\bibinfo{date}{Sept.}
  \bibinfo{year}{1991}), \bibinfo{pages}{410–423}.
\newblock
\showISSN{0098-3500}
\urldef\tempurl%
\url{https://doi.org/10.1145/114697.116813}
\showDOI{\tempurl}


\bibitem[\protect\citeauthoryear{Zou, Zeng, Xiong, Fu, Zhang, and Su}{Zou
  et~al\mbox{.}}{2019}]%
        {Daming:fpe:popl:2020}
\bibfield{author}{\bibinfo{person}{Daming Zou}, \bibinfo{person}{Muhan Zeng},
  \bibinfo{person}{Yingfei Xiong}, \bibinfo{person}{Zhoulai Fu},
  \bibinfo{person}{Lu Zhang}, {and} \bibinfo{person}{Zhendong Su}.}
  \bibinfo{year}{2019}\natexlab{}.
\newblock \showarticletitle{Detecting Floating-Point Errors via Atomic
  Conditions}.
\newblock \bibinfo{journal}{\emph{Proceedings of the ACM on Programming
  Languages}} \bibinfo{volume}{4}, \bibinfo{number}{POPL}, Article
  \bibinfo{articleno}{60} (\bibinfo{date}{Dec.} \bibinfo{year}{2019}),
  \bibinfo{numpages}{27}~pages.
\newblock
\urldef\tempurl%
\url{https://doi.org/10.1145/3371128}
\showDOI{\tempurl}


\end{thebibliography}

%% Appendix
\newpage
\appendix
\section{Details on \tool}
In the appendices, we describe the range reduction technique, special
cases, and the polynomials we generated to create math library
functions in \tool. We use the same range reduction technique for each
family of elementary functions across all types, \ie $ln(x)$,
$log_2(x)$, and $log_{10}(x)$ for all \texttt{bfloat16},
\texttt{posit16}, and \texttt{float} use the same range reduction
technique. Hence, we first describe the range reduction techniques
that we use for each family of elementary functions in
Appendix~\ref{apx:rrtechnique}. In each subsequent section, we
describe the specific special cases, range reduction function, output
compensation function, and the polynomial we use to create each
function for \texttt{bfloat16} (Appendix~\ref{apx:bfloat16}),
\texttt{posit16} (Appendix~\ref{apx:posit16}), and \texttt{float}
(Appendix~\ref{apx:float}).

\section{Range Reduction Techniques}
\label{apx:rrtechnique}
In this section, we explain the general range reduction technique
\tool uses to reduce the input domain for each class of elementary
functions.

\subsection{Logarithm functions ($log_b(x)$)}
\label{rr:log}
We use a slightly modified version of Cody and Waite's range reduction
technique ~\cite{Cody:book:1980} for all logarithm functions. As a
first step, we use the mathematical property of logarithms,
$log_{b}(x) = \frac{log_2(x)}{log_2(b)}$ to approximate logarithm
functions using the approximation of $log_2(x)$. As a second step, we
decompose the input $x$ as $x = t \times 2^m$ where $t$ is the
fractional value represented by the mantissa and $m$ is the exponent
of the input. Then we use the mathematical property of logarithm
functions, $log_2(x \times y^z) = log_2(x) + zlog_2(y)$ to decompose
$log_2(x)$. Thus, any logarithm function $log_b(x)$ can be decomposed
to ,
\[
  log_b(x) = \frac{log_2(t) + m}{log_2(b)}
\]

As a third step, to ease the job of generating an accurate polynomial
for $log_2(t)$, we introduce a new variable $x' = \frac{t - 1}{t + 1}$
and transform the function $log_2(t)$ to a function with rapidly
converging polynomial expansion:
\[
  g(x') = log_2 \left( \frac{1 + x'}{1 - x'} \right)
\]
The function $g(x')$ evaluates to $log_2(t)$. The polynomial expansion
of $g(x')$ is an odd polynomial, \ie $P(x) = c_1 x + c_3 x^3 + c_5 x^5
\dots$. Combining all steps, we decompose
$log_b(x)$ to,
\[
  log_b(x) = \frac{log_2 \left( \frac{1 + x'}{1 - x'} \right) + m}{log_2(b)}
\]

When the input $x$ is decomposed into $x = t \times e$ where
$t \in [1, 2)$ and $e$ is an integer, the range reduction function
$x' = RR(x)$, the output compensation function $y = OC(y', x)$, and
the function that we need to approximate, $y' = g(x')$ can be
summarized as follows,
\[
  RR(x) = \frac{t - 1}{t + 1} \quad\quad
  OC(y', x) = \frac{y' + m}{log_2(b)} \quad\quad
  g(x') = log_2 \left( \frac{1 + x'}{1- x'} \right)
\]
With this range reduction technique, we need to approximate $g(x')$
for the reduced input domain $x' \in [0, \frac{1}{3})$.

\subsection{Exponential Functions ($a^x$)}
\label{rr:exp}
We approximate all exponential functions with $2^{x}$. As a first
step, we use the mathematical property $a^{x} = 2^{x log_2(a)}$ to
decompose any exponential function to a function of $2^{x}$. Second,
we decompose the value $xlog_2(a)$ into the integral part $i$ and the
remaining fractional part $x' \in [0, 1)$, \ie $xlog_2(a) = i +
x'$. We can define $i$ and $x'$ more formally as:
\[
  i = \lfloor x log_2(a)\rfloor, \quad \quad x' = x log_2(a) - i
\]
where $\lfloor x \rfloor$ is a floor function that rounds down $x$ to
an integer. Using the property $2^{x + y} = 2^x2^y$, $a^{x}$ decomposes to
\[
a^{x} = 2^{x log_2(a)} = 2^{i + x'} = 2^{x'}2^{i} = 2^{x log_2(a) -
  \lfloor x log_2(a) \rfloor}2^{\lfloor x log_2(a) \rfloor}
\]
The above decomposition allows us to approximate any exponential
functions by approximating $2^{x}$ for $x \in [0, 1)$. The range
reduction function $x' = RR(x)$, output compensation function
$y = OC(y', x)$, and the function we need to approximate $y' = g(x')$
can be summarized as follows:
\[
  RR(x) = x log_2(b) - \lfloor x log_2(b) \rfloor\quad\quad
  OC(y', x) = y'2^{\lfloor x log_2(b) \rfloor} \quad\quad
  g(x') = 2^{x'}
\]
With this range reduction technique, we need to approximate $2^{x'}$
for the reduced input domain $x' \in [0, 1)$.

\subsection{Square Root Function ($\sqrt{x}$)}
\label{rr:sqrt}
To perform range reduction on $\sqrt{x}$, we first decompose
the input $x$ into $x = x' \times 2^{m}$ where $m$ is an even integer
and $x' = \frac{x}{2^m} \in [1, 4)$. Second, using the mathematical
properties $\sqrt{xy} = \sqrt{x}\sqrt{y}$ and $\sqrt{2^{2x}} = 2^{x}$,
we decompose $\sqrt{x}$ to:
\[
  \sqrt{x} = \sqrt{x' \times 2^m} = \sqrt{x'} \times 2^{\frac{m}{2}}
\]
The above decomposition allows us to approximate the square root
function by approximating $\sqrt{x}$ for $x \in [1, 4)$. Since $m$ is
an even integer, $\frac{m}{2}$ is an integer and multiplication of
$2^{\frac{m}{2}}$ can be performed using integer arithmetic.

When the input $x$ is decomposed into $x = x' \times 2^{m}$ where
$x' \in [1, 4)$ and $m$ is an even integer, the range reduction
function $x' = RR(x)$, output compensation function $y = OC(y', x)$,
and the function we need to approximate $y' = g(x')$ can be summarized
as follows:
\[
  RR(x) = x'\quad\quad
  OC(y', x) = y'2^{\frac{m}{2}}\quad\quad
  g(x') = \sqrt{x}
\]
With this range reduction technique, we need to approximate $\sqrt{x'}$
for the reduced input domain $x' \in [1, 4)$.

\subsection{Cube Root Function ($\sqrt[3]{x}$)}
\label{rr:cbrt}
To perform range reduction on $\sqrt[3]{x}$, we first decompose the
input $x$ into $x = s \times x' \times 2^{m}$. The value
$s \in \{-1, 1\}$ represents the sign of $x$, $m$ is an integer
multiple of 3, and $x' = \frac{x}{2^m} \in [1, 8)$. Second, using the
mathematical properties $\sqrt[3]{xy} = \sqrt[3]{x} \sqrt[3]{y}$ and
$\sqrt[3]{2^{3x}} = 2^x$, we decompose $\sqrt{x}$ to:
\[
  \sqrt[3]{x} = \sqrt[3]{s \times x' \times 2^m} = s \times \sqrt[3]{x'} \times 2^{\frac{m}{3}}
\]
The above decomposition allow us to approximate the cube root function
by approximating $\sqrt[3]{x}$ for $x \in [1, 8)$. Since $m$ is an
integer multiple of 3, $\frac{m}{3}$ is an integer and multiplication
of $2^{\frac{m}{3}}$ can be performed using integer arithmetic.

When we decompose the input $x$ into $x = s \times x' \times 2^{m}$
where $s$ is the sign of the input, $x' \in [1, 8)$, and $m$ is an
integer multiple of 3, the range reduction function $x' = RR(x)$,
output compensation function $y = OC(y', x)$, and the function we need
to approximate $y' = g(x')$ can be summarized as follows:
\[
  RR(x) = x'\quad\quad
  OC(y', x) = s \times y'2^{\frac{m}{3}}\quad\quad
  g(x') = \sqrt[3]{x}
\]
With this range reduction technique, we need to approximate $\sqrt[3]{x'}$
for the reduced input domain $x' \in [1, 8)$.

\subsection{Sinpi Function ($sin(\pi x)$)}
\label{rr:sinpi}

To perform range reduction on $sin(\pi x)$, we use the property of
$sin(\pi x)$ that it is a periodic and odd function. First, using the
property $sin(-\pi x) = -sin(\pi x)$, we decompose the input $x$ into
$x = s \times |x|$ where $s$ is the sign of the input. The function
decomposes to $sin(\pi x) = s \times sin(\pi |x|)$.

Second, we use the properties $sin(\pi (x + 2z)) = sin(\pi x)$ where
$z$ is an integer and $sin(\pi (x + 2z + 1)) = -sin(\pi x)$. We decompose
$|x|$ into $|x| = i + t$ where $i$ is an integer and $t \in [0, 1)$ is
the fractional part, \ie $t = |x| - i$. More formally, we can define
$t$ and $i$ as,
\[
  i = \lfloor |x| \rfloor,  \quad \quad t = |x| - i
\]
If $i$ is an even integer, then $sin(\pi (t + i)) = sin(\pi t)$ (from
the property $sin(\pi (x + 2z)) = sin(\pi x)$). If $i$ is an odd
integer, then $sin(\pi (t + i)) = -sin(\pi t)$ (from the property
$sin(\pi (x + 2z + 1)) = -sin(\pi x)$). Thus, we can decompose the
$sin(\pi x)$ function into,
\[
  sin(\pi x) = s \times sin(\pi |x|) =
  \begin{cases}
    s \times sin(\pi t) & \text{if } i \equiv 0 \: (mod\:2) \\
    -s \times sin(\pi t) & \text{if } i \equiv 1 \: (mod\:2) \\
  \end{cases}
\]

Third, we use the property $sin(\pi t) = sin(\pi (1 - t))$ for
$0.5 < t < 1.0$ and introduce a new variable $x'$,
\[
  x' =
  \begin{cases}
    1 - t & \text{if } 0.5 < t < 1.0 \\
    t & \text{otherwise}
  \end{cases}
\]
Since we perform the subtraction only when $0.5 < t < 1.0$, $x'$ can
be computed exactly due to Sterbenz
Lemma~\cite{Muller:elemfunc:book:2005}. The above decomposition
reduces the input domain to $x' \in [0, 0.5]$ and requires us to
approximate $sin(\pi x')$ for the reduced domain.

In summary, we decompose the input $x$ into $x = s \times (i + t)$
where $s$ is the sign of the input, $i$ is an integer, and
$t \in [0, 1)$ is the fractional part of $|x|$, \ie $|x| = i + t$. The
range reduction function $x' = RR(x)$, the output compensation
function $y = OC(y', x)$, and the function we need to approximate,
$y' = g(x')$ are as follows:
\[
  RR(x) =
  \begin{cases}
    1 - t & \text{if } 0.5 < t < 1.0 \\
    t & \text{otherwise}
  \end{cases},
  \quad
  OC(y', x) =
  \begin{cases}
    s \times y' & \text{if } i \equiv 0 \: (mod\:2) \\
    -s \times y' & \text{if } i \equiv 1 \: (mod\:2) \\
  \end{cases},
  \quad
  g(x') = sin(\pi x')
\]
With this range reduction technique, we need to approximate $sin(\pi
x')$ for the reduced input domain $x' \in [0, 0.5]$.

\subsection{Cospi Function ($cos(\pi x)$)}
\label{rr:cospi}

To perform range reductino on $cos(\pi x)$, we use the property of
$cos(\pi x)$ that it is a periodic and even function. First, using the
property $cos(-\pi x) = cos(\pi x)$, we decompose the input $x$ into
$x = s \times |x|$ where s is the sign of the input. The function
decomposes to $cos(\pi x) = cos(\pi |x|)$.

Second, we use the properties $cos(\pi (x + 2z)) = cos(\pi x)$ where
$z$ is an integer and $cos(\pi (x + 2z + 1)) = -cos(\pi x)$. We
decompose $|x|$ into $|x| = i + t$ where $i$ is an integer and $t \in
[0, 1)$ is the fractional p art, i.e. $t = |x| - i$. More formally, we
can define $t$ and $i$ as,
\[
  i = \lfloor |x| \rfloor, \quad\quad t = |x| - i
\]

If $i$ is an even integer, then $cos(\pi (t + i)) = cos(\pi t)$ (from
the property $cos(\pi (x + 2z)) = cos(\pi x)$). If $i$ is an odd
integer, then $cos(\pi (t + i)) = -cos(\pi t)$ (from the property
$cos(\pi (x + 2z + 1)) = -cos(\pi x)$). Thus, we can decompose
$cos(\pi x)$ into,
\[
  cos(\pi x) = cos(\pi |x|) = (-1)^{i \: (mod \: 2)} \times cos(\pi t)
\]
where $i \: (mod \: 2)$ is the modulus operation in base 2.

Third, we use the property $cos(\pi t) = -cos(\pi (1 - t))$ for $0.5 <
t < 1.0$ and decompose $t$ to,
\[
  x' = 
  \begin{cases}
    1 - t & \text{if } 0.5 < t < 1.0 \\
    t & otherwise
  \end{cases}
\]
Since we perform the subtraction only when $0.5 < t < 1.0$, $1 - t$
can be computed exactly due to Sterbenz Lemma. Consequently,
$cos(\pi x)$ function decomposes to,
\[
  cos(\pi x) =
  \begin{cases}
    -1 \times (-1)^{i \: (mod \: 2)} \times cos(\pi x') & \text{if }
    0.5 < t < 1.0 \\
    (-1)^{i \: (mod \: 2)} \times cos(\pi x') & \text{otherwise}
  \end{cases}
\]
The above decomposition reduces the input domain to $x' \in [0, 0.5]$.

In summary, we decompose the input $x$ into $s \times (i + t)$ where
$s$ is the sign of the input, $i$ is an integer, and $t \in [0, 1)$ is
the fractional part of $|x|$, \ie $|x| = i + t$. The range reduction
function $x' = RR(x)$, the output compensation function $y = OC(y',
x)$, and the function we need to approximate, $y' = g(x')$ are as
follows:
\begin{align}
  RR(x) &=
  \begin{cases}
    1 - t & \text{if } 0.5 < t < 1.0 \\
    t & otherwise
  \end{cases} \nonumber \\
  OC(y', x) &=
  \begin{cases}
    -1 \times (-1)^{i \: (mod \: 2)} \times y' & \text{if }
    0.5 < t < 1.0 \\
    (-1)^{i \: (mod \: 2)} \times y' & \text{otherwise}
  \end{cases} \nonumber \\
  g(x') &= cos(\pi x') \nonumber
\end{align}
With this range reduction technique, we need to approximate $cos(\pi
x')$ for the reduced input domain $x' \in [0, 0.5]$.

\section{Details on Bfloat16 Functions}
\label{apx:bfloat16}

In this section, we explain the bfloat16 functions in \tool. More
specifically, we describe the special cases, the range reduction and
output compensation function, the function we must approximate, and
the polynomials we generated for each bfloat16 math library function
in \tool.

\subsection{$ln(x)$ for Bfloat16}
The elementary function $ln(x)$ is defined over
the input domain $(0, \infty)$. There are three classes of
special case inputs:
\[
  \text{Special case of } ln(x) =
  \begin{cases}
    -\infty & \text{if } x = 0 \\
    \infty & \text{if } x = \infty \\
    NaN & \text{if } x < 0 \text{ or } x = NaN
  \end{cases}
\]

We use the range reduction technique described in
Appendix~\ref{rr:log}. For $ln(x)$, the range reduction function
($x' = RR(x)$), the output compensation function ($y = OC(y', x)$),
and the function to approximate ($y' = g(x')$) can be summarized as
follows:
\[
  RR(x) = \frac{t - 1}{t + 1} \quad\quad
  OC(y', x) = \frac{y' + m}{log_2(e)} \quad\quad
  g(x') = log_2 \left( \frac{1 + x'}{1- x'} \right)
\]
The value $t$ is the fractional value represented by the mantissa of
the input $x$ and $m$ is the exponent, \ie $x = t \times 2^m$. With
this range reduction technique, we need to approximate $g(x')$ for
$x' \in [0, \frac{1}{3})$.

To approximate $g(x')$, we use a $7^{th}$ degree odd polynomial
$P(x) = c_1x + c_3x^3 + c_5x^5 + c_7x^7$ with the coefficients,
\begin{align}
  c_1 & = 2.885102725620722008414986703428439795970916748046875\nonumber\\
  c_3 & = 0.9749438269300123582894457285874523222446441650390625\nonumber\\
  c_5 & =
        0.391172520217394070751737444879836402833461761474609375\nonumber\\
  c_7 & = 1.2722152807088404902202682933420874178409576416015625\nonumber
\end{align}

\subsection{$log_2(x)$ for Bfloat16}
The elementary function $log_2(x)$ is defined over the input domain
$(0, \infty)$. There are three classes of special case inputs:
\[
  \text{Special case of } log_2(x) =
  \begin{cases}
    -\infty & \text{if } x = 0 \\
    \infty & \text{if } x = \infty \\
    NaN & \text{if } x < 0 \text{ or } x = NaN
  \end{cases}
\]

We use the range reduction technique described in
Appendix~\ref{rr:log}. For $log_2(x)$, the range reduction function
($x' = RR(x)$), the output compensation function ($y = OC(y', x)$),
and the function to approximate ($y' = g(x')$) can be summarized as
follows:
\[
  RR(x) = \frac{t - 1}{t + 1} \quad\quad
  OC(y', x) = y' + m \quad\quad
  g(x') = log_2 \left( \frac{1 + x'}{1- x'} \right)
\]
The value $t$ is the fractional value represented by the mantissa of
the input $x$ and $m$ is the exponent, \ie $x = t \times 2^m$. With
this range reduction technique, we need to approximate $g(x')$ for
$x' \in [0, \frac{1}{3})$.

To approximate $g(x')$, we use a $5^{th}$ degree odd polynomial
$P(x) = c_1x + c_3x^3 + c_5x^5$ with the coefficients,
\begin{align}
  c_1 & = 2.885725930059220178947043677908368408679962158203125\nonumber\\
  c_3 & = 0.9477394346709135941608792563783936202526092529296875\nonumber\\
  c_5 & =
        0.7307375337145580740383365991874597966670989990234375\nonumber
\end{align}

\subsection{$log_{10}(x)$ for Bfloat16}
The elementary function $log_{10}(x)$ is defined over the input domain
$(0, \infty)$. There are three classes of special case inputs:
\[
  \text{Special case of } log_{10}(x) =
  \begin{cases}
    -\infty & \text{if } x = 0 \\
    \infty & \text{if } x = \infty \\
    NaN & \text{if } x < 0 \text{ or } x = NaN
  \end{cases}
\]

We use the range reduction technique described in
Appendix~\ref{rr:log}. For $log_{10}(x)$, the range reduction function
($x' = RR(x)$), the output compensation function ($y = OC(y', x)$),
and the function to approximate ($y' = g(x')$) can be summarized as
follows:
\[
  RR(x) = \frac{t - 1}{t + 1} \quad\quad
  OC(y', x) = \frac{y' + m}{log_2(10)} \quad\quad
  g(x') = log_2 \left( \frac{1 + x'}{1- x'} \right)
\]
The value $t$ is the fractional value represented by the mantissa of
the input $x$ and $m$ is the exponent, \ie $x = t \times 2^m$. With
this range reduction technique, we need to approximate $g(x')$ for
$x' \in [0, \frac{1}{3})$.

To approximate $g(x')$, we use a $5^{th}$ degree odd polynomial
$P(x) = c_1x + c_3x^3 + c_5x^5$ with the coefficients,
\begin{align}
  c_1 & = 2.88545942229525831379532974096946418285369873046875\nonumber\\
  c_3 & = 0.956484867363945223672772044665180146694183349609375\nonumber\\
  c_5 & =
        0.6710954935542725596775426311069168150424957275390625\nonumber
\end{align}

\subsection{$e^{x}$ for Bfloat16}
The elementary function $e^{x}$ is defined over the input domain $(-\infty,
\infty)$. There are four classes of special case inputs:
\[
  \text{Special case of } e^{x} =
  \begin{cases}
    0.0 & \text{if } x \leq -93.0 \\
    1.0 & \text{if } -1.953125 \times 10^{-3} \leq x \leq 3.890991
    \times 10^{-3} \\
    \infty & \text{if } x \geq 89.0 \\
    NaN & \text{if } x = NaN
  \end{cases}
\]

We use the range reduction technique described in
Appendix~\ref{rr:exp}. For $e^{x}$, the range reduction function
$x' = RR(x)$, output compensation function $y = OC(y', x)$, and the
function we have to approximate to approximate $y' = g(x')$ is
summarized below:
\[
  RR(x) = x log_2(e) - \lfloor x log_2(e) \rfloor\quad\quad
  OC(y', x) = y'2^{\lfloor x log_2(e) \rfloor} \quad\quad
  g(x') = 2^{x'}
\]
where $\lfloor x \rfloor$ is a floor function that rounds down $x$ to
an integer. With this range reduction technique, we need to
approximate $2^{x'}$ for $x' \in [0, 1)$.

To approximate $2^{x'}$, we use a $4^{th}$ degree polynomial $P(x) = c_0 + c_1x +
c_2x^2 + c_3x^3 + c_4x^4$ with the coefficients,
\begin{align}
  c_0 &=
        1.0000095976211798021182630691328085958957672119140625\nonumber
  \\
  c_1 &=
        0.69279247181322956006255253669223748147487640380859375\nonumber
  \\
  c_2 &=
        0.242560224581628236517616414857911877334117889404296875\nonumber
  \\
  c_3 &=
        5.014719237694532927296364732683287002146244049072265625\times
        10^{-2}\nonumber\\
  c_4 &=
        1.45139853027161404297462610202273936010897159576416015625\times
        10^{-2}\nonumber
\end{align}

\subsection{$2^{x}$ for Bfloat16}
The elementary function $2^{x}$ is defined over the input domain $(-\infty,
\infty)$. There are four classes of special case inputs:
\[
  \text{Special case of } 2^{x} =
  \begin{cases}
    0.0 & \text{if } x \leq -134.0 \\
    1.0 & \text{if } -2.8076171875 \times 10^{-3} \leq x \leq 2.8076171875
    \times 10^{-3} \\
    \infty & \text{if } x \geq 128.0 \\
    NaN & \text{if } x = NaN
  \end{cases}
\]

We use the range reduction technique described in
Appendix~\ref{rr:exp}. For $e^{x}$, the range reduction function
$x' = RR(x)$, output compensation function $y = OC(y', x)$, and the
function we have to approximate to approximate $y' = g(x')$ is
summarized below:
\[
  RR(x) = x - \lfloor x \rfloor\quad\quad
  OC(y', x) = y'2^{\lfloor x \rfloor} \quad\quad
  g(x') = 2^{x'}
\]
where $\lfloor x \rfloor$ is a floor function that rounds down $x$ to
an integer. With this range reduction technique, we need to
approximate $2^{x'}$ for $x' \in [0, 1)$.

To approximate $2^{x'}$, we use a $4^{th}$ degree polynomial $P(x) = c_0 + c_1x +
c_2x^2 + c_3x^3 + c_4x^4$ with the coefficients,
\begin{align}
  c_0 &=
        1.0000091388165410766220020377659238874912261962890625\nonumber
  \\
  c_1 &=
        0.69265463004053107187729665383812971413135528564453125\nonumber
  \\
  c_2 &=
        0.2437159431324379121885925769674940966069698333740234375\nonumber
  \\
  c_3 &=
        4.8046547014740259573528646797058172523975372314453125\times
        10^{-2}\nonumber\\
  c_4 &=
        1.557767964117490015751865684023869107477366924285888671875\times
        10^{-2}\nonumber
\end{align}

\subsection{$10^{x}$ for Bfloat16}
The elementary function $10^{x}$ is defined over the input domain $(-\infty,
\infty)$. There are four classes of special case inputs:
\[
  \text{Special case of } 10^{x} =
  \begin{cases}
    0.0 & \text{if } x \leq -40.5 \\
    1.0 & \text{if } -8.4686279296875 \times 10^{-4} \leq x \leq
    1.68609619140625 \times 10^{-3} \\
    \infty & \text{if } x \geq 38.75 \\
    NaN & \text{if } x = NaN
  \end{cases}
\]

\textbf{Range reduction.}  We use the range reduction technique
described in Appendix~\ref{rr:exp}. The range reduction function
$x' = RR(x)$, output compensation function $y = OC(y', x)$, and the
function we have to approximate to approximate $y' = g(x')$ is
summarized below:
\[
  RR(x) = x log_2(10) - \lfloor x log_2(10) \rfloor\quad\quad
  OC(y', x) = y'2^{\lfloor x log_2(10) \rfloor} \quad\quad
  g(x') = 2^{x'}
\]
where $\lfloor x \rfloor$ is a floor function that rounds down $x$ to
an integer. With this range reduction technique, we need to
approximate $2^{x'}$ for $x' \in [0, 1)$.

To approximate $2^{x'}$, we use a $4^{th}$ degree polynomial $P(x) = c_0 + c_1x +
c_2x^2 + c_3x^3 + c_4x^4$ with the coefficients,
\begin{align}
  c_0 &=
        1.0000778485054981903346060789772309362888336181640625\nonumber
  \\
  c_1 &=
        0.69179740083422547325397999884444288909435272216796875\nonumber
  \\
  c_2 &=
        0.2459833280009494360651700617381720803678035736083984375\nonumber
  \\
  c_3 &=
        4.5758952998196537886865797872815164737403392791748046875\times
        10^{-2}\nonumber\\
  c_4 &=
        1.63907658064124488184187811157244141213595867156982421875\times
        10^{-2}\nonumber
\end{align}

\subsection{$\sqrt{x}$ for Bfloat16}
The elementary function $\sqrt{x}$ is defined over the input domain $[0,
\infty)$. There are three classes of special case inputs:
\[
  \text{Special case of } \sqrt{x} =
  \begin{cases}
    0.0 & \text{if } x = 0.0 \\
    \infty & \text{if } x = \infty\\
    NaN & \text{if } x < 0 \text{ or } x = NaN
  \end{cases}
\]

We use the range reduction technique described in
Appendix~\ref{rr:sqrt}. The range reduction function $x' = RR(x)$, the
output compensation function $y = OC(y', x)$ and the function we have
to approximate $y' = g(x')$ can be summarized as follows:
\[
  RR(x) = x'\quad\quad
  OC(y', x) = y'2^{\frac{m}{2}}\quad\quad
  g(x') = \sqrt{x}
\]
where $x'$ is a value in $[1, 4)$ and $m$ is an even integer such that
$x = x' \times 2^m$ for the input $x$. With this range reduction
technique, we need to approximate $\sqrt{x'}$ for $x' \in [1, 4)$.

To approximate $\sqrt{x'}$, we use a $4^{th}$ degree polynomial
$P(x) = c_0 + c_1x + c_2x^2 + c_3x^3 + c_4x^4$ with the coefficients,
\begin{align}
  c_0 &=
        0.37202139260816802224240973373525775969028472900390625\nonumber
  \\
  c_1 &=
        0.7923315194006106398916244870633818209171295166015625\nonumber
  \\
  c_2 &=
        -0.199230719933062794257949690290843136608600616455078125\nonumber
  \\
  c_3 &=
        3.800384608453956369888970812098705209791660308837890625\times
        10^{-2}\nonumber\\
  c_4 &=
        -3.0848915765425755954043385287377532222308218479156494140625\times
        10^{-3}\nonumber
\end{align}

\subsection{$\sqrt[3]{x}$ for Bfloat16}
The elementary function $\sqrt[3]{x}$ is defined over the input domain $(-\infty,
\infty)$. There are four classes of special case inputs:
\[
  \text{Special case of } \sqrt[3]{x} =
  \begin{cases}
    0.0 & \text{if } x = 0.0 \\
    \infty & \text{if } x = \infty\\
    -\infty & \text{if } x = -\infty\\
    NaN & \text{if } x = NaN
  \end{cases}
\]

We use the range reduction technique described in
Appendix~\ref{rr:cbrt}. The range reduction function $x' = RR(x)$, the
output compensation function $y = OC(y', x)$ and the function we have
to approximate $y' = g(x')$ can be summarized as follows:
\[
  RR(x) = x'\quad\quad
  OC(y', x) = s \times y'2^{\frac{m}{3}}\quad\quad
  g(x') = \sqrt[3]{x}
\]
where $s$ is the sign of the input $x$, $x'$ is a value in $[1, 8)$
and $m$ is integer multiple of 3 such that
$x = s \times x' \times 2^m$. With this range reduction technique, we
need to approximate $\sqrt[3]{x'}$ for $x' \in [1, 8)$.

To approximate $\sqrt[3]{x'}$, we use a $6^{th}$ degree polynomial
$P(x) = c_0 + c_1x + c_2x^2 + c_3x^3 + c_4x^4 + c_5x^5 + c_6x^6$ with
the coefficients,
\begin{align}
  c_0 &= 0.56860957346246798760347473944420926272869110107421875 \nonumber\\
  c_1 &= 0.5752913905623990853399618572439067065715789794921875 \nonumber\\
  c_2 &= -0.180364291120356845521399691278929822146892547607421875 \nonumber\\
  c_3 &= 4.3868412288261666998057108912689727731049060821533203125
        \times 10^{-2}\nonumber\\
  c_4 &= -6.5208421736825845915763721905022975988686084747314453125
        \times 10^{-3}\nonumber\\
  c_5 &= 5.241080546145838146843143334763226448558270931243896484375
        \times 10^{-4}\nonumber\\
  c_6 &=
        -1.7372029717703960593165601888898663673899136483669281005859375
        \times 10^{-5}\nonumber
\end{align}

\subsection{$sin(\pi x)$ for Bfloat16}
The elementary function $sin(\pi x)$ is defined over the input domain $(-\infty,
\infty)$. There are two classes of special case inputs:
\[
  \text{Special case of } \sin(\pi x) =
  \begin{cases}
    NaN & \text{if } x = NaN \text{ or } x = \pm \infty \\
    0 & \text{if } x \geq 256 \text{ or } x \leq -256
  \end{cases}
\]

We use the range reduction technique described in
Appendix~\ref{rr:sinpi}. We decompose the input $x$ into
$x = s \times (i + t)$ where $s$ is the sign of the input, $i$ is an
integer, and $t \in [0, 1)$ is the fractional part of $|x|$, \ie
$|x| = i + t$. The range reduction function $x' = RR(x)$, the output
compensation function $y = OC(y', x)$, and the function we need to
approximate, $y' = g(x')$ can be summarized as follows:
\[
  RR(x) =
  \begin{cases}
    1 - t & \text{if } 0.5 < t < 1.0 \\
    t & \text{otherwise}
  \end{cases},
  \quad
  OC(y', x) =
  \begin{cases}
    s \times y' & \text{if } i \equiv 0 \: (mod\:2) \\
    -s \times y' & \text{if } i \equiv 1 \: (mod\:2) \\
  \end{cases},
  \quad
  g(x') = sin(\pi x)
\]
With this range reduction technique, we need to approximate
$sin(\pi x')$ for $x' \in [0, 0.5]$.

The $sin(\pi x)$ function exhibit a linear-like behavior around $x = 0$.
To approximate $sin(\pi x)$, we use a piecewise
polynomial consisting of two polynomials:
\[
  P(x) =
  \begin{cases}
    c_1x & \text{if } x' \leq 6.011962890625 \times 10^{-3} \\
    d_1x + d_3x^3 + d_5 x^5 + d_7 x^7 & \text{otherwise}
  \end{cases}
\]
with the coefficients,
\begin{align}
  c_1 &= 3.14159292035398163278614447335712611675262451171875
        \nonumber\\
  d_1 &= 3.141515487020253072358855206402949988842010498046875
        \nonumber\\
  d_3 &= -5.16405991738943459523625278961844742298126220703125
        \nonumber\\
  d_5 &= 2.50692180297728217652775128954090178012847900390625
        \nonumber\\
  d_7 &= -0.443008519856437021910977591687696985900402069091796875
        \nonumber
\end{align}

\subsection{$cos(\pi x)$ for Bfloat16}
The elementary function $cos(\pi x)$ is defined over the input domain $(-\infty,
\infty)$. There are two classes of special case inputs:
\[
  \text{Special case of } \cos(\pi x) =
  \begin{cases}
    NaN & \text{if } x = NaN \text{ or } x = \pm \infty \\
    1 & \text{if } x \geq 256 \text{ or } x \leq -256
  \end{cases}
\]

We use the range reduction technique described in
Appendix~\ref{rr:cospi}. We decompose the input $x$ into $x = s \times
(i + t)$ where s is the sign of the input, $i$ is an integer, and $t
\in [0, 1)$ is the fractional part of $|x|$, \ie $|x| = i + t$. The
range reduction function $x' = RR(x)$, the output compensation
function $y = OC(y', x)$, and the function we need to approximate $y'
= g(x')$ can be summarized as follows:
\begin{align}
  RR(x) &=
  \begin{cases}
    1 - t & \text{if } 0.5 < t < 1.0 \\
    t & otherwise
  \end{cases} \nonumber \\
  OC(y', x) &=
  \begin{cases}
    -1 \times (-1)^{i \: (mod \: 2)} \times y' & \text{if }
    0.5 < t < 1.0 \\
    (-1)^{i \: (mod \: 2)} \times y' & \text{otherwise}
  \end{cases} \nonumber \\
  g(x') &= cos(\pi x') \nonumber
\end{align}
With this range reduction technique, we need to approximate
$cos(\pi x')$ for $x' \in [0, 0.5]$.

The $cos(\pi x)$ function exhibit a linear property around $x = 0$. To
approximate $cos(\pi x)$, we use the piecewise
polynomial:
\[
  P(x) =
  \begin{cases}
    c_0 & \text{if } x' \leq 1.98974609375 \times 10^{-2} \\
    d_0 + d_2x^2 + d_4 x^4 + d_6 x^6 & \text{if } 1.98974609375 \times 10^{-2} < x' <
    0.5 \\
    0.0 & \text{if } x' = 0.5
  \end{cases}
\]
with the coefficients,
\begin{align}
  c_0 &= 1.00390625\nonumber\\
  d_0 &= 0.99997996859304827399483883709763176739215850830078125
        \nonumber\\
  d_2 &= -4.9324802047472200428046562592498958110809326171875
        \nonumber\\
  d_4 &= 4.02150995405109146219047033810056746006011962890625
        \nonumber\\
  d_6 &= -1.1640167711700171171429474270553328096866607666015625
        \nonumber
\end{align}

\section{Details on Posit16 functions}
\label{apx:posit16}
In this section, we explain the posit16 functions in \tool. More
specifically, we describe the special cases, the range reduction
technique we used, how we split the reduced domain, and the
polynomials we generated for each posit16 math library function in
\tool.

\subsection{$ln(x)$ for Posit16}
The elementary function $ln(x)$ is defined over the input domain
$(0, \infty)$. There are two classes of special case inputs:
\[
  \text{Special case of } ln(x) =
  \begin{cases}
    NaR & \text{if } x \leq 0 \\
    NaR & \text{if } x = NaR
  \end{cases}
\]

We use the range reduction technique described in
Appendix~\ref{rr:log}. For $ln(x)$, the range reduction function
($x' = RR(x)$), the output compensation function ($y = OC(y', x)$),
and the function to approximate ($y' = g(x')$) can be summarized as
follows:
\[
  RR(x) = \frac{t - 1}{t + 1} \quad\quad
  OC(y', x) = \frac{y' + m}{log_2(e)} \quad\quad
  g(x') = log_2 \left( \frac{1 + x'}{1- x'} \right)
\]
The value $t$ is the fractional value represented by the mantissa of
the input $x$ and $m$ is the exponent, \ie $x = t \times 2^m$. With
this range reduction technique, we need to approximate $g(x')$ for
$x' \in [0, \frac{1}{3})$.

To approximate $g(x')$, we use a $9^{th}$ degree odd polynomial
$P(x) = c_1x + c_3x^3 + c_5x^5 + c_7x^7 + c_9x^9$ with the coefficients,
\begin{align}
  c_1 & = 2.8853901812623536926594169926829636096954345703125\nonumber\\
  c_3 & = 0.96177728824005104257821585633791983127593994140625\nonumber\\
  c_5 & =
        0.57802192858859535729010303839459083974361419677734375\nonumber\\
  c_7 & =
        0.39449243216490248453709455134230665862560272216796875\nonumber\\
  c_9 & =
        0.45254178489671204044242358577321283519268035888671875\nonumber
\end{align}

\subsection{$log_2(x)$ for Posit16}
The elementary function $log_2(x)$ is defined over the input domain
$(0, \infty)$. There are two classes of special case inputs:
\[
  \text{Special case of } log_2(x) =
  \begin{cases}
    NaR & \text{if } x \leq 0 \\
    NaR & \text{if } x = NaR
  \end{cases}
\]

We use the range reduction technique described in
Appendix~\ref{rr:log}. For $log_2(x)$, the range reduction function
($x' = RR(x)$), the output compensation function ($y = OC(y', x)$),
and the function to approximate ($y' = g(x')$) can be summarized as
follows:
\[
  RR(x) = \frac{t - 1}{t + 1} \quad\quad
  OC(y', x) = y' + m \quad\quad
  g(x') = log_2 \left( \frac{1 + x'}{1- x'} \right)
\]
The value $t$ is the fractional value represented by the mantissa of
the input $x$ and $m$ is the exponent, \ie $x = t \times 2^m$. With
this range reduction technique, we need to approximate $g(x')$ for
$x' \in [0, \frac{1}{3})$.

To approximate $g(x')$, we use a $5^{th}$ degree odd polynomial
$P(x) = c_1x + c_3x^3 + c_5x^5 + c_7x^7 + c_9x^9$ with the coefficients,
\begin{align}
  c_1 & = 2.88539115994917327867597123258747160434722900390625\nonumber\\
  c_3 & = 0.9616405555684151007511673014960251748561859130859375\nonumber\\
  c_5 & =
        0.5827497609092706642996972732362337410449981689453125\nonumber\\
  c_7 & =
        0.336729567454907396939489672149647958576679229736328125\nonumber\\
  c_9 & =
        0.68022527114824737903830964569351635873317718505859375\nonumber
\end{align}

\subsection{$log_{10}(x)$ for Posit16}
The elementary function $log_{10}(x)$ is defined over the input domain
$(0, \infty)$. There are two classes of special case inputs:
\[
  \text{Special case of } log_2(x) =
  \begin{cases}
    NaR & \text{if } x \leq 0 \\
    NaR & \text{if } x = NaR
  \end{cases}
\]

We use the range reduction technique described in
Appendix~\ref{rr:log}. For $log_{10}(x)$, the range reduction function
($x' = RR(x)$), the output compensation function ($y = OC(y', x)$),
and the function to approximate ($y' = g(x')$) can be summarized as
follows:
\[
  RR(x) = \frac{t - 1}{t + 1} \quad\quad
  OC(y', x) = \frac{y' + m}{log_2(10)} \quad\quad
  g(x') = log_2 \left( \frac{1 + x'}{1- x'} \right)
\]
The value $t$ is the fractional value represented by the mantissa of
the input $x$ and $m$ is the exponent, \ie $x = t \times 2^m$. With
this range reduction technique, we need to approximate $g(x')$ for
$x' \in [0, \frac{1}{3})$.

We approximate $g(x')$ with a $9^{th}$ degree odd polynomial
$P(x) = c_1x + c_3x^3 + c_5x^5 + c_7x^7 + c_9x^9$ with the coefficients,
\begin{align}
  c_1 & = 2.885392110906054075059046226670034229755401611328125\nonumber\\
  c_3 & = 0.96158476800643521986700079651200212538242340087890625\nonumber\\
  c_5 & =
        0.5837756666515827586039222296676598489284515380859375\nonumber\\
  c_7 & =
        0.330016589138880600540204568460467271506786346435546875\nonumber\\
  c_9 & =
        0.691650888349585102332639507949352264404296875\nonumber
\end{align}

\subsection{$\sqrt{x}$ for Posit16}
The elementary function $\sqrt{x}$ is defined over the input domain $[0,
\infty)$. There are two classes of special case inputs:
\[
  \text{Special case of } \sqrt{x} =
  \begin{cases}
    0.0 & \text{if } x = 0.0 \\
    NaR & \text{if } x < 0 \text{ or } x = NaR
  \end{cases}
\]

We use the range reduction technique described in
Appendix~\ref{rr:sqrt}. The range reduction function $x' = RR(x)$, the
output compensation function $y = OC(y', x)$ and the function we have
to approximate $y' = g(x')$ can be summarized as follows:
\[
  RR(x) = x'\quad\quad
  OC(y', x) = y'2^{\frac{m}{2}}\quad\quad
  g(x') = \sqrt{x}
\]
where $x'$ is a value in $[1, 4)$ and $m$ is an even integer such that
$x = x' \times 2^m$ for the input $x$. With this range reduction
technique, we need to approximate $\sqrt{x'}$ for $x' \in [1, 4)$.

To approximate $\sqrt{x'}$, we use a piecewise polynomial consisting of
two $6^{th}$ degree polynomials
\[
  P(x) =
  \begin{cases}
    c_0 + c_1x + c_2x^2 + c_3x^3 + c_4x^4 + c_5 x^5 + c_6 x^6 &
    \text{if } x' \leq 2.14599609375 \\
    d_0 + d_1x + d_2x^2 + d_3x^3 + d_4x^4 + d_5 x^5 + d_6 x^6 & \text{otherwise}
  \end{cases}
\]
with the coefficients,
\begin{align}
  c_0 &= 0.269593592709484630720595532693550921976566314697265625 \nonumber\\
  c_1 &= 1.129000996028148851024752730154432356357574462890625 \nonumber\\
  c_2 &= -0.64843843364755160418866353211342357099056243896484375 \nonumber\\
  c_3 &= 0.3530868073027828568655195340397767722606658935546875 \nonumber\\
  c_4 &= -0.127171841275129426929169085269677452743053436279296875 \nonumber\\
  c_5 &=
        2.62819293630375920567399106175798806361854076385498046875 \times
        10^{-2}\nonumber\\
  c_6 &= -2.3530402643644897538177662710268123191781342029571533203125
        \times 10^{-3}\nonumber\\
  d_0 &= 0.409156298855834987815427439272752963006496429443359375 \nonumber\\
  d_1 &= 0.74313621747255442784307888359762728214263916015625 \nonumber\\
  d_2 &= -0.1842527001546831189049413524116971530020236968994140625 \nonumber\\
  d_3 &= 4.305139568476913647376846938641392625868320465087890625
        \times 10^{-2}\nonumber\\
  d_4 &= -6.6014424010839810319506426594671211205422878265380859375
        \times 10^{-3}\nonumber\\
  d_5 &= 5.74776888286255573622118841825567869818769395351409912109375
        \times 10^{-4}\nonumber\\
  d_6 &=
        -2.1374405303079146056961790112183052769978530704975128173828125
        \times 10^{-5}\nonumber
\end{align}

\subsection{$sin(\pi x)$ for Posit16}
The elementary function $sin(\pi x)$ is defined over the input domain $(-\infty,
\infty)$. There is one special case input:
\[
  \sin(\pi x) = NaR \: \text{if } x = NaR
\]

We use the range reduction technique described in
Appendix~\ref{rr:sinpi}. We decompose the input $x$ into
$x = s \times (i + t)$ where $s$ is the sign of the input, $i$ is an
integer, and $t \in [0, 1)$ is the fractional part of $|x|$, \ie
$|x| = i + t$. The range reduction function $x' = RR(x)$, the output
compensation function $y = OC(y', x)$, and the function we need to
approximate, $y' = g(x')$ can be summarized as follows:
\[
  RR(x) =
  \begin{cases}
    1 - t & \text{if } 0.5 < t < 1.0 \\
    t & \text{otherwise}
  \end{cases},
  \quad
  OC(y', x) =
  \begin{cases}
    s \times y' & \text{if } i \equiv 0 \: (mod\:2) \\
    -s \times y' & \text{if } i \equiv 1 \: (mod\:2) \\
  \end{cases},
  \quad
  g(x') = sin(\pi x)
\]
With this range reduction technique, we need to approximate
$sin(\pi x')$ for $x' \in [0, 0.5]$.

The $sin(\pi x)$ function exhibit a linear-like behavior around $x = 0$.
To approximate $sin(\pi x)$, we use a piecewise
polynomial consisting of two polynomials:
\[
  P(x) =
  \begin{cases}
    c_1x & \text{if } x' \leq 2.52532958984375 \times 10^{-3} \\
    d_1x + d_3x^3 + d_5 x^5 + d_7 x^7 + d_9x^9 & \text{otherwise}
  \end{cases}
\]
with the coefficients,
\begin{align}
  c_1 &= 3.141577060931899811890843920991756021976470947265625
        \nonumber\\
  d_1 &= 3.141593069399674309494230328709818422794342041015625
        \nonumber\\
  d_3 &= -5.1677486367595673044661452877335250377655029296875
        \nonumber\\
  d_5 &= 2.55098424541712009983029929571785032749176025390625
        \nonumber\\
  d_7 &= -0.60547119473342603246379667325527407228946685791015625
        \nonumber\\
  d_9 &= 9.47599641221426869375221713198698125779628753662109375
        \times 10^{-2} \nonumber
\end{align}

\subsection{$cos(\pi x)$ for Posit16}
The elementary function $cos(\pi x)$ is defined over the input domain $(-\infty,
\infty)$. There are two classes of special case inputs:
\[
   \cos(\pi x) = NaR \: \text{if } x = NaR
\]

We use the range reduction technique described in
Appendix~\ref{rr:cospi}. We decompose the input $x$ into $x = s \times
(i + t)$ where s is the sign of the input, $i$ is an integer, and $t
\in [0, 1)$ is the fractional part of $|x|$, \ie $|x| = i + t$. The
range reduction function $x' = RR(x)$, the output compensation
function $y = OC(y', x)$, and the function we need to approximate $y'
= g(x')$ can be summarized as follows:
\begin{align}
  RR(x) &=
  \begin{cases}
    1 - t & \text{if } 0.5 < t < 1.0 \\
    t & otherwise
  \end{cases} \nonumber \\
  OC(y', x) &=
  \begin{cases}
    -1 \times (-1)^{i \: (mod \: 2)} \times y' & \text{if }
    0.5 < t < 1.0 \\
    (-1)^{i \: (mod \: 2)} \times y' & \text{otherwise}
  \end{cases} \nonumber \\
  g(x') &= cos(\pi x') \nonumber
\end{align}
With this range reduction technique, we need to approximate
$cos(\pi x')$ for $x' \in [0, 0.5]$.

The $cos(\pi x)$ function exhibit a linear property around $x = 0$. To
approximate $cos(\pi x')$, we use the piecewise
polynomial:
\[
  P(x) =
  \begin{cases}
    c_0 & \text{if } x' \leq 3.509521484375 \times 10^{-3} \\
    d_0 + d_2x^2 + d_4 x^4 + d_6 x^6 + d_8 x^8 & \text{if } 3.509521484375 \times 10^{-3} < x' <
    0.5 \\
    0.0 & \text{if } x' = 0.5
  \end{cases}
\]
with the coefficients,
\begin{align}
  c_0 &= 1.0001220703125\nonumber\\
  d_0 &= 1.000000009410458634562246515997685492038726806640625
        \nonumber\\
  d_2 &= -4.93479863229652071510145106003619730472564697265625
        \nonumber\\
  d_4 &= 4.05853647916781223869975292473100125789642333984375
        \nonumber\\
  d_6 &= -1.3327362938689424343152722940430976450443267822265625
        \nonumber\\
  d_8 &= 0.2215338495769658688772096866159699857234954833984375
        \nonumber
\end{align}

\section{$log_2(x)$ for Float}
\label{apx:float}
Our $log_2{x}$ function for float in \tool is guaranteed to produce
the correct results for the inputs in $[1, 2)$. For all other inputs,
the result is undefined.

We use the range reduction technique described in
Appendix~\ref{rr:log} to ease the job of creating the polynomial. For
$log_2(x)$, the range reduction function ($x' = RR(x)$), the output
compensation function ($y = OC(y', x)$), and the function to
approximate ($y' = g(x')$) can be summarized as follows:
\[
  RR(x) = \frac{t - 1}{t + 1} \quad\quad
  OC(y', x) = y' \quad\quad
  g(x') = log_2 \left( \frac{1 + x'}{1- x'} \right)
\]
The value $t$ is the fractional value represented by the mantissa of
the input $x$ when $x$ is decomposed to $x = t \times 2^m$ with an
integer exponent $m$. With this range reduction technique, we need to
approximate $g(x')$ for $x' \in [0, \frac{1}{3})$.

To approximate $g(x')$, we use a $15^{th}$ degree odd polynomial,
\[
  P(x) = c_1x + c_3x^3 + c_5x^5 + c_7x^7 + c_9x^9 + c_{11}x^{11} +
  c_{13}x^{13} + c_{15}x^{15}
\]
with the coefficients,
\begin{align}
  c_1 &= 2.885390081777253090677959335152991116046905517578125\nonumber\\
  c_3 &= 0.9617966943187539197168689497630111873149871826171875\nonumber\\
  c_5 &= 0.57707795150992868826733683818019926548004150390625\nonumber\\
  c_7 &= 0.41220281933294511400589499316993169486522674560546875\nonumber\\
  c_9 &=
        0.320462962813822971330779409981914795935153961181640625\nonumber\\
  c_{11} &=
           0.264665103135787116439558985803159885108470916748046875\nonumber\\
  c_{13} &=
           0.1996122250113066820542684354222728870809078216552734375\nonumber\\
  c_{15} &=
           0.298387164422755202242143468538415618240833282470703125\nonumber
\end{align}

\end{document}